\documentclass[12pt]{article}
\pdfoutput=1

\usepackage{putex}

\usepackage[margin=.9in]{geometry}
\usepackage{graphicx}
\usepackage{dsfont}
\usepackage{amsmath}
\usepackage{amssymb}
\usepackage{array}
\usepackage{mathdots}
\usepackage{epstopdf}
\usepackage{enumerate}
\usepackage{lipsum}
\usepackage{cite}
\usepackage{youngtab}
\usepackage{tensor}
\usepackage{braket}
\usepackage[normalem]{ulem}
\usepackage{slashed}
\usepackage[aligntableaux=center]{ytableau}
\usepackage[utf8]{inputenc}
\usepackage[
colorlinks=true,
linkcolor=blue,
urlcolor=blue,
filecolor=black,
citecolor=red,
]{hyperref}
\usepackage{braket}
\usepackage{mathtools}
\usepackage{rotating}
\usepackage{tabularx}
\normalsize\usepackage{booktabs}
\usepackage{bm}
\usepackage{bbm}
\newcolumntype{Y}{>{\centering\arraybackslash}X}
\newcolumntype{R}{>{\raggedright\arraybackslash}X}
\newcolumntype{C}[1]{>{\centering\arraybackslash}p{#1}}
\usepackage{xcolor}
\definecolor{LightCyan}{rgb}{0.7,1,1}
\definecolor{Gray}{gray}{0.9}
\usepackage{xspace}
\usepackage{caption}
\usepackage{subcaption}
\usepackage{tikz}
\tikzset{
	vertical align/.style={
		baseline=-.5*(height("$+$")-depth("$+$"))
	}
}
\usetikzlibrary{calc}
\usetikzlibrary{snakes}
\usetikzlibrary{arrows}
\usetikzlibrary{patterns}
\usetikzlibrary{positioning}
\usetikzlibrary{decorations.pathmorphing}
\usetikzlibrary{ decorations.markings}
\tikzset{snake it/.style={decorate, decoration={snake, segment length=2mm, amplitude=1mm}}}
\usetikzlibrary{shapes.misc}
\tikzset{cross/.style={cross out, draw=black, minimum size=2*(#1-\pgflinewidth), inner sep=0pt, outer sep=0pt},
	cross/.default={1pt}}
\usetikzlibrary{shapes.geometric}
\definecolor{bleudefrance}{rgb}{0.19, 0.55, 0.91}
\definecolor{candyapplered}{rgb}{1.0, 0.03, 0.0}
\tikzstyle{arrowmid}[0.5]=[decoration={             
	markings, 
	mark=at position 0.5 with {\arrow[xshift=4.7pt]{triangle 45};}
},
postaction={decorate}]

\newcommand{\grp}[1]{\mathrm{#1}}

\newcommand{\grSU}{\grp{SU}}

\newcommand{\grUSp}{\grp{USp}}

\newcommand{\abs}[1]{\left\lvert #1 \right\rvert}

\newcommand {\be} {\begin {equation}}
\newcommand {\ee} {\end {equation}}

\newcommand {\bes} {\begin {equation*}}
\newcommand {\ees} {\end {equation*}}

\newcommand{\es}[2] {\begin{equation} \label{#1} \begin{split} #2 \end{split} \end{equation}}

\newcommand{\Z}{\mathbb{Z}}

\newcommand{\R}{\mathbb{R}}

\newcommand{\beq}{\begin{equation}}
	\newcommand{\eeq}{\end{equation}}

\def\ie{\begin{equation}\begin{aligned}}
		\def\fe{\end{aligned}\end{equation}}

\newcommand{\m}{\mu}
\newcommand{\n}{\nu}

\def\<{\langle}
\def\>{\rangle}

\def\beg{\begin{equation}\begin{gathered}}
		\def\eeg{\end{gathered}\end{equation}}

\def\bea{\begin{equation}\begin{aligned}}
		\def\eea{\end{aligned}\end{equation}}

\newcommand{\hol}{a}
\newcommand{\rt}{\alpha}
\newcommand{\Weyl}{\rho}

\newcommand{\psiL}{\chi}
\newcommand{\psiH}{\Psi}
\newcommand{\cghostHeavy}{C}

\def\Nc{N}

\numberwithin{equation}{section}

\begin{document}

\preprint{PUPT-2653}

\institution{PU}{Joseph Henry Laboratories, Princeton University, Princeton, NJ 08544, USA}
\institution{PCTS}{Princeton Center for Theoretical Science, Princeton University, Princeton, NJ 08544, USA}
\institution{IAS}{Institute for Advanced Study, Princeton, NJ 08540, USA}

\title{Small Circle Expansion for Adjoint QCD$_2$\\ with Periodic Boundary Conditions}

\authors{Ross Dempsey,\worksat{\PU} Igor R.~Klebanov,\worksat{\PU, \PCTS} Silviu S.~Pufu,\worksat{\PU, \PCTS, \IAS} Benjamin T. S\o gaard\worksat{\PU}}

\abstract{We study $1+1$-dimensional $\grSU(N)$ gauge theory coupled to one adjoint multiplet of Majorana fermions on a small spatial circle of circumference $L$.  Using periodic boundary conditions, we derive the effective action for the quantum mechanics of the holonomy and the fermion zero modes in perturbation theory up to order $(gL)^3$.  When the adjoint fermion mass-squared is tuned to $g^2 N/(2\pi)$, the effective action is found to be an example of supersymmetric quantum mechanics with a nontrivial superpotential. We separate the states into the $\mathbb{Z}_N$ center symmetry sectors (universes) labeled by $p=0, \ldots, N-1$ and show that in one of the sectors the supersymmetry is unbroken, while in the others it is broken spontaneously. These results give us new insights into the $(1,1)$ supersymmetry of adjoint QCD$_2$, which has previously been established using light-cone quantization.  When the adjoint mass is set to zero, our effective Hamiltonian does not depend on the fermions at all, so that there are $2^{N-1}$ degenerate sectors of the Hilbert space. This construction appears to provide an explicit realization of the extended symmetry of the massless model, where there are $2^{2N-2}$ operators that commute with the Hamiltonian. We also generalize our results to other gauge groups $G$, for which supersymmetry is found at the adjoint mass-squared  $g^2 h^\vee/(2\pi)$, where $h^\vee$ is the dual Coxeter number of $G$.
}
\date{June 2024}

\maketitle

\tableofcontents

\section{Introduction}

Adjoint QCD$_2$ is a $1+1$-dimensional gauge theory with non-Abelian gauge group $G$ and a multiplet of Majorana fermions in the adjoint representation. This model has been studied extensively in the case of $G = \grSU(\Nc)$, starting with \cite{Dalley:1992yy,Kutasov:1993gq,Bhanot:1993xp}. In particular, following the success of light-cone quantization in deriving the meson spectrum of the 't Hooft model \cite{tHooft:1974pnl}, which has its fermions in the fundamental representation, similar methods have been used to study the large-$\Nc$ limit of adjoint QCD$_2$. Here the spectrum is considerably more complicated, with an infinite number of Regge trajectories of ``gluinoballs," and yet the low-lying bound state masses can be determined numerically with good precision \cite{Bhanot:1993xp, Gross:1997mx, Katz:2013qua, Dempsey:2021xpf,Trittmann:2023dar}. 

There are some particularly interesting values of the fermion mass for which the spectrum reveals special properties of adjoint QCD$_2$. When the fermion mass $m = 0$, the lightest bound states are a fermion with squared mass of approximately $M_f^2 \approx 5.7\, g^2 \Nc/ (2 \pi)$, followed by a boson of squared mass $M_b^2 \approx 10.8\, g^2 \Nc/ (2 \pi)$ \cite{Bhanot:1993xp,Dempsey:2022uie}.\footnote{There is a factor of $2$ difference in the definition of $g^2$ between \cite{tHooft:1974pnl,Bhanot:1993xp,Dempsey:2022uie} and the convention used here, namely $g^2_\text{here} = 2 g^2_\text{there}$. The $g$ used here is the standard definition of Yang-Mills coupling.} The $m=0$ model has a mass gap but no confinement \cite{Gross:1995bp,Komargodski:2020mxz,Dempsey:2021xpf}, which makes it
qualitatively similar to the Schwinger model \cite{Schwinger:1962tp}. However, having a mass gap is atypical for 2D gauge theories with massless fermionic matter (see \cite{Delmastro:2021otj} for a list of gapped such models).  
When the fermion mass is tuned to $m = \pm m_\text{SUSY}$,\footnote{The two values $m = \pm m_\text{SUSY}$ are related to each other by the 
discrete chiral rotation $(\mathbb{Z}_2)_\chi$, which acts as $\psi\rightarrow \gamma^5\psi$.} with 
\begin{equation}
m_\text{SUSY} \equiv g \sqrt{\frac{N}{2\pi}}\,,
\end{equation} 
these lowest states have their masses adjusted such that they form a degenerate boson-fermion pair at mass $M_f^2 = M_b^2 \approx 25\,g^2\Nc/(2\pi)$.\footnote{The coefficients in these expressions show very weak dependence on $\Nc$ \cite{Antonuccio:1998uz,Dempsey:2022uie,Trittmann:2024jkf}, and the values quoted in the main text are accurate for any $\Nc$ ranging from $2$ to infinity.}
The equality of the boson and fermion masses follows from a (1,1) supersymmetry of adjoint QCD$_2$, whose
prior derivations rely in an essential way on light-cone quantization \cite{Kutasov:1993gq,Boorstein:1993nd,Popov:2022vud}. Indeed, while the original space-time action is not manifestly supersymmetric at $m = \pm m_\text{SUSY}$, in light-cone quantization one integrates out the gauge field and the right-handed fermions, obtaining an action purely in terms of the left-handed fermions.   One can then construct explicit supercharges $Q^+$ and $Q^-$ that anti-commute with each other and square to the light-cone momentum $P^+$ and to the light-cone Hamiltonian $P^-$, respectively, thus generating the $(1, 1)$ supersymmetry algebra. Our new results presented here provide strong evidence that the supersymmetry is also present on a spatial circle of any size, with the periodic boundary conditions for the fermions.

Although light-cone quantization has been very useful for studying adjoint QCD$_2$ and other 2D gauge theories \cite{Dalley:1992yy,Kutasov:1993gq,Bhanot:1993xp, Kutasov:1994xq,
Katz:2013qua, Dempsey:2021xpf,Dempsey:2022uie, Trittmann:2023dar}, it does have some significant limitations, at least in its current formulation. In particular, it is difficult to gain access to the different topological sectors of the theory. 
However, in adjoint QCD$_2$ with $\grSU(\Nc)$ gauge group there are $\Nc$ distinct topological sectors \cite{Witten:1978ka}, also referred to as universes \cite{Komargodski:2020mxz}. These are distinguished by a $\mathbb{Z}_{\Nc}$ one-form symmetry (which is the center symmetry of the $\grSU(N)$ gauge theory), whose generator has eigenvalue $e^{2\pi i p/\Nc}$, with $p = 0,1,\ldots,\Nc-1$, in the $p$th universe. (For the theory on a line, one can construct the $p$th universe by placing probe fermions of $\Nc$-ality $p$ and $\Nc-p$ at $x = \mp\infty$, respectively.) Furthermore, it is known from considerations of the topological line operators that when $m = 0$, each universe contains many degenerate vacua such that the total number of vacua is $2^{\Nc-1}$ \cite{Komargodski:2020mxz}. It is not yet known how to construct these vacua using light-cone quantization.

Much of the interesting physics of adjoint QCD$_2$ relies on the presence of the multiple universes. For instance, the $p$th string tension is the energy density of the ground state in the $p$th universe. At $m = 0$, the degeneracy of the vacua implies the theory is not confining \cite{Komargodski:2020mxz}, as was argued early on using other methods \cite{Gross:1995bp}. For any $m>0$, the $p\neq 0$ universes have nonzero energy density and the theory is confining. In particular, at $m = m_\text{SUSY}$, this nonzero energy density indicates the spontaneous breaking of supersymmetry in the $p\neq 0$ universes. We thus expect to see a massless Majorana fermion (i.e.~a Goldstino) in the spectrum of each nontrivial universe \cite{Dubovsky:2018dlk}.

Recently, new approaches to adjoint QCD$_2$ have been developed using lattice gauge theory methods, both in the Hamiltonian \cite{Dempsey:2023fvm} and Euclidean \cite{Bergner:2024ttq} formulations. The $\mathbb{Z}_{\Nc}$ one-form symmetry is preserved on the lattice, and thus the universes can be cleanly separated and studied. So far, the Hamiltonian approach has only been carried out explicitly for the $\grSU(2)$ theory. In this case, one finds exact degeneracy of the $p = 0$ and $p = 1$ universes at $m = 0$ (which in the special case of $\grSU(2)$ is also implied by the mixed anomalies of the theory \cite{Cherman:2019hbq,Dempsey:2023fvm}), and there is clear evidence for the presence of the Goldstino mode in the $p = 1$ universe at $m=m_\text{SUSY}$.

Motivated by the success of the lattice methods, in this paper we study the adjoint QCD$_2$ theory analytically on a small spatial circle with {\it periodic} boundary conditions for the adjoint fermions. 
These boundary conditions have been used extensively in studies of some higher-dimensional gauge theories compactified on a circle \cite{Unsal:2007jx,Poppitz:2012sw}. 
There have been previous studies of the small circle limit of adjoint QCD$_2$, but mostly with anti-periodic boundary conditions \cite{Lenz:1994du,Smilga:1994hc,Smilga:1996dn}.\footnote{See, however, \cite{Cherman:2019hbq}.}
In that case, the effective theory on a small circle has been found to be the quantum mechanics of the Cartan components of the holonomy of the gauge field, with a nontrivial potential resulting from integrating out the heavy fermion modes. There are important physical differences between periodic and anti-periodic boundary conditions. In particular, the anti-periodic case is relevant for studying the theory at finite temperature, as was first done in \cite{Kutasov:1993gq}, and a Hagedorn transition is expected to occur when the adjoint fermions are massive, leading to a deconfined phase for circle radii smaller than some $L_{\rm crit}$ \cite{Semenoff:1996xg}. In contrast, with periodic boundary conditions there is no deconfinement transition, and the theory remains confining down to arbitrarily small radius \cite{Cherman:2019hbq}.

There are also marked differences in the effective theory on a small circle. With periodic boundary conditions for the fermions, the effective quantum mechanics involves not just the Cartan components of the holonomy variable, but also the Cartan components of the zero modes of the fermion \cite{Cherman:2019hbq}.  Like in the anti-periodic case, the holonomy acquires a potential from integrating out the non-zero fermion modes. At leading order in $g L$, one can approximate this potential with that of an $(\Nc-1)$-dimensional bosonic harmonic oscillator of frequency $g \sqrt{\Nc / (2 \pi)}$, and the Cartan components of the fermion zero modes describe an $(\Nc-1)$-dimensional fermionic harmonic oscillator of frequency $|m|$.\footnote{For a general group $G$, the oscillators are $\text{rank}(G)$-dimensional, and the frequency of the bosonic oscillator is $g\sqrt{h^\vee/(2\pi)}$, where $h^\vee$ is the dual Coxeter number of $G$.  See Section~\ref{sec:LOtheory} for more details, including the definition of the dual Coxeter number.} The combined system becomes a model of supersymmetric quantum mechanics \cite{Witten:1981nf,Cooper:1982dm,Tong} when $m^2= g^2 \Nc / (2 \pi)$, which is the same value of $m^2$ where supersymmetry occurs in the light-cone treatment.  This is one of our main results. Furthermore, as we will show in Section~\ref{sec:SUNLeading}, for the $\grSU(\Nc)$ theory one can study how the states of the simple system of harmonic oscillators get split among the $\Nc$ universes. We find that at the supersymmetric mass, supersymmetry is unbroken in the trivial sector, $p=0$, while it is spontaneously broken in all others.

One can systematically compute corrections in $g L$ to the leading order harmonic oscillator picture presented above.  In Section~\ref{sec:effective}, we determine the action of the effective quantum mechanics theory up to order $(g L)^3$.  As we show in more detail in Section~\ref{sec:susy}, this action still takes a supersymmetric form when $m=\pm m_\text{SUSY}$.  This observation leads us to conjecture that, at these values of the mass, adjoint QCD$_2$ preserves supersymmetry when placed on a circle of any length using periodic boundary conditions for the fermions. This is consistent with, and extends, the supersymmetry found in the infinite length limit using the light-cone quantization \cite{Kutasov:1993gq,Bhanot:1993xp,Boorstein:1993nd,Popov:2022vud}.

Another remarkable feature of our small circle expansion is that, for $m=0$, the Hamiltonian does not contain the fermions at all, at least to the order that we have calculated. This provides a simple way to understand the large family of symmetry operators, some of them non-invertible, of adjoint QCD$_2$ on a circle with periodic boundary conditions \cite{Komargodski:2020mxz}. In particular, for gauge group $\grSU(\Nc)$ our results suggest the existence of $2^{\Nc-1}$ degenerate vacua and of $2^{2\Nc - 2}$ operators that commute with the Hamiltonian for any circle size $L$.    (For general gauge group $G$, we have $2^{\text{rank}(G)}$ degenerate vacua and $2^{2\,  \text{rank}(G)}$ such operators.)

The rest of this paper is organized as follows.  In Section~\ref{sec:SUNLeading} we start by describing the $\grSU(\Nc)$ adjoint QCD$_2$ theory on a small circle, emphasizing the supersymmetry at $m^2= g^2 \Nc / (2 \pi)$.  In Section~\ref{sec:LOtheory} we generalize this analysis to an arbitrary gauge group.   In Section~\ref{sec:effective} we go beyond the leading-order analysis and systematically compute corrections in $g L$.  In Section~\ref{Comments} we comment on the number of vacua and the operators that commute with the Hamiltonian at the massless point.   In Section~\ref{sec:susy} we specialize to $m=\pm m_\text{SUSY}$ and show that the effective theory from Section~\ref{sec:effective} can be written in a manifestly supersymmetric form.  We end with a discussion of our results in Section~\ref{sec:discussion}.  Several technical details can be found in the Appendices.

\section{Leading order effective theory for $\grSU(\Nc)$ gauge group}
\label{sec:SUNLeading}

Let us start by analyzing the $\grSU(\Nc)$ adjoint QCD$_2$ theory at leading order in the small circle expansion. We will generalize this analysis to an arbitrary simply-connected gauge group in Section~\ref{sec:LOtheory}.  

\subsection{Effective potential for the holonomy} 

The action governing the dynamics of the adjoint QCD$_2$ theory in Lorentzian signature is\footnote{We do not include the four-fermion operators considered in \cite{Cherman:2019hbq}. When they are not included, there is no danger of them getting generated by quantum effects \cite{Komargodski:2020mxz,Cherman:2024onj}.}
 \es{SLor}{
	S = \int dt\, dx \, \tr \left( - \frac{1}{2 g^2} F_{\mu\nu} F^{\mu\nu} + i \bar \psi \gamma^\mu D_\mu \psi - m \bar \psi \psi \right)  \,,
}
where the gauge covariant derivative and field strength are defined as $D_\mu = \partial_\mu - i[A_\mu,\cdot]$ and $F_{\mu\nu} = \partial_\mu A_\nu -\partial_\nu A_\mu-i[A_\mu,A_\nu]$, respectively, and $\tr$ denotes the trace in the fundamental representation.   To be concrete, we choose the convention $\gamma^0=\sigma_2$, $\gamma^1=-i\sigma_3$, which implies that $\eta^{\mu\nu}=\frac{1}{2}\{\gamma^\mu,\gamma^\nu\} = \operatorname{diag}(1,-1)$. This basis of $\gamma$-matrices is purely imaginary, and the Majorana condition is simply $\psi^*=\psi$, so that $\bar\psi= \psi^T\gamma^0$. The chirality matrix is $\gamma^5 = \gamma^0 \gamma^1 = \sigma_1$. 

When the length of the circle is small, we can perform a Kaluza-Klein reduction on the circle.  Most modes of the fermion and gauge field acquire masses of order $1/L$ and can be integrated out, yielding an effective action for the light modes, which carry energies of order $g$.  With periodic boundary conditions for the adjoint fermion, there are two types of light modes: the holonomy variable 
\es{HolVariableSUN}{
	\Omega = P \exp \left[ i \int_0^L dx\, A_x \right]\,,
} 
and the fermionic modes corresponding to the zero-modes of the covariant derivative operator $D_x$. To derive the effective action, we should carry out the path integral for all the heavy modes while keeping the light modes fixed. To leading order in $gL$, the only contribution from integrating out the heavy modes will be a potential for the holonomy. The effective potential was computed in \cite{Gross:1980br} and takes the form
\es{VeffSUN}{
	V_\text{eff}(\Omega) = \frac{1}{\pi L} \sum_{n=1}^\infty \frac{1}{n^2}\tr_\text{adj}(\Omega^n)\,,
}
where the trace is taken after writing the $\grSU(\Nc)$ group element $\Omega$ in the adjoint representation. A derivation of the above result is reviewed in Appendix \ref{app:1-loop_pot}.
 
The minimum of the potential \eqref{VeffSUN} was determined in \cite{Cherman:2019hbq}.  Since the solution to the minimization problem constitutes the first step in our small circle expansion, let us review it here.  One can pass to a gauge where in the fundamental representation $\Omega$ is a diagonal matrix written as
 \es{OmegaDiag}{
  \Omega = \diag \{ e^{i a_1} , e^{i a_2} \, \ldots, e^{i a_{\Nc}} \} \,, \qquad a_1 + a_2 + \cdots + a_{\Nc} = 0 \,.
 }
We can think of $\Omega$ as arising from the gauge field configuration where
 \es{GaugeDiag}{
  A_t = 0\,, \qquad A_x &= \frac{1}{L}  \diag \{ a_{1} , a_{2} , \ldots , a_{\Nc}  \}  \,.
 }  
Let us denote by ${\bm a}$ the $\Nc$-dimensional vector ${\bm a} = (a_1, a_2, \ldots, a_{\Nc})$;  we restrict it to the hyperplane $a_1 + a_2 + \cdots + a_{\Nc} = 0$ in order for $\Omega$ to be an $\grSU(\Nc)$ matrix, or equivalently for $A$ to be valued in the $\mathfrak{su}(\Nc)$ Lie algebra.  The form \eqref{OmegaDiag}--\eqref{GaugeDiag} does not fully fix the gauge, however.  Indeed, we can still perform gauge transformations that permute the $a_i$, or that shift $(a_j, a_k) \to (a_j + 2 \pi, a_k - 2 \pi)$ for any given $j \neq k$ while leaving all the other $a_i$ the same.  These gauge transformations can be used to bring ${\bm a}$ to a form where the $a_i$ are non-increasing with $i$, $a_1 \geq a_2 \geq \cdots \geq a_{\Nc}$, and the difference between any two $a_i$ does not exceed $2 \pi$ in absolute value.  In other words, $a_i - a_{j} \in [0, 2 \pi)$ for all $i<j$.  Another way of characterizing the space of such configurations is as a simplex with $\Nc$ vertices
 \es{Vk}{
  {\bm v}_k = \biggl( \underbrace{ \frac{2 \pi k}{\Nc}, \ldots, \frac{2 \pi k}{\Nc}}_{\Nc - k}, \underbrace{ \frac{2 \pi (k-\Nc)}{\Nc}, \ldots, \frac{2 \pi (k-\Nc)}{\Nc}}_{k}  \biggr)  \,,
  }
with $k = 0, 1, \ldots, \Nc-1$, that lies within the hyperplane $a_1 + a_2 + \cdots + a_{\Nc} = 0$.  We call this simplex the fundamental domain.  The vertices \eqref{Vk} are identified with each other under gauge transformations.

Noting that the adjoint trace is related to the trace in the fundamental representation via $\tr_\text{adj} \Omega^n =\abs{ \tr \Omega^n}^2 - 1$ for any group element $\Omega$, and evaluating the sum over $n$ in \eqref{VeffSUN}, we can write the effective potential as
 \es{VeffAgain}{
  V_\text{eff} =  \frac{1}{\pi L} \left[\sum_{i< j} \left( \text{Li}_2 (e^{i (a_i - a_j)}) + \text{Li}_2 (e^{i (a_j - a_i)}) \right) + (\Nc-1) \frac{\pi^2}{6}  \right] \,.
 }
Within the fundamental domain, since $a_i - a_j \in [0, 2 \pi)$ for $i<j$, we can use the identity $\text{Li}_2(e^{ix}) + \text{Li}_2(e^{-ix}) = - \frac{\pi^2}{6} + \frac 12 (x - \pi)^2$, which holds for all $x \in [0, 2 \pi]$.  A little bit of algebra then gives the further simplification
 \es{VeffFundSUN}{
  V_\text{eff} = \frac {\Nc}{2 \pi L} \sum_{i=1}^{\Nc} (a_i^2 + 2 \pi i a_i) + \text{const} \,. 
 }
Under the constraint $\sum_i a_i = 0$, the effective potential is minimized for
 \es{astar}{
  {\bm a}_* = \biggl( \pi \frac{\Nc -1 }{\Nc}, \pi \frac{\Nc -3 }{\Nc}, \ldots, -\pi \frac{\Nc -1 }{\Nc}  \biggr) \,.
 }
One can verify that ${\bm a}_*$ is the average of all the vertices \eqref{Vk} of the fundamental domain, ${\bm a}_* = \frac{1}{\Nc} \sum_{k=1}^{\Nc - 1} {\bm v}_k$, so ${\bm a}_*$ lies right at the center of the fundamental domain.

The situation here should be contrasted to the case of anti-periodic boundary conditions for the fermions, in which case the effective potential takes a form that is very similar to \eqref{VeffSUN}:
\es{VeffAP}{
	V_\text{eff}^\text{AP}(\Omega) = \frac{1}{\pi L}\Re\sum_{n=1}^\infty \frac{(-1)^n}{n^2}\tr_\text{adj}(\Omega^n)\,.
}
A similar calculation gives that in the fundamental domain
 \es{VeffAPAgain}{
  V_\text{eff}^\text{AP}= \frac{1}{2 \pi L} \sum_{i<j} \min \bigl(a_i - a_j, 2 \pi - (a_i - a_j) \bigr)^2 + \text{const} \,.
 }
To minimize this expression, we should choose $a_i - a_j = 0$ or $ 2\pi$ for all $i<j$, which is a condition that is obeyed only by the vertices of the fundamental domain.  (Recall that these vertices are identified with each other under gauge transformations.)   Thus, we can take
 \es{astarAP}{
  {\bm a}_*^{\text{AP}} = {\bm v}_1 = (0, 0, \ldots, 0) 
 }
as the minimizing configuration of the potential.\footnote{Even though the minima are equivalent one find $\Nc$ distinct ground states, one localized near each vertex $\bm v_k$.}  In both the periodic and anti-periodic cases, the potential for small fluctuations around the minimum is that of a harmonic oscillator.  

\subsection{Effective action at leading order}\label{sec:lo_action}

To obtain the full leading-order theory for the light modes in the Kaluza-Klein reduction on the circle, we need to add the classical action to the effective potential \eqref{VeffFundSUN}. In the diagonal gauge \eqref{OmegaDiag}, the zero-modes of $D_x$ are spatially constant fields valued in the Cartan, and so the light degrees of freedom for the holonomy and can be embedded in the full fields as
 \es{eq:embedSUN}{
	A_\mu(t,x) &= \frac{\delta_{\mu x}}{L}  \left[ \diag \{ a_{*1} , a_{*2} , \ldots , a_{*\Nc}  \}
	 + g\sqrt{L} \diag \{ q_1(t),  q_2(t), \ldots ,  q_{\Nc}(t) \} \right] \,, \\
	\psi(t,x) &=  \frac{1}{\sqrt{L}} \diag \{ \chi_1(t), \chi_2(t), \ldots , \chi_{\Nc}(t) \} \,,
 }
where the prefactors multiplying the fluctuations $q_i(t)$ and $\chi_i(t)$ were chosen for convenience.   The fluctuations $q_i(t)$ and $\chi_i(t)$ obey $\sum_i q_i = 0$ and $\sum_i \chi_i = 0$ in order for the $A_\mu$ and $\psi$ to be valued in the $\mathfrak{su}(\Nc)$ algebra.  We can use the notation ${\bm q}$ and ${\bm \chi}$ for the $\Nc$-component vectors whose components are $q_i$ and $\chi_i$.

 Inserting \eqref{eq:embedSUN} into the action \eqref{SLor} and adding in the effective potential \eqref{VeffFundSUN} yields the following leading order action
 \es{eq:SLOSUNvec}{
	S_\text{LO}= \int dt\, \left(\dot {\bm q}\cdot \dot {\bm q}
	 - \frac{\Nc g^2}{2\pi} {\bm q} \cdot {\bm q}
	 + i {\bm \chi}^T \cdot \dot {\bm \chi}
	 -m {\bm \chi}^T \cdot \gamma^0  {\bm \chi} \right) \,.
 }
Recall, however, that both ${\bm q}$ and ${\bm \chi}$ are constrained by the requirement that the sums of the components of each of these vectors vanishes.  These constraints can be solved with the help of the weight vectors for the fundamental representation of $\mathfrak{su}(\Nc)$.  The fundamental weight vectors of $\mathfrak{su}(\Nc)$ form a collection of coefficients $w_{ia}$, with $i = 1, \ldots, \Nc$, and $a = 1, \ldots \Nc-1$, that obey the properties
 \es{wProperties}{
  \sum_{i=1}^{\Nc} w_{ia} = 0 \,, \qquad \sum_{i=1}^{\Nc} w_{ia} w_{ib}  = \frac{\delta_{ab}}{2} \,, \qquad
   \sum_{a=1}^{\Nc-1} w_{ia} w_{ja} = \frac{1}{2} \left( \delta_{ij} - \frac{1}{\Nc} \right) \,.
 }
Conceptually, $w_{ia}$ represents the eigenvalue of the $i$th basis vector in $\R^{\Nc}$ under the $T^a$ Cartan element of $\mathfrak{su}(\Nc)$, where the Cartan elements are normalized such that $\tr (T^a T^b) = \delta^{ab} / 2$.  Explicit expressions for $w_{ia}$ can be found in any group theory textbook (see for instance Eq.~(13.9) of \cite{Georgi:1999wka}),\footnote{In our notation, Eq.~(13.9) of \cite{Georgi:1999wka} becomes $w_{ia} = \begin{cases} 0 & \text{if $a \leq i-2$} \,, \\
 - \frac{a}{\sqrt{2a(a + 1)}} & \text{if $a = i-1$} \,, \\
 \frac{1}{\sqrt{2a (a + 1)}} & \text{if $a = i$} \,.
 \end{cases}$} but we will not need them here.  While ${\bm q}$ and ${\bm \chi}$ were defined as constrained $\Nc$-component vectors, we can define $\vec{q}$ and $\vec{\chi}$ to be unconstrained $\Nc - 1$-component vectors
 \es{qchiVec}{
  \vec{q} = 2 \sum_{i=1}^{\Nc} \vec{w}_i q_i  \,, \qquad
   \vec{\chi} = 2 \sum_{i=1}^{\Nc} \vec{w}_i \chi_i  \,, \qquad
    \vec{w}_i \equiv \left(w_{i1}, w_{i2}, \ldots, w_{i(\Nc - 1)} \right) \,,
 }
with components denoted by $q_a$ and $\chi_a$, respectively.  Using the third equation in \eqref{wProperties} we can derive the inverse formulas
 \es{qchiVecInv}{
  q_i = \vec{q} \cdot \vec{w}_i \,, \qquad \chi_i = \vec{\chi} \cdot \vec{w}_i  \,.
 }
Plugging these expressions into \eqref{eq:SLOSUNvec} and using the second equation in \eqref{wProperties}, we find an equivalent form of the leading order action, now written in terms of the unconstrained $\vec{q}$ and~$\vec{\chi}$:
 \es{eq:SLOSUN}{
	S_\text{LO}= \int dt\, \left(\frac{1}{2} \dot{\vec q} \cdot \dot{\vec q}
	 - \frac{\Nc g^2}{4\pi} \vec{q} \cdot \vec{q}
	 + \frac{i}{2} \vec{\chi}^T \cdot \dot{\vec{\chi}}
	 - \frac{m}{2} \vec{\chi}^T \cdot \gamma^0  \vec{\chi} \right) \,.
 }

The expression \eqref{eq:SLOSUN} describes $\Nc-1$ bosonic and $\Nc - 1$ fermionic harmonic oscillators of frequencies $\omega_B =g \sqrt{\frac{\Nc}{2\pi}}$ and $\omega_F = \abs{m}$, respectively. This is immediately suggestive of emergent supersymmetry at $\omega_B=\omega_F$, where the fermionic oscillator gaps line up with their bosonic couterparts giving rise to Fermi/Bose degeneracy.

In addition to the dynamics given by \eqref{eq:SLOSUN}, there is a subtlety related to the parameterization \eqref{eq:embedSUN}. In \cite{Lenz:1994du}, it is shown that the coordinate change from $A_\mu$ to the parameterization \eqref{eq:embedSUN} in the Hilbert space formalism is accompanied by the Jacobian
\begin{align}\label{eq:jacSUN}
	J\left[{\bm a}_*+g\sqrt{L} {\bm q}\right] \propto \prod_{i<j} \sin^2\left(\frac{\left(a_{*i} + g\sqrt{L} q_i\right) - \left(a_{*j} + g\sqrt{L} q_j\right)}{2}\right)\,,
\end{align}
and for the Hamiltonian to be hermitian wave functions must vanish when the Jacobian does. This is reminiscent of transitioning from Cartesian to spherical coordinates in quantum mechanics, where one picks up a Jacobian $\sim r^2$ and the radial wave function must vanish at the origin.

The vanishing locus of the Jacobian \eqref{eq:jacSUN} coincides with the boundaries of the fundamental domain. Thus, to obtain a well-defined quantum mechanical problem, we need to impose Dirichlet boundary conditions on wavefunctions on the fundamental domain. This will not change the conclusions for periodic fermions, as the leading order effective potential \eqref{VeffFundSUN} confines the energy eigenfunctions to the interior of the fundamental domain and corrections from the boundary conditions are exponentially small in $(gL)^{-1}$. Had we instead imposed anti-periodic boundary conditions on the adjoint fermions, the minimum of the effective potential would have been in a corner of the fundamental domain and the Dirichlet boundary conditions would have severely changed the bosonic levels \cite{Lenz:1994du}.

\subsection{Effective Hamiltonian and symmetries}\label{sec:symmetries}

To study the leading order spectrum, let us pass from the action \eqref{eq:SLOSUN} to the Hamiltonian 
\begin{align}\label{eq:HLO}
	H_\text{LO}= \frac{1}{2} \vec p\cdot \vec p+\frac{\Nc g^2}{4\pi}\vec q\cdot \vec q+\frac{1}{2}m\vec\chi^T\cdot \gamma^0\vec\chi \,,
\end{align}
where the canonically conjugate momentum operators $\vec{p}$ obey $[q_a, p_b] = \delta_{ab}$, and we also have $\left\{\chi_a^\alpha, \chi_b^\beta\right\}=\delta^{\alpha\beta}\delta_{ab}$. It is straightforward to determine the spectrum of the Hamiltonian \eqref{eq:HLO}:  we have the tensor product of $\Nc-1$ (bosonic) harmonic oscillators with frequency $\omega_B = g \sqrt{\frac{\Nc}{2 \pi}}$ and $\Nc - 1$ two-level systems (i.e. fermionic harmonic oscillators), each with energies $\pm m/2$.

A nontrivial feature of this limit is the action of the various symmetries of the original adjoint QCD$_2$ theory on the spectrum mentioned above.  The internal symmetries (i.e.~not spacetime symmetries) of SU($\Nc$) adjoint QCD are \cite{Cherman:2019hbq} 
\begin{align}
	m\neq 0 :\qquad \begin{cases}
		\mathbb Z^{[1]}_2\times (\mathbb{Z}_2)_F\,,& \text{for $\Nc=2$} \,, \\
		\left[\mathbb{Z}_{\Nc}^{[1]}\rtimes(\mathbb Z_2)_C\right]\times(\mathbb Z_2)_F\,, & \text{for $\Nc>2$} \,,
	\end{cases}
\end{align}
for non-vanishing mass $m$ and
\begin{align}
	m= 0 :\qquad \begin{cases}
		\mathbb Z^{[1]}_2\times (\mathbb{Z}_2)_F\times (\mathbb{Z}_2)_\chi\,, & \text{for $\Nc=2$} \,, \\
		\left[\mathbb{Z}_{\Nc}^{[1]}\rtimes(\mathbb Z_2)_C\right]\times(\mathbb Z_2)_F\times (\mathbb{Z}_2)_\chi\,, & \text{for $\Nc>2$} \,,
	\end{cases}
\end{align}
in the massless case.  The various factors here act as follows.  The fermion parity $(\mathbb{Z}_2)_F$ acts as $\psi\rightarrow-\psi$, and it leaves $A_\mu$ invariant.  The discrete chiral rotation $(\mathbb{Z}_2)_\chi$ present as a symmetry only when $m=0$, acts as $\psi\rightarrow \gamma^5\psi$, and it also leaves $A_\mu$ invariant. The charge conjugation $(\mathbb{Z}_2)_C$ acts non-trivially only for $\Nc>2$, and it is implemented by $\psi\rightarrow \psi^T$ and $A_\mu \rightarrow -A_\mu^T$, where the transpose is taken in the fundamental representation. Finally, the center symmetry $\mathbb{Z}_{\Nc}^{[1]}$ acts on extended objects rather than on local fields, and, in particular, fundamental Wilson lines carry charge $e^{\frac{2 \pi i}{\Nc}}$ under it. Denoting the operators generating the symmetries $(\mathbb{Z}_2)_F$, $(\mathbb{Z}_2)_\chi$, $(\mathbb{Z}_2)_C$, and $\mathbb{Z}_{\Nc}^{[1]}$ as $\mathcal F$, $\mathcal V$, $\mathcal C$, and $\mathcal U$, respectively, we write down their algebra as derived for periodic fermions in \cite{Cherman:2019hbq}:
 \es{eq:sym_algebra}{
	\mathcal U^{\Nc}&=\mathds{1} \,, \qquad \mathcal C^2 = \mathcal F^2 = \mathds{1} \,, \\
	\mathcal U\mathcal C &= \mathcal C \mathcal{U}^{-1} \,, 
	 \qquad \mathcal U \mathcal F = \mathcal F \mathcal U \,, 
	 \qquad \mathcal F \mathcal C = \mathcal C \mathcal F \,, \\
	\mathcal V^2 &=\mathds 1 \,, 
	 \qquad \mathcal U \mathcal V = (-1)^{\Nc-1}\mathcal V \mathcal U \,, \\
	\mathcal F \mathcal V &= (-1)^{\Nc-1}\mathcal V \mathcal F \,, 
	\qquad \mathcal C \mathcal V = (-1)^{\frac{(\Nc-2)(\Nc-1)}{2}}\mathcal V \mathcal C \,.
 }
The $\Nc$-dependent signs that arise in the relations involving ${\cal V}$ represent mixed anomalies, which imply that the relations involving ${\cal V}$ are realized projectively on the Hilbert space.  While $(\Z_2)_\chi$ is a symmetry only at $m=0$, for $m \neq 0$ it acts by sending $m \to -m$.  In other words, if $H(m)$ is the Hamiltonian for adjoint mass $m$, we have 
 \es{VTransf}{
  {\cal V} H(m) {\cal V}^{-1} = H(-m) \,.
 }
This relation implies that the spectrum of $H(-m)$ is the same as that of $H(m)$, but the corresponding eigenstates of $H(m)$ and $H(-m)$ may have different ${\cal U}$, ${\cal F}$, and ${\cal C}$ quantum numbers, as dictated by \eqref{eq:sym_algebra}.

We will now discuss each of these symmetries in more detail and determine how they act on the effective degrees of freedom.

{\bf Center symmetry.}  Up to a gauge transformation, the generator of the $\Z_{\Nc}$ center symmetry of $\grSU(\Nc)$ acts by multiplying the holonomy group element $\Omega$ in \eqref{OmegaDiag} by $e^{2 \pi i / \Nc}$. Thus, under the action of the $\mathbb{Z}_N$ generator, the set of eigenvalues $\{ e^{i a_i}\}$ should get mapped to the new set $\{ e^{i a_i'} \} = \{ e^{\frac{2 \pi i}{\Nc}} e^{i a_i} \}$. However, determining the individual transformations $a_i \to a_i'$ is complicated by the fact that we chose to work in the gauge where the $a_i$ are ordered in decreasing order $a_1 \geq a_2 \geq \cdots \geq a_{\Nc}$ with no two $a_i$ being farther apart than $2 \pi$.  Nevertheless, it is not hard to check that the transformation
 \es{CenterSym}{
  a_1' &= a_2 + \frac{2\pi}{\Nc} \,, \qquad \qquad \qquad e^{i a_1'} = e^{\frac{2 \pi i}{\Nc}} e^{ i a_2} \,, \\
  a_2' &= a_3 + \frac{2\pi}{\Nc} \,, \qquad \qquad \qquad e^{i a_2'} = e^{\frac{2 \pi i}{\Nc}} e^{i a_3}  \,, \\
  &\ \vdots \\
  a_{\Nc-1}' &= a_{\Nc} + \frac{2\pi}{\Nc} \,, \qquad \qquad e^{i a_{\Nc-1}'} = e^{\frac{2 \pi i}{\Nc}} e^{i a_{\Nc}}  \,, \\
  a_{\Nc}' &= a_1 + \frac{2\pi}{\Nc} - 2 \pi \,, \qquad \quad e^{i a_{\Nc}'} = e^{\frac{2 \pi i}{\Nc}} e^{i a_1}  
 } 
obeys all the desired properties.

From \eqref{CenterSym} one can infer how the center symmetry acts on the fluctuations $q_i$ and $\chi_i$.  The $q_i$ are defined to be proportional to the differences between $a_i$ and the center of the fundamental domain in \eqref{astar}.  Using \eqref{astar} and \eqref{CenterSym} one can verify that $q_i = g^{-1/2}\sqrt{gL}(a_i - a_{*i})$ and $q_i' = g^{-1/2}\sqrt{gL}(a_i' - a_{*i})$ are related by 
 \es{qTransf}{
  q_i \quad \to \quad  q_i' = q_{i + 1}
 } 
for all $i$, with the cyclic identification $q_{\Nc+ 1}  = q_1$.  The geometric significance of this equation is that the $\Z_{\Nc}$ center symmetry acts in the effective theory as a $\Z_{\Nc}$ discrete rotation about the center of the fundamental domain.   As given in \eqref{CenterSym}, the action of the center symmetry generator on the holonomy $\Omega$ involves a multiplication by $e^{2 \pi i / \Nc}$ as well as a permutation of eigenvalues.  The latter is a gauge symmetry transformation that is needed in order to keep $a_i$ in the fundamental domain, and it should accompany the center symmetry transformation when acting on any operators.  In particular, the $\chi_i$ should also transform under the center symmetry as
 \es{chiTransf}{
  \chi_i \quad \to \quad  \chi_i' = \chi_{i + 1} \,.
 } 

In terms of the $\Nc-1$-component vectors $\vec{q}$ and $\vec{\chi}$, the transformations \eqref{qTransf} and \eqref{chiTransf} can be restated as follows.  Using \eqref{qchiVec}--\eqref{qchiVecInv}, we see, for instance, that $\vec{q}$ gets mapped under \eqref{qTransf} to $\vec{q}' = \sum_i 2 \vec{w}_i q_{i+1} = \sum_i 2 \vec{w}_i (\vec{w}_{i+1} \cdot \vec{q})$.  Thus, defining the $(\Nc-1) \times (\Nc-1)$ orthogonal matrix
 \es{UMatrix}{
  U \equiv \sum_{i=1}^{\Nc} 2 \vec{w}_i \vec{w}_{i+1}^T \,,
 }
we have that conjugation by the center symmetry generator ${\cal U}$ gives
 \es{eq:center_sym_action}{
	\mathcal{U}\,  \vec q\,  \mathcal{U}^{-1} = U \vec q \,, \qquad 
	\mathcal{U}\,  \vec p\,  \mathcal{U}^{-1} = U \vec p\,, \qquad 
	\mathcal{U}\,  \vec \chi\,  \mathcal{U}^{-1} = U\vec \chi \,,
 }
where we also included the transformation under the $\Z_{\Nc}$ generator of the canonically conjugate momentum $\vec{p}$.

{\bf Charge conjugation.}  Adjoint QCD also enjoys charge conjugation which, up to gauge transformations, acts on the 2d fields by $\psi\rightarrow \psi^T$ and $A_\mu \rightarrow - A_\mu ^T$, with the transpose being performed in the fundamental representation.  
As the light modes retained in the effective theory are valued in the Cartan, the transpose in the charge conjugation operation can be neglected.   However, a simple change $a_i \to -a_i$ would break the gauge condition $a_1 \geq a_2 \geq \cdots \geq a_{\Nc}$.  To preserve the gauge condition, we should additionally perform the gauge transformation that flips the order of the eigenvalues.  Thus, on $a_i$, charge conjugation should act by sending
 \es{aipCharge}{
  a_i \quad \to \quad a_i' = - a_{\Nc +1-i} \,.
 }
From \eqref{astar} and the fact that we should perform the gauge transformation that reverses the order of the $a_i$ on all quantities, this immediately implies that
 \es{qchiTransf}{
  q_i \quad &\to \quad  q_i' = - q_{\Nc + 1-i} \,, \\
   \chi_i \quad &\to \quad  \chi_i' = \chi_{\Nc + 1-i} \,.
 }
The transformation properties of the $\Nc-1$-component vectors $\vec{q}$ and $\vec{\chi}$, can be inferred as follows.  Using \eqref{qchiVec}--\eqref{qchiVecInv}, we see, for instance, that $\vec{q}$ gets mapped under \eqref{qchiTransf} to $\vec{q}' = -\sum_i 2 \vec{w}_i q_{\Nc + 1 - i} = -\sum_i 2 \vec{w}_i (\vec{w}_{\Nc + i - 1} \cdot \vec{q})$.  Similarly, $\chi$ gets mapped to $\vec{\chi}' = -\sum_i 2 \vec{w}_i \vec{w}_{\Nc + i - 1} \cdot \vec{\chi}$.   Thus, defining the $(\Nc-1) \times (\Nc-1)$ orthogonal matrix
 \es{CMatrix}{
  C \equiv \sum_{i=1}^{\Nc} 2 \vec{w}_i \vec{w}_{\Nc + 1 - i}^T \,,
 }
we find that the charge conjugation generator ${\cal C}$ acts by
 \es{eq:chargeconjugation_action}{
	\mathcal{C}\,  \vec q\,  \mathcal{C}^{-1} = -C \vec q \,, \qquad 
	\mathcal{C}\,  \vec p\,  \mathcal{C}^{-1} = C \vec p\,, \qquad 
	\mathcal{C}\,  \vec \chi\,  \mathcal{C}^{-1} = C \vec \chi \,.
 }

{\bf Fermion parity and chiral rotation.}  Both fermion parity ${\cal F}$ and the chiral rotation ${\cal V}$ act only on the fermions as $\psi \to -\psi$ and $\psi \to \gamma^5 \psi$, while leaving $A_\mu$ invariant.  Thus, we have 
 \es{FVAction}{
  {\cal F}\,  \vec{q}\, {\cal F}^{-1} = \vec{q} \,, \qquad {\cal F}\, \vec{\chi}\,  {\cal F}^{-1} = -\vec{\chi} \,, \\
  {\cal V}\, \vec{q}\, {\cal V}^{-1} = \vec{q} \,, \qquad {\cal V} \, \vec{\chi}\, {\cal V}^{-1} = \gamma^5 \vec{\chi} \,, 
 }
where in the last equation $\gamma^5$ acts on the spinor indices of each of the components of $\vec{\chi}$.

\subsection{Mode expansion basis that diagonalizes ${\cal U}$ and ${\cal F}$} 
\label{UFBasis}

As can be seen from \eqref{FVAction}, the fermion number ${\cal F}$ and the chiral rotation ${\cal V}$ act diagonally on the basis of fluctuations $\vec{q}$ and $\vec{\chi}$, but the center symmetry ${\cal U}$ and charge conjugation ${\cal C}$ non-trivially.  As can be seen from the algebra \eqref{eq:sym_algebra}, it should be possible to pass to another basis where ${\cal U}$ and ${\cal F}$ act diagonally, while ${\cal C}$ and ${\cal V}$ do not.  This basis will be useful for splitting up the states into universes and into fermions and bosons.

For the bosons, let us introduce the creation and annihilation operators $a^\dagger_p$ and $a_p$, respectively, with $p = 1, \ldots, \Nc - 1$, defined by 
 \es{aadaggerDef}{
    a_p &= \sqrt{\frac{\omega}{\Nc}}
	 \sum_{j=1}^{\Nc} e^{\frac{2 \pi i jp}{\Nc}}\vec w_j\cdot \left(\vec q+\frac{i}{\omega}\vec p\right) \,, \\
    a_p^\dagger &= \sqrt{\frac{\omega}{\Nc}}
	 \sum_{j=1}^{\Nc} e^{-\frac{2 \pi i jp}{\Nc}}\vec w_j\cdot \left(\vec q-\frac{i}{\omega}\vec p\right) \,,  
 }
with $\omega \equiv g\sqrt{\frac{\Nc}{2\pi}}$.  It is straightforward to check that $[a_p,a_{p'}^\dagger ] = \delta_{pp'}$.  The inverse relation can be written explicitly
  \es{eq:bose_modes}{
	\vec q &= \sqrt{\frac{1}{\omega \Nc}}
	   \sum_{j,p}e^{-\frac{2 \pi i pj}{\Nc}}\vec w_j (a_p+a_{-p}^\dagger) \,, \\ 
	\vec p &= i\sqrt{\frac{\omega }{\Nc}}
	  \sum_{j,p}e^{\frac{2 \pi i pj}{\Nc}}\vec w_j (a_p^\dagger -a_{-p}) \,,
 }
with $p$ being identified modulo $\Nc$.   The fermion field can be similarly decomposed
 \es{eq:fermi_modes}{
	\vec \chi &= \frac{1}{\sqrt{\Nc}}
	    \sum_{j,p} e^{-\frac{2 \pi i pj}{\Nc}}\vec  w_j
	     \left[\begin{pmatrix}1 \\ i \end{pmatrix} c_p
	     +\begin{pmatrix} 1\\-i \end{pmatrix}c_{-p}^\dagger\right] \,, \\
         c_p &=\frac{1}{\sqrt{\Nc}}\sum_{j=1}^{\Nc}e^{\frac{2 \pi i jp}{\Nc}}
             \begin{pmatrix} 1 \\ i \end{pmatrix}^\dagger\vec \chi \cdot \vec w_j \,,
 }
where the fermionic creation and annihilation operators obey $\{c_p,c_{p'}^\dagger\}=\delta_{pp'}$. 

The symmetry generators ${\cal U}$, ${\cal C}$, ${\cal F}$, ${\cal V}$ act on the creation operators by
 \es{UCFVAction}{
  \mathcal{U}\,  a_p^\dagger \,  \mathcal{U}^{-1} &= e^{\frac{2\pi i p}{\Nc}}a_p^\dagger \,, 
    \qquad \quad\ \  \mathcal{U}\,  c_p^\dagger\,  \mathcal{U}^{-1}  = e^{\frac{2 \pi i p}{\Nc}} c_p^\dagger \,, \\
  \mathcal{F}\, a_p^\dagger\, \mathcal{F}^{-1} &= a_p^\dagger \,, 
    \qquad \quad \ \qquad\mathcal{F}\, c_p^\dagger \, \mathcal{F}^{-1} = -c_p^\dagger \,, \\  
  \mathcal{C} \, a_p^\dagger \,  \mathcal{C}^{-1} &= -e^{-\frac{2\pi i p}{\Nc}}a_{-p}^\dagger \,, 
   \qquad  \mathcal{C} \, c_p^\dagger \,  \mathcal{C}^{-1} = e^{-\frac{2\pi i p}{\Nc}}c_{-p}^\dagger   \,, \\
  \mathcal{V}\,  a_p^\dagger\, \mathcal{V}^{-1} &= a^\dagger_{p}\,,
    \qquad \qquad \quad \ \  \mathcal{V}\,  c_p^\dagger \, \mathcal{V}^{-1}=ic_{-p}  \,.
 }
The action on the annihilation operators can be found by conjugation of the relations in \eqref{UCFVAction}.  

In terms of the creation and annihilation operators, the leading order Hamiltonian \eqref{eq:HLO} is
 \es{eq:HLO_wb}{
	H_\text{LO}=\sum_{p=1}^{\Nc-1}\biggl[ \omega \left( a^\dagger_p a_p + \frac 12 \right) +m \left( c^\dagger_pc_p - \frac 12 \right) \biggr] \,.
 }
One can check that, as expected, ${\cal U}$, ${\cal F}$, ${\cal C}$ commute with $H_\text{LO}$, while ${\cal V}$ commutes with $H_\text{LO}$ only when $m=0$.  (More generally, ${\cal V} H_{\text{LO}} {\cal V}^{-1} = H_{\text{LO}} \big|_{m \to -m}$.)  At the supersymmetric points $m = \pm\omega$, the Hamiltonian \eqref{eq:HLO_wb} can be written as a square of a supercharge
 \es{QSquared}{
  H_\text{LO} \bigg|_{m=\omega} = \frac{1}{2} \{Q, Q^\dagger\}  \,,
 }
with $Q = \sqrt{2 \omega} \sum_{p=1}^{\Nc-1} a_p c_p^\dagger$ for $m = \omega$ and $Q = \sqrt{2 \omega} \sum_{p=1}^{\Nc-1} a_p c_{-p}$ for $m = -\omega$.

\subsection{Universes and leading order spectrum}

We can construct the spectrum of the Hamiltonian by acting on the Fock vacuum $\ket{0}$, which is annihilated by all $a_p$ and $c_p$, with the creation operators $a_p^\dagger$ and $c_p^\dagger$.  Since per \eqref{UCFVAction} these creation operators have definite ${\cal U}$ and ${\cal F}$ charges, the states constructed this way will be simultaneous eigenstates of $H_\text{LO}$, ${\cal U}$, and ${\cal F}$.   Let us assume that the Fock vacuum is invariant under ${\cal U}$ and ${\cal F}$:
 \es{UFVacuum}{
	\mathcal U \ket{0} = \ket{0} \,,
	 \qquad \mathcal F \ket 0 = \ket 0 \,.
 }

At the supersymmetric values $m = \pm \omega$, center symmetry and the supersymmetry commute $[\mathcal U,Q]=0$ which implies that we have a genuine supersymmetry where each universe transforms into itself.  One can further compute the Witten index ${\cal I}_p \equiv \tr_p [(-1)^F]$ in the $p$th universe as follows.  Consider the weighted sums of the ${\cal I}_p$ graded by $e^\frac{2 \pi i p z}{N}$,
 \es{calIz}{
  {\cal I}(z) \equiv \sum_{p=0}^{N-1} e^\frac{2 \pi i p z}{N} {\cal I}_p \,,
 } 
where $z \in \{0, 1, \ldots, N-1\}$ is a parameter.  (${\cal I}(z)$ is the discrete Fourier transform of ${\cal I}_p$.)  The refined index ${\cal I}(z)$ can be identified with ${\cal I}(z) = \tr[(-1)^F\mathcal U^z e^{-\beta H}]$, where the factor of $\mathcal U^z$ gives the grading $e^\frac{2 \pi i p z}{N}$, and $e^{-\beta H}$ serves as a regulator at intermediate steps.  The refined index ${\cal I}(z)$ is independent of $\beta$ because $H$ is $Q$-exact.  The refined index can be computed as a product of bosonic and fermionic contributions, ${\cal I}(z) = {\cal I}_B(z) {\cal I}_F(z)$.  The bosonic contribution is itself a product over the $p$ harmonic oscillators.  For the $p$th harmonic oscillator, the state $(a_p^\dagger)^n \ket{0}$ has $H = \omega (n + \frac 12)$ and ${\cal U} = e^{2 \pi i p n /N}$, where the latter follows from the fact that the vacuum has $\Z_N$ charge $0$ and $a_p^\dagger$ has $\Z_N$ charge $p$.  Summing over all the states and taking the product over all $p$, we obtain 
 \es{GotIB}{
  {\cal I}_B(z) = \prod_{p=1}^{\Nc-1} \left[ \sum_{n=0}^\infty e^{-\beta\omega(n+\frac 12)+\frac{2\pi i p z n}{\Nc}} \right]
   = \prod_{p=1}^{N-1} \frac{1}{e^{\frac{\beta \omega}{2}}-e^{-\frac{\beta \omega}{2}+\frac{2 \pi i z p}{\Nc}}} \,.
 }
The contribution of the fermionic oscillators is also a product over the $p$ fermionic oscillators, where each oscillator contributes $e^{\frac{\beta m}{2}}-e^{-\frac{\beta m}{2}+\frac{2 \pi i z p}{\Nc}}$, where the first term corresponds to $\ket{0}$ and the second to $c_p^\dagger \ket{0}$.  Thus,
 \es{GotIF}{
  {\cal I}_F(z) = \prod_{p=1}^{\Nc-1} \left[ e^{\frac{\beta m}{2}}-e^{-\frac{\beta m}{2}+\frac{2 \pi i z p}{\Nc}} \right] \,.
 }
It is clear from \eqref{GotIB} and \eqref{GotIF} that when $m = \omega$ the refined index ${\cal I}(z) = 1$.  Comparing this expression with \eqref{calIz}, we conclude that
 \es{IperUniverse}{
  {\cal I}_p = \begin{cases} 1\,, & \text{for $p=0$} \,, \\
  0 \,, & \text{for $p = 1, \ldots, N-1$} \,.
  \end{cases}
 } 
As we will now see in examples, this is consistent with the fact that supersymmetry is spontaneously broken in all the non-trivial universes, and only in the $p=0$ universe is supersymmetry unbroken.  When $m=-\omega$, one can show that ${\cal I}(z) = (-1)^{(N-1) z}$, which implies that \eqref{IperUniverse} also holds when $N$ is odd, but when $N$ is even we have
 \es{IperUniverseNegMassEvenN}{
  {\cal I}_p = \begin{cases} -1\,, & \text{for $p=\frac{N}{2}$} \,, \\
  0 \,, & \text{for $p \neq \frac{N}{2}$} \,.
  \end{cases}
 } 
The supersymmetry now is unbroken only in the $p = N/2$ universe and unbroken in the others.  The relation between the indices at $m = \pm \omega$ is consistent with the algebra \eqref{eq:sym_algebra} as well as the fact that the axial symmetry transformation changes the sign of $m$.

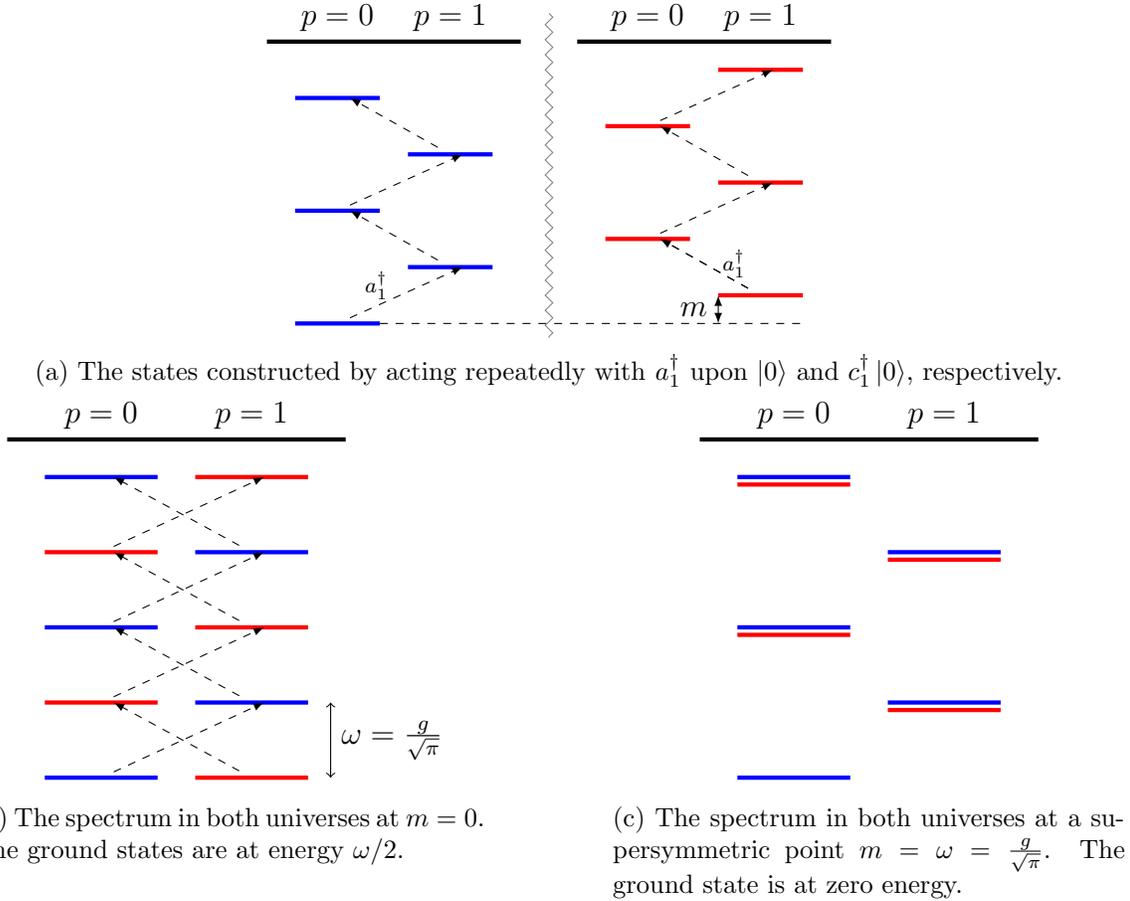
\begin{figure}
	\centering
	\begin{subfigure}[t]{\linewidth}
		\centering
		\begin{tikzpicture}[scale=.75]
			\foreach \n in {0,2,4} {
				\draw[ultra thick,blue] (0,\n) -- node (l\n) {} (1.5,\n);
			};
			\draw[dashed] (1.5,0) -- ++(7.5,0);
			\foreach \n in {1,3} {
				\draw[ultra thick,blue] (2,\n) -- node (r\n) {} (3.5,\n);
			};
			\foreach \n in {0} {
				\pgfmathsetmacro{\m}{\n+1}
				\draw[-latex,black,dashed] (l\n) -- node[left,shift={(-.1,.1)}] {\scriptsize $a_1^\dagger$} (r\m);
			};
			\foreach \n in {2} {
				\pgfmathsetmacro{\m}{\n+1}
				\draw[-latex,black,dashed] (l\n) -- (r\m);
			};
			\foreach \n in {1,3} {
				\pgfmathsetmacro{\m}{\n+1}
				\draw[-latex,black,dashed] (r\n) -- (l\m);
			};
			\draw[ultra thick,black] (-.5,5) -- (4,5);
			\node[above] at (.75,5) {$p = 0$};
			\node[above] at (2.75,5) {$p = 1$};
			
			\begin{scope}[xshift=5.5cm]
			\draw[gray,decoration = {zigzag,segment length = 2mm, amplitude = .5mm}, decorate] (-1, 5.5) -- (-1,-.25);
			\begin{scope}[yshift=.5cm]
			\foreach \n in {0,2,4} {
				\draw[ultra thick,red] (2,\n) -- node (r\n) {} (3.5,\n);
			};
			\draw[latex-latex] (2,0) -- node[left] {$m$} (2,-.5);
			\foreach \n in {1,3} {
				\draw[ultra thick,red] (0,\n) -- node (l\n) {} (1.5,\n);
			};
			\foreach \n in {0} {
				\pgfmathsetmacro{\m}{\n+1}
				\draw[-latex,black,dashed] (r\n) -- node[right,shift={(.1,0)}] {\scriptsize $a_1^\dagger$} (l\m);
			};
			\foreach \n in {0,2} {
				\pgfmathsetmacro{\m}{\n+1}
				\draw[-latex,dashed] (r\n) -- (l\m);
			};
			\foreach \n in {1,3} {
				\pgfmathsetmacro{\m}{\n+1}
				\draw[-latex,dashed] (l\n) -- (r\m);
			};
			\end{scope}
			\draw[ultra thick,black] (-.5,5) -- (4,5);
			\node[above] at (.75,5) {$p = 0$};
			\node[above] at (2.75,5) {$p = 1$};
			\end{scope}
		\end{tikzpicture}
		\caption{The states constructed by acting repeatedly with $a^\dagger_1$ upon $\ket{0}$ and $c^\dagger_1\ket{0}$, respectively.}
	\end{subfigure}\\
	\begin{subfigure}[t]{0.4\linewidth}
		\centering
		\begin{tikzpicture}
			\foreach \n in {0,2,4} {
				\draw[ultra thick,blue] (0,\n) -- node (l\n) {} (1.5,\n);
				\draw[ultra thick,red] (2,\n) -- node (r\n) {} (3.5,\n);
			};
			\foreach \n in {1,3} {
				\draw[ultra thick,red] (0,\n) -- node (l\n) {} (1.5,\n);
				\draw[ultra thick,blue] (2,\n) -- node (r\n) {} (3.5,\n);
			};
			\foreach \n in {0,2} {
				\pgfmathsetmacro{\m}{\n+1}
				\draw[-latex,dashed] (l\n) -- (r\m);
				\draw[-latex,dashed] (r\n) -- (l\m);
			};
			\foreach \n in {1,3} {
				\pgfmathsetmacro{\m}{\n+1}
				\draw[-latex,dashed] (l\n) -- (r\m);
				\draw[-latex,dashed] (r\n) -- (l\m);
			};
			\draw[ultra thick,black] (-.5,4.5) -- (4,4.5);
			\node[above] at (.75,4.5) {$p = 0$};
			\node[above] at (2.75,4.5) {$p = 1$};
			
			\draw[<->] (3.8,0) -- node[right] {$\omega = \frac{g}{\sqrt{\pi}}$} (3.8,1);
		\end{tikzpicture}
		\caption{The spectrum in both universes at $m = 0$. The ground states are at energy $\omega/2$.}
		\label{fig:su2_nonsusy}
	\end{subfigure}%
	\hspace{.1\linewidth}%
	\begin{subfigure}[t]{0.4\linewidth}
		\centering
		\begin{tikzpicture}
			\foreach \n in {0,2} {
				\draw[ultra thick,blue] (0,\n) -- node (l\n) {} (1.5,\n);
				\draw[ultra thick,red] (2,{\n+.9}) -- node (r\n) {} (3.5,{\n+.9});
			};
			\foreach \n in {1,3} {
				\draw[ultra thick,red] (0,{\n+.9}) -- node (l\n) {} (1.5,{\n+.9});
				\draw[ultra thick,blue] (2,\n) -- node (r\n) {} (3.5,\n);
			};
			\draw[ultra thick,blue] (0,4) -- node (l4) {} (1.5,4);
			\draw[ultra thick,black] (-.5,4.5) -- (4,4.5);
			\node[above] at (.75,4.5) {$p = 0$};
			\node[above] at (2.75,4.5) {$p = 1$};
		\end{tikzpicture}
		\caption{The spectrum in both universes at a supersymmetric point $m = \omega = \frac{g}{\sqrt{\pi}}$. The ground state is at zero energy.}
		\label{fig:su2_susy}
	\end{subfigure}
	\caption{The spectrum for the leading-order $\grSU(2)$ theory where blue and red lines signify bosonic and fermionic states, respectively. The dashed lines correspond to excitations created by $a^\dagger_1$ acting upon $\ket{0}$ and $c^\dagger_1 \ket 0$.}
\end{figure}

In the examples below we focus only on the supersymmetric point at $m = \omega$, since, as we just mentioned, the spectrum at $m = -\omega$ can be obtained from \eqref{eq:sym_algebra}--\eqref{VTransf}.  For the $\grSU(2)$ theory, the supersymmetry at $m=\omega$ manifests in a particularly simple way. There is a single bosonic creation operator $a^\dagger_1$, and a single fermionic creation operator $c^\dagger_1$. Starting from either $\ket{0}$ and $c^\dagger_1\ket{0}$, we can act repeatedly with $a^\dagger_1$, generating two ladders of states which each alternate between the $p = 0$ and $p = 1$ universes. These ladders are depicted with $m = 0$ in figure \ref{fig:su2_nonsusy}. When we set $m = \omega$, the fermionic states are shifted up one rung in the ladder, which leads to a supersymmetric spectrum as shown in Figure \ref{fig:su2_susy}. In particular, in the $p=0$ universe the supersymmetry is unbroken. There is a unique bosonic ground state at zero energy, and then all excited states exhibit a boson-fermion degeneracy, so that the Witten index equals $1$. 
In the $p=1$ universe the supersymmetry is spontaneously broken, and so the $p = 1$ ground state also exhibits boson-fermion degeneracy, with the fermionic ground state interpreted as a goldstino mode. In this universe, the Witten index clearly vanishes.

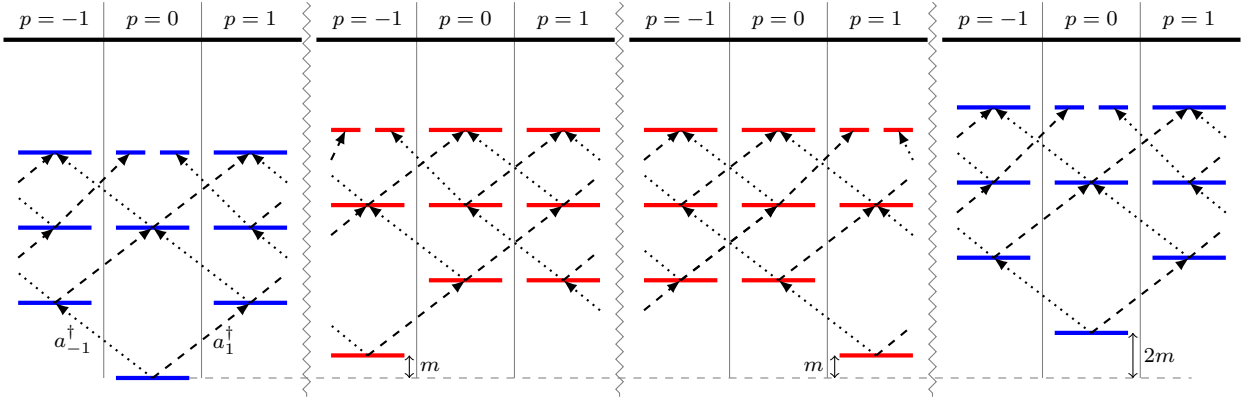
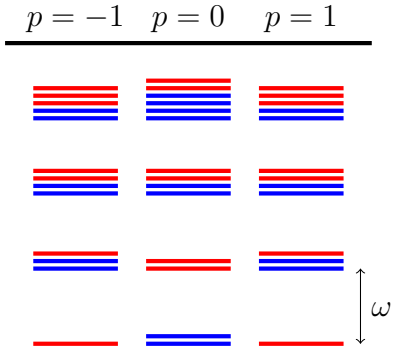
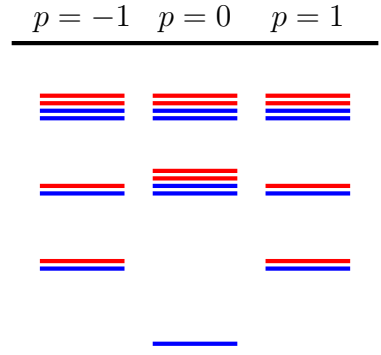
\begin{figure}[t]
	\centering
	\begin{subfigure}[t]{\linewidth}%
		\centering
		\begin{tikzpicture}[xscale=1.3]
			\draw[thin,gray] (0.875,0) -- (0.875,5) (1.875,0) -- (1.875,5);
			\draw[ultra thick,blue] (1,0) -- ++(.75, 0);
			\draw[dashed,gray] (1.75,0) -- (12,0);
			\draw[ultra thick,blue] (0,1) -- ++(.75, 0);
			\draw[ultra thick,blue] (2,1) -- ++(.75, 0);
			\draw[ultra thick,blue] (0,2) -- ++(.75, 0);
			\draw[ultra thick,blue] (1,2) -- ++(.75, 0);
			\draw[ultra thick,blue] (2,2) -- ++(.75, 0);
			\draw[ultra thick,blue] (0,3) -- ++(.75, 0);
			\draw[ultra thick,blue] (1,3) -- ++(.3,0);
			\draw[ultra thick,blue] (1.45,3) -- ++(.3, 0);
			\draw[ultra thick,blue] (2,3) -- ++(.75,0);
			\draw[thick,dotted,-latex] (1.375,0) -- node[left] {\scriptsize $a^\dagger_{-1}$} (0.375,1);
			\draw[thick,dashed,-latex] (0.375,1) -- (1.375, 2);
			\draw[thick,dashed,-latex] (1.375,0) -- node[right] {\scriptsize $a^\dagger_{1}$} (2.375,1);
			\draw[thick,dotted,-latex] (2.375,1) -- (1.375, 2);
			\draw[thick,dotted,-latex] (0.375,1) -- (0,1.4) (2.75,1.6) -- (2.375,2);
			\draw[thick,dashed,-latex] (2.375,1) -- (2.75,1.4) (0,1.6) -- (0.375,2);
			
			\draw[thick,dashed,-latex] (1.375,2) -- (2.375,3);
			\draw[thick,dashed,-latex] (0.375,2) -- (1.15,3);
			\draw[thick,dashed,-latex] (2.375,2) -- (2.75,2.4) (0,2.6) -- (0.375,3);
			\draw[thick,dotted,-latex] (2.375,2) -- (1.6,3);
			\draw[thick,dotted,-latex] (1.375,2) -- (0.375,3);
			\draw[thick,dotted,-latex] (0.375,2) -- (0,2.4) (2.75,2.6) -- (2.375,3);
			
			\draw[ultra thick,black] (-.15,4.5) -- (2.9,4.5);
			\node[above] at (.375,4.5) {\scriptsize $p = -1$};
			\node[above] at (1.375,4.5) {\scriptsize $p =0$};
			\node[above] at (2.375,4.5) {\scriptsize $p = 1$};
			
			\begin{scope}[xshift=3.2cm]
			\draw[gray,decoration = {zigzag,segment length = 2mm, amplitude = .5mm}, decorate] (-.25, 5) -- (-.25,-.25);
			\draw[thin,gray] (0.875,0) -- (0.875,5) (1.875,0) -- (1.875,5);
			\begin{scope}[yshift=.3cm]
			\draw[ultra thick,red] (0,0) -- ++(.75, 0);
			\draw[<->] (0.8,0) -- node[right] {\scriptsize $m$} ++(0,-0.3);
			\draw[ultra thick,red] (1,1) -- ++(.75, 0);
			\draw[ultra thick,red] (2,1) -- ++(.75, 0);
			\draw[ultra thick,red] (0,2) -- ++(.75, 0);
			\draw[ultra thick,red] (1,2) -- ++(.75, 0);
			\draw[ultra thick,red] (2,2) -- ++(.75, 0);
			\draw[ultra thick,red] (2,3) -- ++(.75, 0);
			\draw[ultra thick,red] (1,3) -- ++(.75,0);
			\draw[ultra thick,red] (0,3) -- ++(.3,0);
			\draw[ultra thick,red] (0.45,3) -- ++(.3, 0);
			\draw[thick,dashed,-latex] (0.375,0) -- (1.375,1);
			\draw[thick,dotted,-latex] (0.375,0) -- (0,.4) (2.75,.6) -- (2.375,1);
			\draw[thick,dashed,-latex] (1.375,1) -- (2.375, 2);
			\draw[thick,dotted,-latex] (1.375,1) -- (0.375, 2);
			\draw[thick,dashed,-latex] (2.375,1) -- (2.75,1.4) (0,1.6) -- (0.375,2);
			\draw[thick,dotted,-latex] (2.375,1) -- (1.375,2);
			
			\draw[thick,dashed,-latex] (1.375,2) -- (2.375,3);
			\draw[thick,dashed,-latex] (0.375,2) -- (1.375,3);
			\draw[thick,dashed,-latex] (2.375,2) -- (2.75,2.4) (0,2.6) -- (0.15,3);
			\draw[thick,dotted,-latex] (2.375,2) -- (1.375,3);
			\draw[thick,dotted,-latex] (1.375,2) -- (0.6,3);
			\draw[thick,dotted,-latex] (0.375,2) -- (0,2.4) (2.75,2.6) -- (2.375,3);
			\end{scope}
			
			\draw[ultra thick,black] (-.15,4.5) -- (2.9,4.5);
			\node[above] at (.375,4.5) {\scriptsize $p = -1$};
			\node[above] at (1.375,4.5) {\scriptsize $p =0$};
			\node[above] at (2.375,4.5) {\scriptsize $p = 1$};
			\end{scope}
			
			\begin{scope}[xshift=6.4cm]
			\draw[gray,decoration = {zigzag,segment length = 2mm, amplitude = .5mm}, decorate] (-.25, 5) -- (-.25,-.25);
			\draw[thin,gray] (0.875,0) -- (0.875,5) (1.875,0) -- (1.875,5);
			\begin{scope}[yshift=.3cm]
			\draw[ultra thick,red] (2,0) -- ++(.75, 0);
			\draw[<->] (1.95,0) -- node[left] {\scriptsize $m$} ++(0,-0.3);
			\draw[ultra thick,red] (0,1) -- ++(.75, 0);
			\draw[ultra thick,red] (1,1) -- ++(.75, 0);
			\draw[ultra thick,red] (0,2) -- ++(.75, 0);
			\draw[ultra thick,red] (1,2) -- ++(.75, 0);
			\draw[ultra thick,red] (2,2) -- ++(.75, 0);
			\draw[ultra thick,red] (0,3) -- ++(.75, 0);
			\draw[ultra thick,red] (1,3) -- ++(.75,0);
			\draw[ultra thick,red] (2,3) -- ++(.3,0);
			\draw[ultra thick,red] (2.45,3) -- ++(.3, 0);
			\draw[thick,dotted,-latex] (2.375,0) -- (1.375,1);
			\draw[thick,dashed,-latex] (0.375,1) -- (1.375, 2);
			\draw[thick,dashed,-latex] (2.375,0) -- (2.75,.4) (0,.6) -- (0.375,1);
			\draw[thick,dotted,-latex] (1.375,1) -- (0.375, 2);
			\draw[thick,dotted,-latex] (0.375,1) -- (0,1.4) (2.75,1.6) -- (2.375,2);
			\draw[thick,dashed,-latex] (0.375,1) -- (1.375,2);
			\draw[thick,dashed,-latex] (1.375,1) -- (2.375,2);
			
			\draw[thick,dashed,-latex] (1.375,2) -- (2.15,3);
			\draw[thick,dashed,-latex] (0.375,2) -- (1.375,3);
			\draw[thick,dashed,-latex] (2.375,2) -- (2.75,2.4) (0,2.6) -- (0.375,3);
			\draw[thick,dotted,-latex] (2.375,2) -- (1.375,3);
			\draw[thick,dotted,-latex] (1.375,2) -- (0.375,3);
			\draw[thick,dotted,-latex] (0.375,2) -- (0,2.4) (2.75,2.6) -- (2.6,3);
			\end{scope}
			
			\draw[ultra thick,black] (-.15,4.5) -- (2.9,4.5);
			\node[above] at (.375,4.5) {\scriptsize $p = -1$};
			\node[above] at (1.375,4.5) {\scriptsize $p =0$};
			\node[above] at (2.375,4.5) {\scriptsize $p = 1$};
			\end{scope}
			
			\begin{scope}[xshift=9.6cm]
			\draw[gray,decoration = {zigzag,segment length = 2mm, amplitude = .5mm}, decorate] (-.25, 5) -- (-.25,-.25);
			\draw[thin,gray] (0.875,0) -- (0.875,5) (1.875,0) -- (1.875,5);
			\begin{scope}[yshift=.6cm]
			\draw[ultra thick,blue] (1,0) -- ++(.75, 0);
			\draw[<->] (1.8,0) -- node[right] {\scriptsize $2m$} ++(0,-0.6);
			\draw[ultra thick,blue] (0,1) -- ++(.75, 0);
			\draw[ultra thick,blue] (2,1) -- ++(.75, 0);
			\draw[ultra thick,blue] (0,2) -- ++(.75, 0);
			\draw[ultra thick,blue] (1,2) -- ++(.75, 0);
			\draw[ultra thick,blue] (2,2) -- ++(.75, 0);
			\draw[ultra thick,blue] (0,3) -- ++(.75, 0);
			\draw[ultra thick,blue] (1,3) -- ++(.3,0);
			\draw[ultra thick,blue] (1.45,3) -- ++(.3, 0);
			\draw[ultra thick,blue] (2,3) -- ++(.75,0);
			\draw[thick,dotted,-latex] (1.375,0) -- (0.375,1);
			\draw[thick,dashed,-latex] (0.375,1) -- (1.375, 2);
			\draw[thick,dashed,-latex] (1.375,0) -- (2.375,1);
			\draw[thick,dotted,-latex] (2.375,1) -- (1.375, 2);
			\draw[thick,dotted,-latex] (0.375,1) -- (0,1.4) (2.75,1.6) -- (2.375,2);
			\draw[thick,dashed,-latex] (2.375,1) -- (2.75,1.4) (0,1.6) -- (0.375,2);
			
			\draw[thick,dashed,-latex] (1.375,2) -- (2.375,3);
			\draw[thick,dashed,-latex] (0.375,2) -- (1.15,3);
			\draw[thick,dashed,-latex] (2.375,2) -- (2.75,2.4) (0,2.6) -- (0.375,3);
			\draw[thick,dotted,-latex] (2.375,2) -- (1.6,3);
			\draw[thick,dotted,-latex] (1.375,2) -- (0.375,3);
			\draw[thick,dotted,-latex] (0.375,2) -- (0,2.4) (2.75,2.6) -- (2.375,3);
			\end{scope}
			
			\draw[ultra thick,black] (-.15,4.5) -- (2.9,4.5);
			\node[above] at (.375,4.5) {\scriptsize $p = -1$};
			\node[above] at (1.375,4.5) {\scriptsize $p =0$};
			\node[above] at (2.375,4.5) {\scriptsize $p = 1$};
			\end{scope}
		\end{tikzpicture}
		\caption{The states constructed by acting repeatedly with $a^\dagger_{1}$ (dashed) or $a^\dagger_{-1}$ (dotted) upon $\ket{0}$, $c^\dagger_{-1}\ket{0}$, $c^\dagger_{1}\ket{0}$, or $c^\dagger_1 c^\dagger_{-1}\ket{0}$, respectively.}
	\end{subfigure}\\[1em]
	\begin{subfigure}[t]{0.4\linewidth}
		\centering
		\begin{tikzpicture}[xscale=.75]
			\draw[ultra thick,blue] (0,0) -- ++(1.5,0);
			\draw[ultra thick,blue] (2,1) -- ++(1.5,0);
			\draw[ultra thick,blue] (-2,1) -- ++(1.5,0);
			\draw[ultra thick,blue] (0,2) -- ++(1.5,0);
			\draw[ultra thick,blue] (2,2) -- ++(1.5,0);
			\draw[ultra thick,blue] (-2,2) -- ++(1.5,0);
			\draw[ultra thick,blue] (0,3) -- ++(1.5,0);
			\draw[ultra thick,blue] (0,3.1) -- ++(1.5,0);
			\draw[ultra thick,blue] (2,3) -- ++(1.5,0);
			\draw[ultra thick,blue] (-2,3) -- ++(1.5,0);
			
			\draw[ultra thick,red] (2,0) -- ++(1.5,0);
			\draw[ultra thick,red] (0,1) -- ++(1.5,0);
			\draw[ultra thick,red] (-2,1.2) -- ++(1.5,0);
			\draw[ultra thick,red] (0,2.2) -- ++(1.5,0);
			\draw[ultra thick,red] (2,2.2) -- ++(1.5,0);
			\draw[ultra thick,red] (-2,2.2) -- ++(1.5,0);
			\draw[ultra thick,red] (2,3.2) -- ++(1.5,0);
			\draw[ultra thick,red] (2,3.3) -- ++(1.5,0);
			\draw[ultra thick,red] (0,3.4) -- ++(1.5,0);
			\draw[ultra thick,red] (-2,3.2) -- ++(1.5,0);
			
			\draw[ultra thick,red] (-2,0) -- ++(1.5,0);
			\draw[ultra thick,red] (0,1.1) -- ++(1.5,0);
			\draw[ultra thick,red] (2,1.2) -- ++(1.5,0);
			\draw[ultra thick,red] (0,2.3) -- ++(1.5,0);
			\draw[ultra thick,red] (2,2.3) -- ++(1.5,0);
			\draw[ultra thick,red] (-2,2.3) -- ++(1.5,0);
			\draw[ultra thick,red] (-2,3.3) -- ++(1.5,0);
			\draw[ultra thick,red] (-2,3.4) -- ++(1.5,0);
			\draw[ultra thick,red] (0,3.5) -- ++(1.5,0);
			\draw[ultra thick,red] (2,3.4) -- ++(1.5,0);
			
			\draw[ultra thick,blue] (0,0.1) -- ++(1.5,0);
			\draw[ultra thick,blue] (2,1.1) -- ++(1.5,0);
			\draw[ultra thick,blue] (-2,1.1) -- ++(1.5,0);
			\draw[ultra thick,blue] (0,2.1) -- ++(1.5,0);
			\draw[ultra thick,blue] (2,2.1) -- ++(1.5,0);
			\draw[ultra thick,blue] (-2,2.1) -- ++(1.5,0);
			\draw[ultra thick,blue] (0,3.2) -- ++(1.5,0);
			\draw[ultra thick,blue] (0,3.3) -- ++(1.5,0);
			\draw[ultra thick,blue] (2,3.1) -- ++(1.5,0);
			\draw[ultra thick,blue] (-2,3.1) -- ++(1.5,0);

			\draw[ultra thick,black] (-2.5,4) -- (4,4);
			\node[above] at (.75,4) {$p = 0$};
			\node[above] at (2.75,4) {$p = 1$};
			\node[above] at (-1.25,4) {$p = -1$};
			
			\draw[<->] (3.8,0) -- node[right] {$\omega = g\sqrt{\frac{3}{2\pi}}$} ++(0,1);
		\end{tikzpicture}
		\caption{The spectrum in the three universes at $m = 0$. The ground states are at energy $\omega$.}
		\label{fig:su3_massless}
	\end{subfigure}%
	\hspace{.1\linewidth}%
	\begin{subfigure}[t]{0.4\linewidth}
		\centering
		\begin{tikzpicture}[xscale=.75]
			\draw[ultra thick,blue] (0,0) -- ++(1.5,0);
			\draw[ultra thick,blue] (2,1) -- ++(1.5,0);
			\draw[ultra thick,blue] (-2,1) -- ++(1.5,0);
			\draw[ultra thick,blue] (0,2) -- ++(1.5,0);
			\draw[ultra thick,blue] (2,2) -- ++(1.5,0);
			\draw[ultra thick,blue] (-2,2) -- ++(1.5,0);
			\draw[ultra thick,blue] (0,3) -- ++(1.5,0);
			\draw[ultra thick,blue] (0,3.1) -- ++(1.5,0);
			\draw[ultra thick,blue] (2,3) -- ++(1.5,0);
			\draw[ultra thick,blue] (-2,3) -- ++(1.5,0);
			
			\draw[ultra thick,red] (2,1.1) -- ++(1.5,0);
			\draw[ultra thick,red] (0,2.3) -- ++(1.5,0);
			\draw[ultra thick,red] (-2,2.1) -- ++(1.5,0);
			\draw[ultra thick,red] (0,3.2) -- ++(1.5,0);
			\draw[ultra thick,red] (2,3.2) -- ++(1.5,0);
			\draw[ultra thick,red] (-2,3.2) -- ++(1.5,0);
			
			\draw[ultra thick,red] (-2,1.1) -- ++(1.5,0);
			\draw[ultra thick,red] (0,2.2) -- ++(1.5,0);
			\draw[ultra thick,red] (2,2.1) -- ++(1.5,0);
			\draw[ultra thick,red] (0,3.3) -- ++(1.5,0);
			\draw[ultra thick,red] (2,3.3) -- ++(1.5,0);
			\draw[ultra thick,red] (-2,3.3) -- ++(1.5,0);
			
			\draw[ultra thick,blue] (0,2.1) -- ++(1.5,0);
			\draw[ultra thick,blue] (2,3.1) -- ++(1.5,0);
			\draw[ultra thick,blue] (-2,3.1) -- ++(1.5,0);
			
			\draw[ultra thick,black] (-2.5,4) -- (4,4);
			\node[above] at (-1.25,4) {$p = -1$};
			\node[above] at (0.75,4) {$p = 0$};
			\node[above] at (2.75,4) {$p = 1$};
		\end{tikzpicture}
		\caption{The spectrum in the three universes at the supersymmetric point $m = \omega = g\sqrt{\frac{3}{2\pi}}$ The ground state is at zero energy.}
		\label{fig:su3_susy}
	\end{subfigure}
	\caption{The leading order spectrum of the $\grSU(3)$ theory. Blue and red lines denote bosonic and fermionic states, respectively.}
	\label{fig:su3}
\end{figure}

To illustrate the more general case, we can consider the $\grSU(3)$ theory. The construction of states is depicted in Figure \ref{fig:su3}. To make the diagrams more symmetric, we refer to the $p = 2$ universe equivalently as $p = -1$. We can generate bosonic states by acting with the two bosonic creation operators $a^\dagger_{\pm 1}$ on either $\ket{0}$ or $c^\dagger_1 c^\dagger_{-1}\ket{0}$, and fermionic states by acting with the bosonic creation operators on $c^\dagger_{\pm 1}\ket{0}$. When $m = 0$, there is boson-fermion degeneracy but not within each universe, as shown in Figure \ref{fig:su3_massless}. When we set $m = \omega$, one can check explicitly that the states all line up such that once again we have an unbroken supersymmetric spectrum in the $p = 0$ universe, and broken supersymmetric spectra with goldstinos in the $p = \pm 1$ universes.

\section{Leading order effective theory for arbitrary gauge group}
\label{sec:LOtheory}

The analysis of the previous section can be generalized to the adjoint QCD$_2$ theory with an arbitrary simply-connected gauge group $G$. Both the gauge potential $A_\mu$ and the Majorana fermion are valued in the Lie algebra $\mathfrak{g}=\operatorname{Lie}(G)$. For classical Lie groups, the fundamental and adjoint traces are related via the dual Coxeter number $h^\vee$ as $\tr_\text{fund} = \frac{1}{2h^\vee}\tr_\text{adj}$. The latter form is well-defined for any algebra, so we use it to write the Lorentzian action as
 \es{SLorGen}{
	S = \frac{1}{2h^\vee} \int dt\, dx \, \tr_\text{adj} \left( - \frac{1}{2 g^2} F_{\mu\nu} F^{\mu\nu} + i \bar \psi \gamma^\mu D_\mu \psi - m \bar \psi \psi \right)  \,.
}
See Table~\ref{tab:GTable} for key data of simply-connected Lie groups, including $h^\vee$.\footnote{The dual Coxeter number is defined in general as follows.  If $\vec{\theta}$ is the largest root and $\vec{\rho}$ is the Weyl vector defined as half the sum of all positive roots, then the dual Coxeter number is $h^\vee \equiv 1 + \frac{2 \vec{\theta} \cdot \vec{\rho}}{\abs{\vec{\theta}}^2}$. }

As in the $\grSU(\Nc)$ case, on a small circle most modes of the fermion and the gauge field acquire masses of order $1/L$, and one can write down an effective theory in terms of the holonomy and the zero modes of the covariant derivative operator $D_x$ acting on the fermion.  The holonomy variable $\Omega$ defined just as in \eqref{HolVariableSUN} acquires the potential given by the same expression as in \eqref{VeffSUN}.  To determine the minimum of this potential, one can then take $A_t = 0$ and $A_x$ to be a constant in the Cartan subalgebra as in \eqref{GaugeDiag}, restricted to lie within an appropriately defined fundamental domain.

\begin{table}[htp]
\begin{center}
\begin{tabular}{c|c|c|c}
 $G$  & dim $G$ & rank & $h^\vee$ \\
 \hline 
 $\grSU(\Nc)$  & $\Nc^2 - 1$ & $\Nc - 1$ & $\Nc$ \\
 $\grp{Spin}(\Nc)$  & $\frac{1}{2} \Nc ( \Nc - 1)$ &  $\lfloor \frac 12 \Nc \rfloor$ & $\Nc - 2$ \\
 $\grUSp(2\Nc)$  & $\Nc (2 \Nc + 1)$ & $\Nc$ & $\Nc + 1$ \\
 $G_2$  & $14$ & $2$ & $4$ \\
 $F_4$  & $52$ & $4$ & $9$\\
 $E_6$  & $78$ & $6$ & $12$  \\
 $E_7$  & $133$ & $7$ & $18$ \\
 $E_8$  & $248$ & $8$ & $30$ \\
\end{tabular}
\end{center}
\caption{Simply connected gauge groups and some of their properties.}
\label{tab:GTable}
\end{table}%

To define the notion of a fundamental domain for arbitrary $G$, let us establish some notation.  Let $H^a$, with $a = 1, \ldots, \text{rank}(G)$, be a basis of Hermitian operators ($(H^a)^\dagger = H^a$) for the Cartan subalgebra of $\mathfrak{g}$.  We can write $\vec{H}$ for the vector of operators whose components are $H^a$.  More generally, for a quantity $X$ that is valued in the Cartan subalgebra, we will write $X^a$ for the corresponding components and $\vec{X}$ for the $\text{rank}(G)$-dimensional vector of components.  We have $X = \vec{X} \cdot \vec{H}$, with the inner product being the standard one on $\R^{\text{rank}(G)}$.

The non-zero $\vec{H}$ eigenvalues of the states of the adjoint representation are the roots $\vec{\rt}$, which are vectors in $\R^{\text{rank}(G)}$, $\vec{H} \ket{\vec{\rt}} = \vec{\rt} \ket{\vec{\rt}}$.  To each $\vec{\rt}$ there corresponds a Lie algebra element $E^{\vec{\rt}}$ obeying the commutation relations
\es{Relns}{
	[H^a, H^{b}] &= 0\,, \qquad \qquad
	[H^a, E^{\vec \rt}] = \rt^a E^{\vec \rt} \,, \\
	[E^{\vec \rt}, E^{-\vec \rt}] &= \vec{\rt} \cdot \vec{H} \,, \qquad
	[E^{\vec \rt}, E^{\vec \rt'}] = N_{{\vec \rt}, \vec \rt'}^{\vec \rt + \vec{\rt}'} E^{{\vec \rt} + \vec{\rt}'} \,,
}
where $\rt^a$ is the $a$th component of $\vec{\rt}$, as well as $(E^{\vec{\rt}})^\dagger = E^{-\vec{\rt}}$.  A general Lie algebra element $Y$ can be decomposed into Cartan and root components $\vec{Y}$ and $Y^{\vec{\rt}}$, respectively, as in
 \es{YDecomp}{
  Y = \vec{Y} \cdot \vec{H} + \sum_{\vec{\rt}} Y^{\vec{\rt}} E^{\vec{\rt}} \,.
 }
Finally, let us normalize the generators such that 
\es{NormGen}{
	\tr_\text{adj} (H^a H^b) =  
	 \sum_{\vec{\alpha}} \alpha^a \alpha^b = h^\vee \delta^{ab} \,, \quad
	\tr_\text{adj} (H^a E^{\vec \rt}) = 0 \,, \quad
	\tr_\text{adj} (E^{\vec \rt} E^{-\vec \rt'}) =  h^\vee \delta^{\vec \rt, \vec \rt'} \,,
}
where $h^\vee$ is the dual Coxeter number as in \eqref{SLorGen}.  This normalization ensures that the largest root has unit length, since in general one can show that $\tr_\text{adj} H^a H^b = |\vec{\theta}|^2 h^\vee \delta^{ab}$, where $\vec{\theta}$ is the largest root.  See Section~13.2.4~of \cite{DiFrancesco:1997nk}.

For example, in the $G = \grSU(\Nc)$ case, there are $\Nc (\Nc - 1)$ roots $\vec{\rt}_{ij} =  \vec{w}_i - \vec{w}_j $, with $i \neq j$, and where $\vec{w}_i$ are the fundamental weight vectors introduced around \eqref{wProperties}.   For a diagonal matrix $X = \diag\{X_1, X_2, \ldots, X_n\}$, we have $\vec{X} = 2 \sum_i X_i w_i$ as in \eqref{qchiVec}.  We also have $h^\vee=\Nc$ in \eqref{NormGen}, which is consistent with \eqref{wProperties}.

With this notation, the analogs of \eqref{OmegaDiag} and \eqref{GaugeDiag} for general $G$ can be written as
 \es{AOmega}{
  A_t = 0 \,, \qquad A_x = \frac{1}{L} \vec{a} \cdot \vec{H} \,, \qquad \Omega = e^{i \vec{a} \cdot \vec{H}} \,,
 }
where $\vec{a}$ is a vector in $\R^{\text{rank}(G)}$. Let us now discuss the restriction to the fundamental domain.  In the $\grSU(\Nc)$ case, this restriction arose as a consequence of the gauge transformations that send $(a_i, a_j) \to (a_i + 2 \pi, a_j - 2 \pi)$ for some $i \neq j$ and those that permute the $a_i$.  In terms of $\vec{a} = 2 \sum_i a_i w_i$, a gauge transformation of the first type can also be written as
 \es{Gauge1}{
  \vec{a}  \to \vec{a} + 4 \pi \vec{\rt}_{ij} \,,
 }
where we used $\vec{\rt}_{ij} = \vec{w}_i - \vec{w}_j$.  This is a translation by $4 \pi$ times the root vector $\vec{\rt}_{ij}$.  More generally, a translation of $\vec{a}$ by $4 \pi$ times any vector of the root lattice would also be a gauge symmetry.  Similarly, the transformation that interchanges $a_i$ with $a_j$ while keeping all the other diagonal entries the same can be written as 
 \es{Gauge2}{
  \vec{a} \to \vec{a} + 2(a_j - a_i) \vec{w}_i + 2(a_i - a_j)  \vec{w}_j =  \vec{a}  -  2 ( \vec{a} \cdot \vec{\rt}_{ij} ) \vec{\rt}_{ij} \,, 
 }
where we used $a_i = \vec{a} \cdot \vec{w}_i$.  This is a reflection of $\vec{a}$ about the unit vector $\vec{\rt}_{ij}$.

Both types of transformations generalize to arbitrary $G$ almost with no change.  For arbitrary $G$, the (time-independent) gauge transformations act as 
\es{GaugeTransf}{
	A_\mu \to g A_\mu g^{-1} - i (\partial_\mu g) g^{-1} \,, \qquad \psi \to g \psi g^{-1} \,.
}
For the first type of gauge transformation, we consider $g(t, x) = e^{i K x/L}$, for some Cartan element $K = \vec{K} \cdot \vec{H}$.  In order for $g(t, 0) = g(t, L)$, $K$ needs to obey $e^{iK} =\mathds 1$.  Such transformations translate $A_x$ by $K/L$ and they translate $\hol$ by $\vec{K}$:
\es{AxTranslation}{
 \vec{\hol} \to \vec{\hol} + \vec{K} \,.
}
It was shown in \cite{Goddard:1976qe} that, when $G$ is simply connected, the condition $e^{iK} = \mathds 1$ implies that $\vec{K}$ belongs to the lattice generated by $2 \pi \vec{\rt}^\vee$, where $\vec{\rt}^\vee = 2 \vec{\rt} / \abs{\vec{\rt}}^2$ are the coroots of $G$.  By identifying points differing by such a shift, the space of $\vec \hol$ then splits into gauge-equivalent unit cells, and it will be most convenient to restrict ourselves to the Wigner-Seitz unit cell, obeying the property that $\vec{\hol}$ is closer to the origin than it is to any other lattice point. One can show that the generators $2 \pi \vec{\rt}^\vee$ include the minimum-norm elements of the lattice, and so this condition implies
\es{alpharhoFirst}{
 	\vec{\hol} \in \text{Wigner-Seitz cell} \quad \Longleftrightarrow \quad	
	\abs{\vec{\hol} \cdot 2 \pi \vec{\rt}^\vee} < \frac 12 \abs{2 \pi \vec{\rt}^\vee}^2   
	\qquad 
	\text{for all $\vec \rt$}\,.
}
We can use $\vec{\rt}^\vee = 2 \vec{\rt} / \abs{\vec{\rt}}^2$ on the LHS, as well as $\abs{\vec{\rt}^\vee} \abs{\vec{\rt}} = 2$ to simplify this condition to
\es{alpharho}{
 	\vec{\hol} \in \text{Wigner-Seitz cell} \quad \Longleftrightarrow \quad	\abs{\vec{\hol} \cdot \vec \rt} < 2 \pi  \qquad 
	\text{for all $\vec \rt$}\,.
}

The second type of gauge transformations, which generalize \eqref{Gauge2} to arbitrary $G$, are those with
\es{gWeyl}{
	g = \exp\left[ i \frac{\pi  }{2 \abs{\vec \rt}} (E^{\vec \rt} + E^{-\vec \rt}) \right] \,.
}
Such transformations generate the Weyl reflections:
\es{WeylRefl}{
	A_x \to A_x  - 2 \vec \rt \frac{\vec{A}_x \cdot \vec \rt}{\abs{\vec \rt}^2} \,, \qquad 
	\vec \hol \to \vec \hol  - 2 \vec \rt \frac{\vec{\hol} \cdot \vec \rt}{\abs{\vec \rt}^2} \,.
}
This transformation splits the $\vec{\hol}$ space into gauge-equivalent Weyl chambers.  The fundamental Weyl chamber is that for which $\vec \hol \cdot \vec \rt > 0$ for all positive roots $\vec \rt > 0$.

\begin{figure}
	\centering
	\begin{subfigure}[b]{0.5\textwidth}
		\centering
		\begin{tikzpicture}[scale=.75]
			\draw[thick,gray,fill=gray!30,dashed] (0,0) -- (30:{2/sqrt(3)}) -- (-30:{2/sqrt(3)}) -- cycle;
			\draw[thick] (30:{2/sqrt(3)}) -- (90:{2/sqrt(3)})-- (150:{2/sqrt(3)})-- (210:{2/sqrt(3)})-- (270:{2/sqrt(3)})-- (330:{2/sqrt(3)})-- (30:{2/sqrt(3)});
			\node at (0:2)[circle,fill,red,inner sep=1.5pt] {};
			\node at (60:2)[circle,fill,red,inner sep=1.5pt] {};
			\node at (-60:2)[circle,fill,red,inner sep=1.5pt] {};
			\node at (180:2)[circle,fill,blue,inner sep=1.5pt] {};
			\node at (240:2)[circle,fill,blue,inner sep=1.5pt] {};
			\node at (120:2)[circle,fill,blue,inner sep=1.5pt] {};
			\node at ({2/3},0)[cross out,draw,inner sep=1.5pt] {};
			\path (0,{-2*sqrt(3)}) -- (0,{2*sqrt(3)});
		\end{tikzpicture}
		\caption{SU(3)}
	\end{subfigure}
	\begin{subfigure}[b]{0.5\textwidth}
		\centering
		\begin{tikzpicture}[scale=.75]
			\draw[thick,gray,fill=gray!30,dashed] (0,0) -- (30:{2/sqrt(3)}) -- (1,0) -- cycle;
			\draw[thick] (30:{2/sqrt(3)}) -- (90:{2/sqrt(3)})-- (150:{2/sqrt(3)})-- (210:{2/sqrt(3)})-- (270:{2/sqrt(3)})-- (330:{2/sqrt(3)})-- (30:{2/sqrt(3)});
			\node at (0:2)[circle,fill,red,inner sep=1.5pt] {};
			\node at (60:2)[circle,fill,red,inner sep=1.5pt] {};
			\node at (-60:2)[circle,fill,red,inner sep=1.5pt] {};
			\node at (90:{2*sqrt(3)})[circle,fill,red,inner sep=1.5pt] {};
			\node at (30:{2*sqrt(3)})[circle,fill,red,inner sep=1.5pt] {};
			\node at (-30:{2*sqrt(3)})[circle,fill,red,inner sep=1.5pt] {};
			\node at (150:{2*sqrt(3)})[circle,fill,blue,inner sep=1.5pt] {};
			\node at (210:{2*sqrt(3)})[circle,fill,blue,inner sep=1.5pt] {};
			\node at (270:{2*sqrt(3)})[circle,fill,blue,inner sep=1.5pt] {};
			\node at (180:2)[circle,fill,blue,inner sep=1.5pt] {};
			\node at (240:2)[circle,fill,blue,inner sep=1.5pt] {};
			\node at (120:2)[circle,fill,blue,inner sep=1.5pt] {};
			\node at ({3/4},{1/(4*sqrt(3))})[cross out,draw,inner sep=1.5pt] {};
		\end{tikzpicture}
		\caption{G$_2$}
	\end{subfigure}
	\caption{The coroots of $\grSU(3)$ and $G_2$, with red dots denoting positive co-roots and blue dots denoting negative coroots. The solid line is the boundary of the Wigner-Seitz cell \eqref{alpharho}, the dashed line is the boundary of the fundamental domain \eqref{FundDomain}, and the minimum of the effective potential within the fundamental domain is denoted with a cross.}
	\label{fig:fund_dom}
\end{figure}
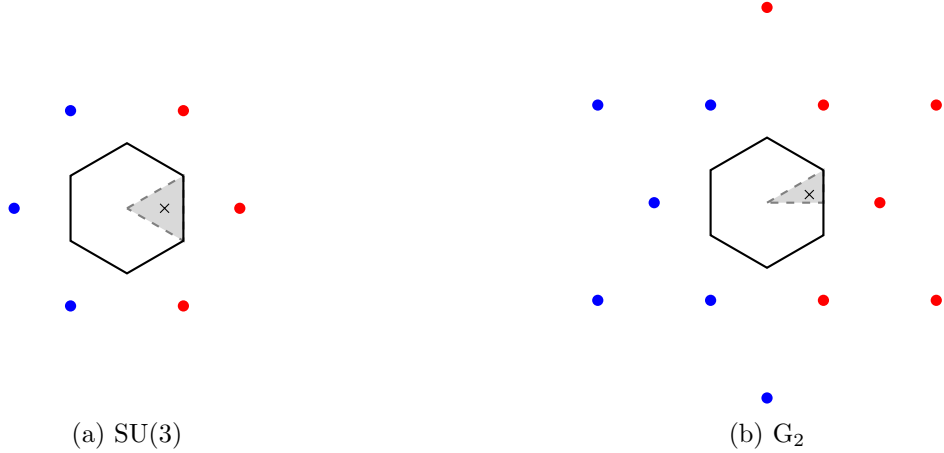

Thus, combining the two transformations \eqref{AxTranslation} and \eqref{WeylRefl}, we can restrict our attention to the intersection between the fundamental Weyl chamber and the Wigner-Seitz unit cell.  We call this region the fundamental domain ${\cal F}$. Examples of ${\cal F}$ for $\grSU(3)$ and $G_2$ are given in Figure \ref{fig:fund_dom}. Explicitly, it is defined by
\es{FundDomain}{
	{\cal F} =  \{ \vec{\hol}:\,  \vec{\hol} \cdot \vec{\rt} \in [0, 2 \pi) \text{ for all } \vec{\rt} > 0\} \,.
}

With $\vec{a} \in {\cal F}$, we have
\es{VeffFund}{
	 V_\text{eff}(\vec{a})
	 = \frac{1}{\pi L} \sum_{\vec\alpha}\sum_{n=1}^\infty \frac{1}{n^2}e^{in \vec a\cdot \vec \alpha} = \frac{2 \pi}{L} \sum_{\vec \rt >0} \left( \frac{\vec \hol \cdot \vec \rt}{2 \pi}   - \frac 12 \right)^2 \,.
}
We can evaluate the sum over roots using $\sum_{\vec \rt > 0} \rt^a \rt^b =   \frac{h^\vee}{2} \delta^{ab} $ (see \eqref{NormGen}) and $\sum_{\vec \rt>0} \vec{\rt} = 2 \vec{\Weyl}$ (by definition of the Weyl vector $\vec{\Weyl}$). This gives
\es{VeffFund2}{
	 V_\text{eff}(\vec{a}) = \frac{2 \pi}{L}  \left( \frac{h^\vee}{8 \pi^2} \vec{\hol}^2   - \frac{1}{ \pi} \vec{\hol} \cdot \vec{\Weyl} + \text{const} \right) \,.
}
Thus, in the fundamental domain, the minimum of the effective potential is unique and located at\footnote{We can verify that $\vec \hol_*$ indeed belongs to the fundamental domain as follows.   Let $\vec{\alpha} > 0$ be a positive root.  We need to show that $\vec{a}_* \cdot \vec{\alpha} \in (0, 2 \pi)$.  That $\vec{a}_* \cdot \vec{\alpha} > 0$ follows from the fact that $\vec{\rho} \cdot \vec{\alpha} \geq 0$, which in turn follows from $\vec{\rho} \cdot \vec{\alpha}_j =  \abs{\vec{\alpha}_j}^2/2 > 0$ for every simple root $\vec{\alpha}_j$ and that the positive root $\vec{\alpha}$ can be written as a linear combination of simple roots with non-negative coefficients.  To show that $\vec{a}_* \cdot \vec{\alpha}<2 \pi$, first note that if $\vec{\theta}$ is the largest root, then $\vec{\theta} - \vec{\alpha}$ is also a non-negative root, so $\vec{\rho} \cdot (\vec{\theta} - \vec{\alpha}) \geq 0$, or equivalently $\vec{\rho} \cdot \vec{\alpha} \leq \vec{\rho} \cdot  \vec{\theta}$.  This implies $\vec a_* \cdot \vec\alpha = \frac{4\pi}{h^\vee}\vec\rho\cdot \vec\alpha \leq\frac{4\pi}{h^\vee}\vec\rho\cdot \vec\theta = \frac{2\pi}{h^\vee}(h^\vee-1)<2\pi$, where we used the definition of the dual Coxeter number $h^\vee = 1 + 2\vec \theta\cdot \vec \rho$ in the convention where $\vec{\theta}$ has unit length.  We have thus shown that $\vec{a}_* \cdot \vec{\alpha} \in (0, 2 \pi)$, so $\vec{a}_*$ belongs to the interior of the fundamental domain ${\cal F}$.\label{FDFootnote}
}
\es{VMin}{
	 \vec \hol_* = \frac{4 \pi}{h^\vee} \vec{\Weyl}  \,.
}
This expression generalizes \eqref{astar} to arbitrary $G$.

Finally, we can obtain the leading order effective action by considering fluctuations of $\vec{a}$ around $\vec{a}_*$, and restricting to the Cartan modes of the adjoint fermion $\psi$:
 \es{AFluct}{
  A_t = 0 \,, \qquad A_x = \frac{1}{L} (\vec{a}_* + g\sqrt{L}\,  \vec{q}(t) ) \cdot \vec{H} \,, \qquad
   \psi = \frac{1}{\sqrt{L}} \vec{\chi}(t) \cdot \vec{H} \,.
 }
From \eqref{VeffFund2}, the potential for $\vec{q}$ is $V_\text{eff} = \frac{h^\vee g^2}{4\pi} \vec{q} \cdot \vec{q} $.  To obtain the full effective action, we should substitute the ansatz \eqref{AFluct} into \eqref{SLor} and add the contribution of the effective potential.  The result is just \eqref{eq:SLOSUN} with $\Nc$ replaced by the dual Coxeter number $h^\vee$:
 \es{eq:SLO}{
	S_\text{LO}= \int dt\, \left(\frac{1}{2} \dot{\vec q} \cdot \dot{\vec q}
	 - \frac{h^\vee g^2}{4\pi} \vec{q} \cdot \vec{q}
	 + \frac{i}{2} \vec{\chi}^T \cdot \dot{\vec{\chi}}
	 - \frac{m}{2} \vec{\chi}^T \cdot \gamma^0  \vec{\chi} \right) \,.
 }
The action \eqref{eq:SLO} is that of $\text{rank}(G)$ bosonic and fermionic harmonic oscillators of frequencies $\omega_B =g \sqrt{\frac{h^\vee}{2\pi}}$ and $\omega_F=\abs{m}$, respectively.  As discussed around \eqref{eq:jacSUN} in the $\grSU(\Nc)$ case, all wavefunctions should obey Dirichlet boundary conditions at the boundaries of the fundamental domain.  Since the minimum of the potential is within the fundamental domain, the Dirichlet boundary conditions have an exponentially small effect on the energy levels.

\section{Subleading effective theory}
\label{sec:effective}

The effective action \eqref{eq:SLO} should be viewed as the leading-order piece of an effective action for the adjoint QCD$_2$ theory on a small spatial circle of circumference $L$. In this section we will develop the subleading corrections and write the next several terms of the effective action as a power series in $gL \ll 1$.

We will begin by passing to Euclidean signature with $t = -i\tau$, which will simplify the remainder of the computation. Identifying $-iS$ with the Euclidean action $S_E$, we find that \eqref{SLor} becomes
\begin{equation}\label{eq:SEuc}
	S_E = \frac{1}{2 h^\vee}\int d\tau\,dx\,\tr_\text{adj}\left(\frac{1}{g^2}F_{\tau x}^2 + \psi^T\left(D_\tau - i\gamma^5 D_x\right)\psi + m\psi^T \gamma^0 \psi\right)\,.
\end{equation}
Our procedure will be to fix a gauge, add the requisite ghost contribution to the action, and then expand around the solution corresponding to the minimum \eqref{VMin} of the leading-order effective potential for the holonomy. We will find an effective theory of bosons and fermions, coming from holonomy degrees of freedom and adjoint fermion zero modes respectively. In Section~\ref{sec:susy}, we will show that it exhibits supersymmetry order-by-order when the adjoint fermion mass is $m = \pm m_\text{SUSY}$, with
 \es{mSUSYGen}{
	m_\text{SUSY} = g\sqrt{\frac{h^\vee}{2\pi}} \,.
 }

\subsection{Gauge fixing}\label{sec:gauge_fixing}

On our spacetime $S^1\times\mathbb{R}$, we can make a gauge transformation to fix the spatial component of the gauge field to be independent of $x$ and aligned with the Cartan subalgebra:
\begin{align}\label{eq:gauge_choice}
	\partial_x A^a_x =0,\qquad A^{\vec \alpha}_x =0\,.
\end{align}
The remaining zero mode of $A_x$ leads to a $\tau$-dependent holonomy of the gauge field around the spatial circle,
\begin{align}
	\Omega(\tau) = e^{i\int_0^L dx\, A_x(\tau,x)} = e^{iLA_x(\tau)}\,.
\end{align}

This choice of gauge should be accompanied by a ghost action via the standard Faddeev-Popov procedure. Our gauge-fixing condition can be represented as
\begin{equation}\label{eq:gauge_fix}
	f(A_\mu) = 0, \qquad\text{with}\qquad f(A_\mu) = L\sum_a \left(\partial_x A_x^a\right)H^k+\sum_{\vec\alpha} A^{\vec\alpha}_x E^{\vec \alpha}\,.
\end{equation}
To restrict the path integral to gauge configurations with $f(A_\mu) = 0$, we insert a factor of
\begin{equation}
	1 = \int \mathcal{D}\Lambda\,\delta\left(f(A^{(\Lambda)}\right)\det\left(\frac{\delta f(A^{(\Lambda)}_\mu)}{\delta\Lambda}\right)\,,
\end{equation}
where $\Lambda(\tau,x) \in\mathfrak{g}$ is the parameter of a gauge transformation and $A^{(\Lambda)}_\mu = A_\mu + D_\mu\Lambda$ is the transformed gauge field. To evaluate the functional determinant, we note that $\delta A_\mu = D_\mu \Lambda = \partial_\mu \Lambda-i[A_\mu,\Lambda]$ can be written in the Cartan-Weyl basis as
\begin{align}
	\delta A_\mu = \sum_a (\partial_\mu \Lambda^a)H^a+\sum_{\vec \alpha} \left(\partial_\mu-i(\vec A_\mu \cdot \vec \alpha)\Lambda^{\vec \alpha}\right)E^{\vec \alpha}\,.
\end{align}
Evaluating $f(A_\mu + \delta A_\mu)$ using \eqref{eq:gauge_fix} gives
\begin{align}
	\delta f(A_\mu) = L\sum_a(\partial^2_x\Lambda^a)H^a+\sum_{\vec{\alpha}}\left(\partial_x \Lambda^{\vec \alpha}-i(\vec A_x\cdot \vec \alpha)\Lambda^{\vec \alpha}\right)E^{\vec \alpha}\,.
\end{align}
We can read off that, up to constant factors, the Fadeev-Popov determinant is the product of $\det (-\partial_x^2)$ for each of the Cartan directions and $\det \left( i \partial_x +  \vec{A}_x \cdot \vec{\rt}  \right)$ for each root direction $\vec \rt$. These determinants can be reproduced from a Gaussian path integral of adjoint-valued anticommuting ghosts $c$ and $\bar c$. The ghost action can be taken to be
\es{Sghost}{
	S_\text{ghost} = 
	\int d\tau\, dx\, 
	\left[ 2i \tr \left( \bar c D_x c \right)  + \sum_a \bar c^a (-\partial_x^2 - i \partial_x) c^a   \right]  \,.
}
The second term does not couple to the dynamical gauge field and can be dropped for the purposes of this calculation. 

\subsection{Small circle scaling}

We will now rewrite the action on $S^1\times\mathbb{R}$ in terms of dimensionless fields in a manner that facilitates deriving the effective action order-by-order in $gL$. For the gauge field, we want to expand around the configuration $A_x = \frac{a_*}{L}$, with $a_*$ as in \eqref{VMin}, and $A_\tau = 0$. Furthermore, following \eqref{AFluct} we set
\begin{equation}
	A_x(\tau, x) = \frac{1}{L}\left(\vec{a}_* + \sqrt{gL} \vec{\bar{q}}(\tau)\right)\cdot\vec{H}\,,
\end{equation}
where $\bar q = g^{1/2} q$ is dimensionless, so that $\mathcal{O}(1)$ values of $\bar q$ correspond to $\mathcal{O}(g)$ energies. Likewise, we set
\begin{equation}
	A_\tau(\tau, x) = g\Phi(\tau, x)
\end{equation}
so that $\Phi$ has an analogous scaling behavior. For the ghost field and the fermion, we rescale by $L^{-1/2}$ to get dimensionless fields, and we separate out the zero mode $\chi(\tau)$ of the fermion:
\es{psiScaling}{
	\psi(\tau, x) &=\frac{1}{\sqrt{L}} \psiL(\tau) + \frac{1}{\sqrt{L}}  \psiH(\tau, x) \,,\\
	c(\tau, x) &= \frac{1}{\sqrt{L}} \cghostHeavy(\tau, x)\,.
}
In terms of these fields, our goal will be to integrate out the ``heavy'' modes $\Phi$, $C$, and $\Psi$ and obtain an effective theory of the ``light'' modes $q$ and $\chi$. The light modes are independent of $x$ and valued in the Cartan, while the heavy modes either have nonzero momentum along the $S^1$ direction or are valued in the roots of $\mathfrak{g}$.

We can also pass to dimensionless versions of the spacetime coordinates. Since the momenta are of order $L^{-1}$ and we are interested in energies of order $g$, we will use
\begin{equation}
	\bar\tau \equiv g\tau, \qquad \bar x \equiv x/L\,.
\end{equation}

In terms of these rescaled fields and coordinates, we can write the full action as
\begin{equation}
	S_E + S_\text{ghost} = S_L + S_H + S_{HL}\,,
\end{equation}
where the three terms are as follows. The action for the light modes is
\begin{equation}\label{SL}
	S_L = \int d\bar\tau \tr\left((\partial_{\bar\tau}\bar q_a)(\partial_{\bar\tau}\bar q_a) + \chi^T_a \partial_{\bar\tau}\chi_a + \frac{m}{g}\chi^T_a \gamma^0\chi_a\right)\,.
\end{equation}
All terms in this piece are $\mathcal{O}\left((gL)^0\right)$ (in particular, we treat $m/g$ as $\mathcal{O}(1)$). The action for the heavy modes is
\begin{equation}\label{SH}
\begin{split}
	S_H = \int d\bar\tau\,d\bar x\,\tr\Bigg(&\frac{1}{L}\Big( (\partial_{\bar x} \Phi - i[a_*,\Phi])^2 - i\Psi^T \Gamma^5 (\partial_{\bar x}\Psi - i[a_*, \Psi]) + 2i\bar C(\partial_{\bar x} C - i[a_*,C])\Big) \\
	&+\Psi^T \partial_{\bar\tau}\Psi + \frac{m}{g}\Psi^T\gamma^0 \Psi - i\Psi^T[\Phi,\Psi]\Bigg)\,.
\end{split}
\end{equation} 
The terms on the second line are subleading in $gL$. Finally, there are terms involving interactions between heavy and light modes:
\begin{equation}\label{Sint}
\begin{split}
	S_{HL} = \int d\bar\tau\,d\bar x\,\tr\Bigg(&\frac{1}{\sqrt{gL}}\Big( -\Psi^T \gamma^5[\bar q,\Psi] + 2i(\partial_{\bar x}\Phi - i[a_*,\Phi])[\Phi,\bar q] + 2\bar C[\bar q, C]\Big) \\
	&-2i\chi^T[\Phi,\Psi] - [\Phi, \bar q]^2\Bigg)\,.
\end{split}
\end{equation}
Again, terms on the second line are subleading in $gL$.

\subsection{Feynman rules}\label{sec:feyn_rules}

The effective action for the light modes $q$ and $\chi$ can be expressed as
\begin{align}\label{eq:cumulant_exp}
	S_\text{eff}[q,\chi] &=S_L[q,\chi] -\log \int \mathcal D[\Phi,\Psi,C]\, e^{-S_H[\Phi,\Psi,C]-S_{HL}[\Phi,\Psi,C;q,\chi]}\nonumber\\
	&=S_L+ \braket{S_{HL}}_c-\frac{1}{2}\braket{S_{HL}^2}_c+\frac{1}{6}\braket{S_{HL}^3}_c+\ldots\,,
\end{align}
where $\braket{\ldots}_c$ denotes a connected correlator calculated from the path integral over heavy modes.

To compute these connected correlators, we will write down explicit Feynman rules in terms of frequency and momentum modes of the fields expanded in the Weyl-Cartan basis. We will denote frequency modes proportional to $e^{i\omega \bar\tau}$ by $\omega\in\mathbb{R}$ and momentum modes proportional to $e^{2\pi i n \bar x}$ by $n\in\mathbb{Z}$. Edges will be labeled either by a root $\alpha$ or an index $a$ of the Cartan subalgebra. 

The propagators for the heavy fields can be read off from \eqref{SH}. The propagator for the heavy mode $\Phi$ of the gauge field is given by either\footnote{
The kinetic term for the heavy fields does not include a quadratic term for the spatially constant part of $\Phi$ valued in the Cartan. It appears linearly in the second line of \eqref{SH}, and this term should be interpreted as a Lagrange multiplier imposing the constraint of charge neutrality
\begin{align}
	\int dx\sum_{\vec \alpha>0} \vec \alpha^k\Psi^{-\vec \alpha T}\Psi^{\vec \alpha} = 0.
\end{align}
In our perturbative calculation, this contraint will be automatically satisfied because the unperturbed vacuum as well as all the interaction vertices are uncharged.}
\begin{align}
	\begin{tikzpicture}[vertical align]
		\draw [very thick, snake it] (-1,0) -- node[above] {$a$} (1,0);
		\draw [->] (-.6,-.3) -- node[below] {$\omega,n$} (.6,-.3);
	\end{tikzpicture} &= D_0(\omega,n),\qquad n\neq 0\,,\\
	\text{or}\qquad\begin{tikzpicture}[vertical align]
		\draw [very thick, snake it] (-1,0) --  (1,0);
		\draw [->] (-.6,.3) -- node[above] {$\vec{\alpha}$} (.6,.3);
		\draw [->] (-.6,-.3) -- node[below] {$\omega,n$} (.6,-.3);
	\end{tikzpicture} &= D_{\vec a_*\cdot \vec \alpha}(\omega,n)\,
\end{align}
depending on whether the propagating field is aligned with a Cartan direction $a$ or a root $\alpha$. The function on the right hand side is
\begin{align}
	D_Q(\omega,n) = \frac{gL}{(2\pi n -Q)^2}\,.
\end{align}
The ghost fields can be taken to lie only in the roots, as explained after \eqref{Sghost}, and the ghost propagator is
\begin{align}
	\begin{tikzpicture}[vertical align]
		\draw [very thick, dashed] (-1,0) --  (1,0);
		\draw [->] (-.6,.2) -- node[above] {$\vec{\alpha}$} (.6,.2);
		\draw [->] (-.6,-.2) -- node[below] {$\omega,n$} (.6,-.2);
	\end{tikzpicture} &= H_{\vec a_*\cdot \vec \alpha}(\omega,n)
\end{align}
where
\begin{equation}
	H_Q(\omega, n) = -\frac{gL}{2\pi n-Q}\,.
\end{equation}
For the heavy modes of the fermion, $\Psi$, we will keep the subleading quadratic terms from \eqref{SH} in the propagator in order to regulate the short time behavior, and expand in $gL$ after evaluating diagrams. The propagator for $\Psi$ is then given by
\begin{align}
	\begin{tikzpicture}[vertical align]
		\draw [very thick] (-1,0) -- node[above] {$a$} (1,0);
		\draw [->] (-.6,-.2) -- node[below] {$\omega,n$} (.6,-.2);
	\end{tikzpicture} &= G_{0}(\omega,n)\,,\\
	\begin{tikzpicture}[vertical align]
		\draw [very thick] (-1,0) --  (1,0);
		\draw [->] (-.6,.2) -- node[above] {$\vec{\alpha}$} (.6,.2);
		\draw [->] (-.6,-.2) -- node[below] {$\omega,n$} (.6,-.2);
	\end{tikzpicture} &= G_{\vec a_* \cdot \vec \alpha}(\omega,n)\,,
\end{align}
with
\begin{align}
	G_Q(\omega,n) = gL \frac{-i(gL\omega)+(2\pi n-Q) \gamma^5+mL\gamma^0}{(gL\omega)^2 + (2\pi n-Q)^2 +(mL)^2}\,.
\end{align}

The interaction terms between $\Phi$ and $\Psi$ in the second line of \eqref{SH} correspond to the vertices
\begin{align}
	\begin{tikzpicture}[vertical align]
		\draw [very thick] (60:1) -- (0,0) -- node[right] {$a$} (-60:1);
		\draw [very thick, snake it] (-1,0) -- (0,0); 
		\draw [->] (-.9,.25) ..controls (-.1,.25) and ($(60:.1)+(150:.15)$).. ($(60:.9)+(150:.15)$) node[left] {$\vec\alpha$};
	\end{tikzpicture}&=-i\vec\alpha_a, &  \begin{tikzpicture}[vertical align]
	\draw [very thick] (60:1)-- (0,0) -- (-60:1);
	\draw [->] ($(-60:.9)+(30:.15)$) ..controls ($(-60:.1)+(30:.15)$) and ($(60:.1)+(-30:.15)$).. ($(60:.9)+(-30:.15)$) node[right] {$\vec\alpha$};
	\draw [very thick, snake it] (-1,0) -- node[above] {$a$} (0,0); 
	\end{tikzpicture}&= -\frac{i}{2}\vec \alpha_a\, ,\\
	\begin{tikzpicture}[vertical align]
		\draw [very thick] (60:1) -- (0,0) -- (-60:1);
		\draw [->,shift={(-30:.2)}] (60:.2) -- node[right] {$\vec\alpha''$} (60:.9);
		\draw [->,shift={(30:.2)}] (-60:.9) -- node[right] {$\vec\alpha$} (-60:.2);
		\draw [very thick, snake it] (-1,0) -- (0,0); 
		\draw [->,shift={(90:.3)}] (-.9,0) -- node[above] {$\vec\alpha'$} (-.2,0);
	\end{tikzpicture}&=-\frac{i}{2}N_{\vec \alpha,\vec \alpha'}^{\vec \alpha''}\,,
\end{align}
where the structure constants $N_{\vec \alpha,\vec \alpha'}^{\vec \alpha''}$ are those appearing in \eqref{Relns}, and momentum/frequency labels and momentum/frequency-conserving delta functions have been dropped for clarity.

The terms in \eqref{Sint} couple the light modes to the heavy modes via the following vertices (blue external lines will always denote light modes):
\begin{align}\label{eq:external_vertices}
	\begin{tikzpicture}[vertical align]
		\draw [very thick] (60:1) -- (0,0) -- (-60:1);
	\draw [->] ($(-60:.9)+(30:.15)$) ..controls ($(-60:.1)+(30:.15)$) and ($(60:.1)+(-30:.15)$).. ($(60:.9)+(-30:.15)$) node[right] {$\vec\alpha$};
		\draw [very thick, snake it,color=bleudefrance] (-1,0) node[left] {$a$} -- (0,0); 
		\draw[->] (-.8,.3) -- node[above] {$\omega$} (-.2,.3);
	\end{tikzpicture}&=-\frac{(gL)^{-\frac 12}}{2} \gamma^5 \vec \alpha_a \bar q_a(\omega)\,,&\begin{tikzpicture}[vertical align]
		\draw [very thick, snake it] (60:1) -- (0,0) -- (-60:1);
	\draw [->] ($(-60:.9)+(30:.2)$) ..controls ($(-60:.1)+(30:.2)$) and ($(60:.1)+(-30:.25)$).. ($(60:.9)+(-30:.25)$) node[right] {$\vec\alpha, n$};
		\draw [very thick, snake it,color=bleudefrance] (-1,0) node[left] {$a$} -- (0,0); 
		\draw[->] (-.8,.3) -- node[above] {$\omega$} (-.2,.3);
	\end{tikzpicture}&=-(gL)^{-\frac 12} (2\pi n-\vec a_*\cdot \vec \alpha)\vec \alpha_a \bar q_a(\omega)\,,\nonumber \\
	\begin{tikzpicture}[vertical align]
		\draw [very thick, dashed] (60:1) -- (0,0) -- (-60:1);
	\draw [->] ($(-60:.9)+(30:.15)$) ..controls ($(-60:.1)+(30:.15)$) and ($(60:.1)+(-30:.15)$).. ($(60:.9)+(-30:.15)$) node[right] {$\vec\alpha$};
		\draw [very thick, snake it,color=bleudefrance] (-1,0) node[left] {$a$} -- (0,0); 
		\draw[->] (-.8,.3) -- node[above] {$\omega$} (-.2,.3);
	\end{tikzpicture}&=(gL)^{-\frac 12}\vec \alpha_a \bar q_a(\omega)\, , &
	\begin{tikzpicture}[vertical align]
		\draw [very thick] (60:1) -- (0,0);
		\draw [very thick, snake it] (0,0) -- (-60:1);
	\draw [->] ($(-60:.9)+(30:.2)$) ..controls ($(-60:.1)+(30:.2)$) and ($(60:.1)+(-30:.15)$).. ($(60:.9)+(-30:.15)$) node[right] {$\vec\alpha$};
		\draw [very thick,color=bleudefrance] (-1,0) node[left] {$a$} -- (0,0); 
		\draw[->] (-.8,.2) -- node[above] {$\omega$} (-.2,.2);
	\end{tikzpicture} &= i\vec \alpha_a \chi_a(\omega)\,,\\
	\begin{tikzpicture}[vertical align]
		\draw [very thick, snake it] (45:1) -- (0,0);
		\draw [very thick, snake it] (-45:1) -- (0,0);
	\draw [->] ($(-45:.9)+(45:.25)$) ..controls ($(-45:.1)+(45:.25)$) and ($(45:.1)+(-45:.25)$).. ($(45:.9)+(-45:.25)$) node[right] {$\vec\alpha$};
		\draw [very thick, snake it,color=bleudefrance] (135:1) node[left]{$a$} -- (0,0);
		\draw [shift={(45:.2)},->] (135:.9) -- node[above right,shift={(-135:.1)}] {$\omega_1$} (135:.3);
		\draw [shift={(-45:.2)},->] (-135:.9) -- node[below right,shift={(135:.1)}] {$\omega_2$} (-135:.3);
		\draw [very thick, snake it,color=bleudefrance] (-135:1) node[left]{$b$} -- (0,0);
	\end{tikzpicture} &=\frac{1}{2}\vec \alpha_a \vec \alpha_b \bar q_a(\omega_1)\bar q_b(\omega_2)\,.\nonumber
\end{align}
Again we have dropped most of the momentum/frequency labels, and all of the momentum/frequency-conserving delta functions.

By following these Feynman rules, we would find divergent integrals for some diagrams. See Appendix \ref{sec:tadpole_diagram} for an example. To remedy this, we point-split the interaction vertices connecting a gauge boson with two fermions. In momentum space, the point-splitting procedure corresponds to multiplying the vertex by a factor
\begin{align}\label{eq:point_splitting}
	\begin{tikzpicture}[vertical align]
		\draw [very thick, gray,snake it] (-1,0) --(0,0);
		\draw [thick,->](-.8,0.3) --node[above] {$\omega,\vec\alpha$}   (-.2,.3);
		\draw [very thick,gray] (0,0) --  (60:1);
		\draw [very thick,gray] (0,0) --  (-60:1);
	\end{tikzpicture} &\rightarrow e^{i\frac{\epsilon}{2}\sgn(\vec\alpha)\omega},
	&
	\begin{tikzpicture}[vertical align]
		\draw [very thick, gray,snake it] (-1,0) -- node[below] {$a$} (0,0);
		\draw [thick,->](-.8,0.3) --node[above] {$\omega$}   (-.2,.3);
		\draw [very thick,gray] (0,0) --  (60:1);
		\draw [thick,<-] (60:.2)   ++(-30:.2) -- node[right] {$\omega_2,\vec\alpha_2$} ++ (60:.6);
		\draw [very thick,gray] (0,0) --  (-60:1);
		\draw [thick,<-] (-60:.2)   ++(30:.2) -- node[right] {$\omega_1,\vec\alpha_1$}++ (-60:.6);
	\end{tikzpicture} &\rightarrow e^{i\frac{\epsilon}{2}\sgn(\vec \alpha_1-\vec\alpha_2)(\omega_1- \omega_2)}.
\end{align}
The gray lines indicate that the same procedure applies whether the lines are for light or heavy fields.

\subsection{Perturbative calculation}\label{sec:diagrams}

We will now use the Feynman rules derived in the previous section to compute the effective action up to order $(gL)^3$, following \eqref{eq:cumulant_exp}. We group the terms according to their order in $gL$ and the light fields they involve.

From the Feynman rules, we see that heavy propagators all carry factors of $gL$, and we can effectively treat all external gluon legs as carrying factors of $(gL)^{-1/2}$ provided we keep track of an extra factor of $gL$ for the four-point vertex with two external gluon lines in \eqref{eq:external_vertices}. Furthermore, each loop integral will contribute a factor of $(gL)^{-1}$. 

For a particular Feynman diagram, let $n_g$ be the number of external gluon legs, $n_f$ be the number of external fermion legs, $n_4$ be the number of four-point vertices, $n_i$ be the number of internal vertices (all of which are three-point), and $n_p$ be the number of heavy propagators. The number of loops $n_\ell$ is then given by
\begin{equation}
	n_\ell = n_p - ((n_g - n_4) + n_f + n_i) + 1\,.
\end{equation}
Thus, the lowest order at which such a diagram could contribute is
\begin{equation}
	(gL)^{n_p + n_4 - n_g/2 - n_\ell} = (gL)^{n_f + n_g/2 - 1} (gL)^{n_i}\,,
\end{equation}
and so for a given set of external legs we can simply organize the diagrams by the number of internal vertices. For some diagrams, the leading-order contribution suggested by this argument will turn out to vanish.

\subsubsection{The cancellation of diagrams without fermions}

We can construct a family of diagrams with no external fermion legs or internal heavy fermions. Equivalently, these diagrams have only external gluon legs and a single internal loop of either the heavy gluon fields or the ghost fields. These are the diagrams we would calculate in order to derive the effective action of a pure gauge theory on a small circle. The pure gauge theory would have no corrections beyond the leading-order effective action, and we can test our setup by showing this explicitly.

As an example, consider the two diagrams with a single external gluon leg and only ghost and gluon internal propagators:
\begin{equation}
	D_\text{glue} = \begin{tikzpicture}[vertical align]
		\draw[very thick, snake it, color=bleudefrance] (-1.2,0) -- (-0.4,0);
		\draw[very thick, snake it] (0,0) circle (0.4);
	\end{tikzpicture}\,,\qquad
	D_\text{ghost} = \begin{tikzpicture}[vertical align]
		\draw[very thick, snake it, color=bleudefrance] (-1.2,0) -- (-0.4,0);
		\draw[very thick, dashed] (0,0) circle (0.4);
	\end{tikzpicture}\,.
\end{equation}
Following our Feynman rules, we find
\begin{equation}
\begin{split}
	D_\text{glue} &= -\frac{\bar q_a(\omega)}{\sqrt{gL}} \sum_{\vec\alpha} \vec{\alpha}_a \left(\sum_{n=-\infty}^\infty \int \frac{d\omega'}{2\pi}(2\pi n - \vec{a}_*\cdot\vec\alpha) D_{\vec{a}_*\cdot\vec{\alpha}}(\omega', n)\right)\,,\\
	D_\text{ghost} &= -\frac{\bar q_a(\omega)}{\sqrt{gL}} \sum_{\vec\alpha} \vec{\alpha}_a \left(\sum_{n=-\infty}^\infty \int \frac{d\omega'}{2\pi} H_{\vec{a}_*\cdot\vec{\alpha}}(\omega', n)\right)\,.
\end{split}
\end{equation}
Note there is an extra factor of $-1$ in $D_\text{ghost}$ due to the loop of the anticommuting ghost. Since $(2\pi n-Q)D_Q(\omega', n) = -H_Q(\omega', n)$, we see that $D_\text{glue} + D_\text{ghost} = 0$.

We can generalize this reasoning to diagrams with $n_g$ external gluon legs. There is only one way we can draw a diagram with ghosts, and we need to include a factor of $(n_g-1)!$ for the different contractions and another factor of $\frac{1}{n_g!}$ for the identical vertices. Thus, we find
\begin{equation}\label{eq:ghost_loop}
	\begin{tikzpicture}[vertical align]
		\foreach \t in {-135,-90,...,135}
			\draw[very thick, snake it, color=bleudefrance] (\t:1.2) -- (\t:.4);
		\draw[very thick, dashed] (0,0) circle (0.4);
		\foreach \t in {150,165,...,210} {
			\node[fill,circle,minimum size=2,inner sep=0] at (\t:.9) {};
		};
	\end{tikzpicture} = \frac{(-1)^{n_g+1}}{n_g}\frac{\bar q_{a_1}(\omega_1)\cdots \bar q_{a_{n_g}}(\omega_{m_g})}{\sqrt{gL}} \sum_{\vec\alpha} \left(\vec{\alpha}_{a_1}\cdots \vec{\alpha}_{a_{n_g}} \sum_{n=-\infty}^\infty \int \frac{d\omega'}{2\pi} \frac{1}{(2\pi n-\vec{a}_*\cdot\vec{\alpha})^{n_g}}\right)\,.
\end{equation}
For the diagrams with gluon loops, we can have $0\le n_4 \le \lfloor n_g/2\rfloor$ four-point vertices. The power of $(2\pi n-\vec{a}_*\cdot\vec{\alpha})$ in the integrand will be $-2(n_g-n_4) + (n_g-2n_4) = -n_g$, and so each diagram will yield an expression with the same structure as \eqref{eq:ghost_loop} but with a different prefactor. There is a factor of $\frac{1}{n_4! (n_g-2n_4)!}$ for identical vertices, a factor of $2^{n_g-n_4-1}(n_g-n_4-1)!$ for different contractions, a sign of $(-1)^{n_g-n_4}$ from the three-point vertex factors and a factor of $(1/2)^{n_4}$ for the four-point vertex factors. Combining these and summing over $n_4$ gives
\begin{equation}
	\sum_{n_4 = 0}^{\lfloor n_g/2\rfloor} \frac{(-1)^{n_g-n_4}2^{n_g - 2n_4 - 1}}{n_g-n_4} \binom{n_g-n_4}{n_4} = \frac{(-1)^{n_g}}{n_g}\,.
\end{equation}
Thus, this sum cancels the contribution from the ghost diagram, and so at any order in $gL$ we can ignore all diagrams that do not involve fermion lines.

\subsubsection{The vanishing of one-particle reducible diagrams}
If a diagram can be disconnected by cutting a single internal fermion or gluon line, then it must vanish. Indeed, consider a generic such diagram:
\begin{align}
	\begin{tikzpicture}[vertical align]
		\draw [very thick, snake it] (-1,0) -- (0,0);
		\draw[very thick,pattern=north east lines] (-1.5,0) circle (.5);
		\draw[very thick,pattern=north east lines] (.5,0) circle (.5);
		\draw[very thick,bleudefrance,snake it,xshift=-1.5cm] (120:.5) -- (120:1.5);
		\draw[very thick,bleudefrance,snake it,xshift=-1.5cm] (180:.5) -- (180:1.5);
		\draw[very thick,bleudefrance,snake it,xshift=-1.5cm] (240:.5) -- (240:1.5);
		\draw[very thick,bleudefrance,xshift=.5cm] (60:.5) -- (60:1.5);
		\draw[very thick,bleudefrance,snake it,xshift=.5cm] (0:.5) -- (0:1.5);
		\draw[very thick,bleudefrance,xshift=.5cm] (-60:.5) -- (-60:1.5);
	\end{tikzpicture}
\end{align}
The light external fields all have zero momentum and are aligned with a Cartan direction. Conservation of momentum and the structure of the vertices would then require that the internal line also has zero momentum and is aligned with a Cartan direction, but then it would not be one of the heavy fields we are integrating out and the diagram should be discarded.

\subsubsection{Non-vanishing diagrams}

We will now give the values of diagrams that contribute to the effective action up to order $(gL)^3$. The diagrams that make nonzero contributions are summarized in Table~\ref{tab:diagrams}. We have checked that diagrams with any other set of external legs only give nonzero contributions at $\mathcal{O}\left((gL)^{7/2}\right)$ or higher.

\begingroup
\renewcommand{\arraystretch}{1.5}
\begin{table}[!htp]
    \centering
    \begin{tabular}{>{\centering\arraybackslash}m{3cm}>{\centering\arraybackslash}m{6cm}>{\centering\arraybackslash}m{6cm}}
        \toprule
        \textbf{External legs} & \multicolumn{2}{c}{\textbf{Internal vertices}} \\ 
        ~ & \textbf{0} & \textbf{2} \\
        \midrule
        \( q \) & \begin{tikzpicture}[scale=.75]\label{eq:linear_dia}
		\draw [very thick, snake it, color=bleudefrance] (-1,0) -- (0,0);
		\draw[very thick] (1,0) arc (0:360:0.5);
	\end{tikzpicture} & \begin{tikzpicture}[scale=.75]
		\draw [very thick, snake it, color=bleudefrance] (-1,0) -- (0,0);
		\draw[very thick,snake it] (.5,0) circle (0.5);
		\filldraw[very thick,fill=white,draw=black] (1,0) circle (0.25);
		\begin{scope}[xshift=2.5cm]
			\draw [very thick, snake it, color=bleudefrance] (-1,0) -- (0,0);
			\draw[very thick] (.5,0) circle (0.5);
			\draw[very thick,snake it] (.5,.5) -- ++ (0,-1);
		\end{scope}
	\end{tikzpicture} \\ 
        \( q^2 \) & \begin{tikzpicture}[scale=.75]
		\draw [very thick, snake it, color=bleudefrance] (-1,0) -- (0,0);
		\draw [very thick, snake it, color=bleudefrance] (1,0) -- (2,0);
		\draw[very thick] (1,0) arc (0:360:0.5);
	\end{tikzpicture} & \begin{tikzpicture}[scale=.75]
		\draw[bleudefrance, very thick, snake it] (0,0) -- (1,0) (3,0) -- (2,0); 
		\draw[very thick, snake it] (1.5,0) circle (.5); 
		\draw[very thick, fill=white] (1.5,.5) circle (.25);
	\end{tikzpicture}\\
        \( \chi^2 \) & \begin{tikzpicture}[scale=.75]
		\draw [very thick, color=bleudefrance] (-1,0) -- (0,0);
		\draw[very thick, snake it] (1,0) arc (0:180:0.5);
		\draw[very thick] (1,0) arc (0:-180:0.5);
		\draw [very thick, color=bleudefrance] (1,0) -- (2,0);
	\end{tikzpicture} & ~ \\
        \( q^3 \) & \begin{tikzpicture}[scale=.75]
		\draw[very thick, color=bleudefrance,snake it] (120:0.5) -- (120:1.2);
		\draw[very thick, color=bleudefrance,snake it] (240:0.5) -- (240:1.2);
		\draw[very thick, color=bleudefrance,snake it] (0:0.5) -- (0:1.2);
		\draw [very thick] (0,0) circle (0.5);
	\end{tikzpicture} & \begin{tikzpicture}[scale=.75]
		\draw[very thick, color=bleudefrance,snake it] (120:0.5) -- (120:1.2);
		\draw[very thick, color=bleudefrance,snake it] (240:0.5) -- (240:1.2);
		\draw[very thick, color=bleudefrance,snake it] (0:0.5) -- (0:1.2);
		\draw [very thick,snake it] (0,0) circle (0.5);
		\draw[very thick,fill=white] (-.5,0) circle (0.25);
		\begin{scope}[xshift=2.5cm]
		\draw[very thick, color=bleudefrance,snake it] (-0.5,0) -- ++(135:.8);
		\draw[very thick, color=bleudefrance,snake it] (-0.5,0) -- ++(225:.8);
		\draw[very thick, color=bleudefrance,snake it] (0:0.5) -- (0:1.2);
		\draw [very thick,snake it] (0,0) circle (0.5);
		\draw[very thick,fill=white] (0,0.5) circle (0.25);
		\end{scope}
	\end{tikzpicture}\\
        \( q\chi^2 \) & \begin{tikzpicture}[scale=.75]
		\draw[very thick, color=bleudefrance] (110:0.5) -- (110:1.2);
		\draw[very thick, color=bleudefrance] (250:0.5) -- (250:1.2);
		\draw[very thick, color=bleudefrance,snake it] (0:0.5) -- (0:1.2);
		\draw [very thick] (110:0.5) arc (110:250:0.5);
		\draw [very thick,snake it] (110:0.5) arc (110:-110:0.5);
		\begin{scope}[xshift=2.5cm]
		\draw[very thick, color=bleudefrance] (110:0.5) -- (110:1.2);
		\draw[very thick, color=bleudefrance] (250:0.5) -- (250:1.2);
		\draw[very thick, color=bleudefrance,snake it] (0:0.5) -- (0:1.2);
		\draw [very thick,snake it] (250:0.5) arc (250:110:0.5);
		\draw [very thick] (110:0.5) arc (110:-110:0.5);
		\end{scope}
	\end{tikzpicture}& \\
        \( q^4 \) & \begin{tikzpicture}[scale=.75]
		\draw [very thick] (-.5,.5) -- (.5,.5);
		\draw [very thick] (-.5,-.5) -- (-.5,.5);
		\draw [very thick] (.5,.5) -- (.5,-0.5);
		\draw [very thick] (-.5,-.5) -- +(1,-0);
		\draw [very thick, snake it, color=bleudefrance] (-.5,.5) -- +(135:1);
		\draw [very thick, snake it, color=bleudefrance] (.5,.5) -- +(45:1);
		\draw [very thick, snake it, color=bleudefrance] (-.5,-.5) -- +(225:1);
		\draw [very thick, snake it, color=bleudefrance] (.5,-.5) -- +(-45:1);
	\end{tikzpicture} & \begin{tikzpicture}[scale=.75]
		\draw [very thick, snake it] (-.5,.5) -- (.5,.5);
		\draw [very thick, snake it] (-.5,-.5) -- (-.5,.5);
		\draw [very thick, snake it] (.5,.5) -- (.5,-0.5);
		\draw [very thick, snake it] (-.5,-.5) -- +(1,-0);
		\draw [very thick, snake it, color=bleudefrance] (-.5,.5) -- +(135:1);
		\draw [very thick, snake it, color=bleudefrance] (.5,.5) -- +(45:1);
		\draw [very thick, snake it, color=bleudefrance] (-.5,-.5) -- +(225:1);
		\draw [very thick, snake it, color=bleudefrance] (.5,-.5) -- +(-45:1);
		\draw[very thick,fill=white] (0,.5) circle (.3);
		\begin{scope}[xshift=3cm]
		\draw [very thick, snake it] (-.75,-.75) -- +(60:1.5);
		\draw [very thick, snake it] (.75,-.75) -- +(120:1.5);
		\draw [very thick, snake it] (-.75,-.75) -- +(1.5,-0);
		\draw [very thick, snake it, color=bleudefrance] (-.75,-.75)++(60:1.5) -- +(135:1);
		\draw [very thick, snake it, color=bleudefrance] (-.75,-.75)++(60:1.5) -- +(45:1);
		\draw [very thick, snake it, color=bleudefrance] (-.75,-.75) -- +(225:1);
		\draw [very thick, snake it, color=bleudefrance] (.75,-.75) -- +(-45:1);
		\draw[very thick,fill=white] (0,-.75) circle (.3);
		\end{scope}
		\begin{scope}[xshift=3cm,yshift=-3cm]
		\draw [very thick, snake it] (-.75,-.75) -- +(60:1.5);
		\draw [very thick, snake it] (.75,-.75) -- +(120:1.5);
		\draw [very thick, snake it] (-.75,-.75) -- +(1.5,-0);
		\draw [very thick, snake it, color=bleudefrance] (-.75,-.75)++(60:1.5) -- +(135:1);
		\draw [very thick, snake it, color=bleudefrance] (-.75,-.75)++(60:1.5) -- +(45:1);
		\draw [very thick, snake it, color=bleudefrance] (-.75,-.75) -- +(225:1);
		\draw [very thick, snake it, color=bleudefrance] (.75,-.75) -- +(-45:1);
		\draw[very thick,fill=white] (-0.75,-0.75)++(60:.75) circle (.3);
		\end{scope}
		\begin{scope}[yshift=-3cm]
		\draw [very thick, snake it] (0:.5) arc (0:180:.5);
		\draw [very thick, snake it] (180:.5) arc (180:360:.5);
		\draw[very thick,fill=white] (0,.5) circle (.3);
		\draw [very thick, snake it, color=bleudefrance] (.5,0) -- +(45:1);
		\draw [very thick, snake it, color=bleudefrance] (.5,0) -- +(-45:1);
		\draw [very thick, snake it, color=bleudefrance] (-.5,0) -- +(135:1);
		\draw [very thick, snake it, color=bleudefrance] (-.5,0) -- +(225:1);
		\end{scope}
	\end{tikzpicture} \\
        \( q^2\chi^2 \) & \begin{tikzpicture}[scale=.75]
		\draw [very thick, snake it] (-.5,.5) -- (.5,.5);
		\draw [very thick, snake it] (-.5,-.5) -- (-.5,.5);
		\draw [very thick, snake it] (.5,.5) -- (.5,-0.5);
		\draw [very thick] (-.5,-.5) -- +(1,-0);
		\draw [very thick, snake it, color=bleudefrance] (-.5,.5) -- +(135:1);
		\draw [very thick, snake it, color=bleudefrance] (.5,.5) -- +(45:1);
		\draw [very thick, color=bleudefrance] (-.5,-.5) -- +(225:1);
		\draw [very thick, color=bleudefrance] (.5,-.5) -- +(-45:1);
		\begin{scope}[xshift=3cm]
		\draw [very thick, snake it] (-.5,-.5) -- +(60:1);
		\draw [very thick, snake it] (.5,-.5) -- +(120:1);
		\draw [very thick] (-.5,-.5) -- +(1,-0);
		\draw [very thick, snake it, color=bleudefrance] (-.5,-.5)++(60:1) -- +(45:1);
		\draw [very thick, snake it, color=bleudefrance] (-.5,-.5)++(60:1) -- +(135:1);
		\draw [very thick, color=bleudefrance] (-.5,-.5) -- +(225:1);
		\draw [very thick, color=bleudefrance] (.5,-.5) -- +(-45:1);
		\end{scope}
		\begin{scope}[yshift=-3cm]
		\draw [very thick, snake it] (-.5,.5) -- (.5,.5);
		\draw [very thick] (-.5,-.5) -- (-.5,.5);
		\draw [very thick, snake it] (.5,.5) -- (.5,-0.5);
		\draw [very thick] (-.5,-.5) -- +(1,-0);
		\draw [very thick, color=bleudefrance] (-.5,.5) -- +(135:1);
		\draw [very thick, snake it, color=bleudefrance] (.5,.5) -- +(45:1);
		\draw [very thick, snake it, color=bleudefrance] (-.5,-.5) -- +(225:1);
		\draw [very thick, color=bleudefrance] (.5,-.5) -- +(-45:1);
		\end{scope}
		\begin{scope}[xshift=3cm,yshift=-3cm]
		\draw [very thick] (-.5,.5) -- (.5,.5);
		\draw [very thick] (-.5,-.5) -- (-.5,.5);
		\draw [very thick] (.5,.5) -- (.5,-0.5);
		\draw [very thick, snake it] (-.5,-.5) -- +(1,-0);
		\draw [very thick, snake it, color=bleudefrance] (-.5,.5) -- +(135:1);
		\draw [very thick, snake it, color=bleudefrance] (.5,.5) -- +(45:1);
		\draw [very thick, color=bleudefrance] (-.5,-.5) -- +(225:1);
		\draw [very thick, color=bleudefrance] (.5,-.5) -- +(-45:1);
		\end{scope}
	\end{tikzpicture} & \\
        \bottomrule
    \end{tabular}
    \caption{The diagrams with nonzero contributions to the effective action \eqref{SLAgain}. Several other diagrams with two internal vertices can be drawn, but only contribute at $\mathcal{O}\left((gL)^{7/2}\right)$ or higher.}
    \label{tab:diagrams}
\end{table}
\endgroup

The diagrams will involve the following tensor structures constructed from the roots:
\begin{equation}\label{eq:tensors}
\begin{split}
	A_a&\equiv \sum_{\vec \alpha >0}\alpha ^a \sum_n \frac{\sgn(2 \pi n - \vec{\hol}_* \cdot \vec{\rt})}{\abs{2 \pi n - \vec{\hol}_* \cdot \vec{\rt}}^2} \,, \\
	B_{ab} &\equiv \sum_{\vec \rt>0} \rt^a \rt^{b} \sum_n \frac{1}{\abs{2 \pi n - \vec{\hol}_* \cdot \vec{\rt}}^3}\,, \\
	C_{abc} &\equiv \sum_{\vec \rt>0} \rt^a \rt^{b} \rt^c  \sum_n \frac{\sgn(2 \pi n - \vec{\hol}_* \cdot \vec{\rt})}{\abs{2 \pi n - \vec{\hol}_* \cdot \vec{\rt}}^4}\,, \\
	D_{abcd} &\equiv \sum_{\vec \rt>0} \rt^a \rt^{b} \rt^c \rt^d \sum_n \frac{1}{\abs{2 \pi n - \vec{\hol}_* \cdot \vec{\rt}}^5} \,.
\end{split}
\end{equation}

The two contributing diagrams with one external gluon leg are
\begin{equation}\label{eq:q_diagrams}
\begin{split}
	\begin{tikzpicture}[vertical align]
		\draw [very thick, snake it, color=bleudefrance] (-1,0) -- (0,0);
		\draw[very thick] (1,0) arc (0:360:0.5);
	\end{tikzpicture}&=-\frac{(gL)^{\frac 32}m^2}{2 g^2}\int d\tau\,  A_a \bar q_a+\mathcal O\left((gL)^{7/2}\right)\,,\\
	\begin{tikzpicture}[vertical align]
		\draw [very thick, snake it, color=bleudefrance] (-1,0) -- (0,0);
		\draw[very thick,snake it] (.5,0) circle (0.5);
		\filldraw[very thick,fill=white,draw=black] (1,0) circle (0.25);
	\end{tikzpicture}&=\frac{(gL)^{\frac 32}h^\vee}{2 \pi}\int d\tau\,  A_a \bar q_a +\mathcal O\left((gL)^{7/2}\right)\,.
\end{split}
\end{equation}
See Appendix \ref{app:feynman} for the details of how these diagrams are computed. The other diagram we could draw with two internal vertices vanishes at $\mathcal{O}\left((gL)^{3/2}\right)$, and its subleading terms are too suppressed to contribute to our effective action:
\begin{equation}
	\begin{tikzpicture}[vertical align]
		\draw [very thick, snake it, color=bleudefrance] (-1,0) -- (0,0);
		\draw[very thick] (.5,0) circle (0.5);
		\draw[very thick,snake it] (.5,.5) -- ++ (0,-1);
	\end{tikzpicture}=\mathcal O ((gL)^{\frac 72})\,.
\end{equation}

The three contributing diagrams with two external gluon legs are
\begin{equation}\label{eq:lo_pot_dia}
\begin{split}
	\begin{tikzpicture}[vertical align]
		\draw [very thick, snake it, color=bleudefrance] (-1,0) -- (0,0);
		\draw [very thick, snake it, color=bleudefrance] (1,0) -- (2,0);
		\draw[very thick] (1,0) arc (0:360:0.5);
	\end{tikzpicture}&=\int d\tau\left[\frac{h^\vee}{4\pi}\bar q_a \bar q_a-(gL)^2\frac{m^2}{2g^2} B_{ab}\bar q_a \bar q_b\right]+\mathcal{O}((gL)^4)\,,\\
	\begin{tikzpicture}[vertical align]
		\draw[bleudefrance, very thick, snake it] (0,0) -- (1,0) (3,0) -- (2,0); 
		\draw[very thick, snake it] (1.5,0) circle (.5); 
		\draw[very thick, fill=white] (1.5,.5) circle (.25);
	\end{tikzpicture}&=\frac{(gL)^2h^\vee}{\pi}\int d\tau\, B_{ab}\bar q_a \bar q_b+\mathcal{O}((gL)^4)\,,\\
	\begin{tikzpicture}[vertical align]
		\draw[bleudefrance,very thick, snake it] (-1,-.5) -- (1,-0.5); 
		\draw[very thick, snake it] (0,0) circle (.5); 
		\draw[very thick,fill=white] (0,.5) circle (.25);
	\end{tikzpicture} &=-\frac{(gL)^2h^\vee}{4\pi}\int d\tau\, B_{ab}\bar q_a \bar q_b+\mathcal{O}((gL)^4)\,.
\end{split}
\end{equation}
Note that the leading order term in \eqref{eq:lo_pot_dia} reproduces the harmonic potential derived in \eqref{VeffFund2}. There are two other diagrams we could draw with two internal vertices, but they end up vanishing at $\mathcal{O}\left((gL)^2\right)$:
\begin{align}
	\begin{tikzpicture}[vertical align]
		\draw[bleudefrance,very thick, snake it] (0,0) -- (1,0); 
		\draw[very thick] (1.5,0) circle (.5); \draw[xshift=1.5cm,very thick, snake it] (150:.5) ..controls (120:1.5) and (60:1.5).. (30:.5); 
		\draw[bleudefrance,very thick, snake it] (2,0) -- (3,0);
	\end{tikzpicture}&=
	\begin{tikzpicture}[vertical align]
		\draw[bleudefrance, very thick, snake it] (0,0) -- (1,0) (3,0) -- (2,0); 
		\draw[very thick, snake it] (1.5,0) circle (.5); 
		\draw[very thick,fill=white] (2,0) circle (.25);
	\end{tikzpicture} = \mathcal{O}((gL)^4)\,.
\end{align}

The only relevant diagram with two external fermion legs is
\begin{align}
	\begin{tikzpicture}[vertical align]
		\draw [very thick, color=bleudefrance] (-1,0) -- (0,0);
		\draw[very thick, snake it] (1,0) arc (0:180:0.5);
		\draw[very thick] (1,0) arc (0:-180:0.5);
		\draw [very thick, color=bleudefrance] (1,0) -- (2,0);
	\end{tikzpicture}&=\frac{(gL)^2m}{2g}\int d\tau \, B_{ab}\chi_a^T \gamma^0 \chi _b+\mathcal{O}((gL)^4)\,.
\end{align}

The three contributing diagrams with three external gluon legs are
\begin{equation}
\begin{split}
	\begin{tikzpicture}[vertical align]
		\draw[very thick, color=bleudefrance,snake it] (120:0.5) -- (120:1.2);
		\draw[very thick, color=bleudefrance,snake it] (240:0.5) -- (240:1.2);
		\draw[very thick, color=bleudefrance,snake it] (0:0.5) -- (0:1.2);
		\draw [very thick] (0,0) circle (0.5);
	\end{tikzpicture}&=-\frac{(gL)^{\frac 52}m^2}{2g^2}\int d\tau\, C_{abc}\bar q_a\bar q_b\bar q_c+\mathcal O((gL)^{\frac{9}{2}})\,,\\
	\begin{tikzpicture}[vertical align]
		\draw[very thick, color=bleudefrance,snake it] (120:0.5) -- (120:1.2);
		\draw[very thick, color=bleudefrance,snake it] (240:0.5) -- (240:1.2);
		\draw[very thick, color=bleudefrance,snake it] (0:0.5) -- (0:1.2);
		\draw [very thick,snake it] (0,0) circle (0.5);
		\draw[very thick,fill=white] (-.5,0) circle (0.25);
	\end{tikzpicture}&=\frac{2(gL)^{\frac 52}h^\vee}{\pi}\int d\tau\, C_{abc}\bar q_a\bar q_b\bar q_c+\mathcal O((gL)^{\frac{9}{2}})\,,\\
	\begin{tikzpicture}[vertical align]
		\draw[very thick, color=bleudefrance,snake it] (-0.5,0) -- ++(135:.8);
		\draw[very thick, color=bleudefrance,snake it] (-0.5,0) -- ++(225:.8);
		\draw[very thick, color=bleudefrance,snake it] (0:0.5) -- (0:1.2);
		\draw [very thick,snake it] (0,0) circle (0.5);
		\draw[very thick,fill=white] (0,0.5) circle (0.25);
	\end{tikzpicture}&=-\frac{(gL)^{\frac 52}h^\vee}{\pi}\int d\tau\, C_{abc}\bar q_a \bar q_b \bar q_c+\mathcal O((gL)^{\frac{9}{2}})\,.
\end{split}
\end{equation}
There are six more diagrams with two internal vertices which end up not contributing:
\begin{equation}
\begin{split}
	&\begin{tikzpicture}[vertical align]
		\draw[very thick, color=bleudefrance,snake it] (120:0.5) -- (120:1.2);
		\draw[very thick, color=bleudefrance,snake it] (240:0.5) -- (240:1.2);
		\draw[very thick, color=bleudefrance,snake it] (0:0.5) -- (0:1.2);
		\draw [very thick,snake it] (0,.5) arc (90:270:0.5);
		\draw [very thick] (0,-.5) arc (270:450:0.5);
		\draw [very thick] (0,-.5) -- (0,.5);
	\end{tikzpicture}=
	\begin{tikzpicture}[vertical align]
		\draw[very thick, color=bleudefrance,snake it] (120:0.5) -- (120:1.2);
		\draw[very thick, color=bleudefrance,snake it] (240:0.5) -- (240:1.2);
		\draw[very thick, color=bleudefrance,snake it] (0:0.5) -- (0:1.2);
		\draw [very thick] (0,.5) arc (90:270:0.5);
		\draw [very thick,snake it] (0,-.5) arc (270:450:0.5);
		\draw [very thick] (0,-.5) -- (0,.5);
	\end{tikzpicture}=
	\begin{tikzpicture}[vertical align]
		\draw[very thick, color=bleudefrance,snake it] (120:0.5) -- (120:1.2);
		\draw[very thick, color=bleudefrance,snake it] (240:0.5) -- (240:1.2);
		\draw [very thick] (0,0) circle (0.5);
		\draw [line width = 1.3mm, snake it, color=white] (0,0) .. controls (.5,0.5) .. (1.2,0);
		\draw [very thick, snake it, color=bleudefrance] (0,0) .. controls (.5,0.5) .. (1.2,0);
		\draw [very thick, snake it] (-.5,0) -- (.5,0);
	\end{tikzpicture}\\
	={}&\begin{tikzpicture}[vertical align]
		\draw[very thick, color=bleudefrance,snake it] (120:0.5) -- (120:1.2);
		\draw[very thick, color=bleudefrance,snake it] (240:0.5) -- (240:1.2);
		\draw[very thick, color=bleudefrance,snake it] (0:0.5) -- (0:1.2);
		\draw [very thick] (0,0) circle (0.5);
		\filldraw[fill=white,color=white] (-.5,0) circle (0.25);
		\draw[very thick] (-.5,.25) arc (90:-90:0.25);
		\draw[very thick,snake it] (-.5,.25) arc (90:270:0.25);
	\end{tikzpicture}
	\begin{tikzpicture}[vertical align]
		\draw[very thick, color=bleudefrance,snake it] (120:0.5) -- (120:1.2);
		\draw[very thick, color=bleudefrance,snake it] (240:0.5) -- (240:1.2);
		\draw[very thick, color=bleudefrance,snake it] (0:0.5) -- (0:1.2);
		\draw [very thick] (0,0) circle (0.5);
		\draw[very thick, snake it] (0,.5) -- (0,-.5);
	\end{tikzpicture}=
	\begin{tikzpicture}[vertical align]
		\draw[very thick, color=bleudefrance,snake it] (-0.5,0) -- ++(135:.8);
		\draw[very thick, color=bleudefrance,snake it] (-0.5,0) -- ++(225:.8);
		\draw[very thick, color=bleudefrance,snake it] (0:0.5) -- (0:1.2);
		\draw [very thick,snake it] (0,.5) arc (90:270:0.5);
		\draw [very thick] (0,-.5) arc (270:450:0.5);
		\draw [very thick] (0,-.5) -- (0,.5);
	\end{tikzpicture}=\mathcal O((gL)^{\frac{9}{2}})\,.
\end{split}
\end{equation}

The two relevant diagrams with a single external gluon leg and two external fermion legs are
\begin{align}
	\begin{tikzpicture}[vertical align]
		\draw[very thick, color=bleudefrance] (120:0.5) -- (120:1.2);
		\draw[very thick, color=bleudefrance] (240:0.5) -- (240:1.2);
		\draw[very thick, color=bleudefrance,snake it] (0:0.5) -- (0:1.2);
		\draw [very thick] (120:0.5) arc (120:240:0.5);
		\draw [very thick,snake it] (120:0.5) arc (120:-120:0.5);
	\end{tikzpicture}&=\frac{(gL)^{\frac{5}{2}}m}{g}\int d\tau\, C_{abc} \bar q_c \chi_a^T \gamma^0 \chi_b+\mathcal{O}((gL)^{\frac{9}{2}})\,,\\
	\begin{tikzpicture}[vertical align]
		\draw[very thick, color=bleudefrance] (120:0.5) -- (120:1.2);
		\draw[very thick, color=bleudefrance] (240:0.5) -- (240:1.2);
		\draw[very thick, color=bleudefrance,snake it] (0:0.5) -- (0:1.2);
		\draw [very thick,snake it] (120:0.5) arc (120:240:0.5);
		\draw [very thick] (120:0.5) arc (120:-120:0.5);
	\end{tikzpicture} &= \frac{(gL)^{\frac{5}{2}}m}{2g}\int d\tau\, C_{abc} \bar q_c \chi_a^T \gamma^0 \chi_b+\mathcal{O}((gL)^{\frac{9}{2}})\,.
\end{align}

The five contributing diagrams with four external gluon legs are
\begin{align}
	\begin{tikzpicture}[vertical align]
		\draw [very thick] (-.5,.5) -- (.5,.5);
		\draw [very thick] (-.5,-.5) -- (-.5,.5);
		\draw [very thick] (.5,.5) -- (.5,-0.5);
		\draw [very thick] (-.5,-.5) -- +(1,-0);
		\draw [very thick, snake it, color=bleudefrance] (-.5,.5) -- +(135:1);
		\draw [very thick, snake it, color=bleudefrance] (.5,.5) -- +(45:1);
		\draw [very thick, snake it, color=bleudefrance] (-.5,-.5) -- +(225:1);
		\draw [very thick, snake it, color=bleudefrance] (.5,-.5) -- +(-45:1);
	\end{tikzpicture} &= -\frac{(gL)^{3}m^2}{2g^2}\int d\tau\, D_{abcd}\bar q_a\bar q_b\bar q_c\bar q_d +\mathcal O((gL)^5)\,,\\
	\begin{tikzpicture}[vertical align]
		\draw [very thick, snake it] (-.5,.5) -- (.5,.5);
		\draw [very thick, snake it] (-.5,-.5) -- (-.5,.5);
		\draw [very thick, snake it] (.5,.5) -- (.5,-0.5);
		\draw [very thick, snake it] (-.5,-.5) -- +(1,-0);
		\draw [very thick, snake it, color=bleudefrance] (-.5,.5) -- +(135:1);
		\draw [very thick, snake it, color=bleudefrance] (.5,.5) -- +(45:1);
		\draw [very thick, snake it, color=bleudefrance] (-.5,-.5) -- +(225:1);
		\draw [very thick, snake it, color=bleudefrance] (.5,-.5) -- +(-45:1);
		\draw[very thick,fill=white] (0,.5) circle (.3);
	\end{tikzpicture} &=\frac{4(gL)^{3}h^\vee}{\pi}\int d\tau\, D_{abcd}\bar q_a\bar q_b\bar q_c\bar q_d +\mathcal O((gL)^5)\,,\\
	\begin{tikzpicture}[vertical align]
		\draw [very thick, snake it] (-.75,-.75) -- +(60:1.5);
		\draw [very thick, snake it] (.75,-.75) -- +(120:1.5);
		\draw [very thick, snake it] (-.75,-.75) -- +(1.5,-0);
		\draw [very thick, snake it, color=bleudefrance] (-.75,-.75)++(60:1.5) -- +(135:1);
		\draw [very thick, snake it, color=bleudefrance] (-.75,-.75)++(60:1.5) -- +(45:1);
		\draw [very thick, snake it, color=bleudefrance] (-.75,-.75) -- +(225:1);
		\draw [very thick, snake it, color=bleudefrance] (.75,-.75) -- +(-45:1);
		\draw[very thick,fill=white] (0,-.75) circle (.3);
	\end{tikzpicture}&=-\frac{(gL)^{3}h^\vee}{\pi}\int d\tau\, D_{abcd}\bar q_a\bar q_b\bar q_c\bar q_d +\mathcal O((gL)^5)\,,\\
	\begin{tikzpicture}[vertical align]
		\draw [very thick, snake it] (-.75,-.75) -- +(60:1.5);
		\draw [very thick, snake it] (.75,-.75) -- +(120:1.5);
		\draw [very thick, snake it] (-.75,-.75) -- +(1.5,-0);
		\draw [very thick, snake it, color=bleudefrance] (-.75,-.75)++(60:1.5) -- +(135:1);
		\draw [very thick, snake it, color=bleudefrance] (-.75,-.75)++(60:1.5) -- +(45:1);
		\draw [very thick, snake it, color=bleudefrance] (-.75,-.75) -- +(225:1);
		\draw [very thick, snake it, color=bleudefrance] (.75,-.75) -- +(-45:1);
		\draw[very thick,fill=white] (-0.75,-0.75)++(60:.75) circle (.3);
	\end{tikzpicture}&=-\frac{2(gL)^{3}h^\vee}{\pi}\int d\tau\, D_{abcd}\bar q_a\bar q_b\bar q_c\bar q_d +\mathcal O((gL)^5)\,,\\
	\begin{tikzpicture}[vertical align]
		\draw [very thick, snake it] (0:.5) arc (0:180:.5);
		\draw [very thick, snake it] (180:.5) arc (180:360:.5);
		\draw[very thick,fill=white] (0,.5) circle (.3);
		\draw [very thick, snake it, color=bleudefrance] (.5,0) -- +(45:1);
		\draw [very thick, snake it, color=bleudefrance] (.5,0) -- +(-45:1);
		\draw [very thick, snake it, color=bleudefrance] (-.5,0) -- +(135:1);
		\draw [very thick, snake it, color=bleudefrance] (-.5,0) -- +(225:1);
	\end{tikzpicture}&=\frac{(gL)^{3}h^\vee}{4\pi}\int d\tau\, D_{abcd}\bar q_a\bar q_b\bar q_c\bar q_d +\mathcal O((gL)^5)\,,
\end{align}
There are eleven more diagrams with two internal vertices that do not contribute:
\begin{align}
	&\begin{tikzpicture}[vertical align]
		\draw [very thick, snake it] (-.5,.5) -- (.5,.5);
		\draw [very thick, snake it] (-.5,-.5) -- (-.5,.5);
		\draw [very thick, snake it] (.5,.5) -- (.5,-0.5);
		\draw [very thick, snake it] (-.5,-.5) -- +(1,-0);
		\draw [very thick, snake it, color=bleudefrance] (-.5,.5) -- +(135:1);
		\draw [very thick, snake it, color=bleudefrance] (.5,.5) -- +(45:1);
		\draw [very thick, snake it, color=bleudefrance] (-.5,-.5) -- +(225:1);
		\draw [very thick, snake it, color=bleudefrance] (.5,-.5) -- +(-45:1);
		\draw[very thick,fill=white] (.5,.5) circle (.3);
	\end{tikzpicture}=
	\begin{tikzpicture}[vertical align]
		\draw [very thick, snake it] (.75,-.75) arc (0:180:.75);
		\draw [very thick, snake it] (-.75,-.75) -- +(1.5,-0);
		\draw [very thick, snake it, color=bleudefrance] (0,0) -- +(135:1.4);
		\draw [very thick, snake it, color=bleudefrance] (0,0) -- +(45:1.4);
		\draw [very thick, snake it, color=bleudefrance] (-.75,-.75) -- +(225:1);
		\draw [very thick, snake it, color=bleudefrance] (.75,-.75) -- +(-45:1);
		\draw[very thick,fill=white] (0,0.) circle (.4);
	\end{tikzpicture}=
	\begin{tikzpicture}[vertical align]
		\draw [very thick, snake it] (.75,-.75) arc (0:180:.75);
		\draw [very thick, snake it] (-.75,-.75) -- +(1.5,-0);
		\draw [very thick, snake it, color=bleudefrance] (0,0) -- +(135:1.4);
		\draw [very thick, snake it, color=bleudefrance] (-.75,-.75) -- +(225:1);
		\draw [very thick, snake it, color=bleudefrance] (.75,-.75) -- +(-45:1);
		\draw[very thick,fill=white] (0,0.) circle (.4);
		\draw [line width = 1.3mm, snake it, color=white] (0,-.4) .. controls (0,0.2) .. (45:1.4);
		\draw [very thick, snake it, color=bleudefrance] (0,-.4) .. controls (0,0.2) .. (45:1.4);
		\draw[very thick] (0.4,0.) arc (0:-180:.4);
	\end{tikzpicture}=
	\begin{tikzpicture}[vertical align]
		\draw [very thick] (-.5,.5) -- (.5,.5);
		\draw [very thick] (-.5,-.5) -- (-.5,.5);
		\draw [very thick] (.5,.5) -- (.5,-0.5);
		\draw [very thick] (-.5,-.5) -- +(1,-0);
		\draw [very thick, snake it, color=bleudefrance] (-.5,.5) -- +(135:1);
		\draw [very thick, snake it, color=bleudefrance] (.5,.5) -- +(45:1);
		\draw [very thick, snake it, color=bleudefrance] (-.5,-.5) -- +(225:1);
		\draw [very thick, snake it, color=bleudefrance] (.5,-.5) -- +(-45:1);
		\draw[very thick, snake it,fill=white] (.5,.5)++(45:.4) arc (45:405:.4);
	\end{tikzpicture}\\
	&\quad=\begin{tikzpicture}[vertical align]
		\draw [very thick] (.75,-.75) arc (0:50:.75);
		\draw [very thick] (-.75,-.75) arc (180:130:.75);
		\draw [very thick] (-.75,-.75) -- +(1.5,-0);
		\draw [very thick, snake it, color=bleudefrance] (135:.5) -- +(135:1);
		\draw [very thick, snake it, color=bleudefrance] (-.75,-.75) -- +(225:1);
		\draw [very thick, snake it, color=bleudefrance] (.75,-.75) -- +(-45:1);
		\draw[very thick, snake it] (-19:0.5) arc (-19:135:.5);
		\draw[very thick, snake it] (135:0.5) arc (135:199:.5);
		\draw [line width = 1.3mm, snake it, color=white] (0,-.5) .. controls (0,0.1) .. (45:1.4);
		\draw [very thick, snake it, color=bleudefrance] (0,-.5) .. controls (0,0.1) .. (45:1.4);
		\draw[very thick, ] (-19:0.5) arc (-19:-161:.5);
	\end{tikzpicture}=
	\begin{tikzpicture}[vertical align]
		\draw [very thick] (.3,.5) arc (0:-180:.3);
		\draw [very thick,snake it] (.3,.5) arc (0:180:.3);
		\draw [very thick] (-.5,.5) -- (-.3,.5);
		\draw [very thick] (.5,.5) -- (.3,.5);
		\draw [very thick] (-.5,-.5) -- (-.5,.5);
		\draw [very thick] (.5,.5) -- (.5,-0.5);
		\draw [very thick] (-.5,-.5) -- +(1,-0);
		\draw [very thick, snake it, color=bleudefrance] (-.5,.5) -- +(135:1);
		\draw [very thick, snake it, color=bleudefrance] (.5,.5) -- +(45:1);
		\draw [very thick, snake it, color=bleudefrance] (-.5,-.5) -- +(225:1);
		\draw [very thick, snake it, color=bleudefrance] (.5,-.5) -- +(-45:1);
	\end{tikzpicture}=
	\begin{tikzpicture}[vertical align]
		\draw [very thick] (-.5,.5) -- (.5,.5);
		\draw [very thick] (-.5,-.5) -- (-.5,.5);
		\draw [very thick] (.5,.5) -- (.5,-0.5);
		\draw [very thick] (-.5,-.5) -- +(1,-0);
		\draw [very thick, snake it] (0,.5) arc (180:270:0.5);
		\draw [very thick, snake it, color=bleudefrance] (-.5,.5) -- +(135:1);
		\draw [very thick, snake it, color=bleudefrance] (.5,.5) -- +(45:1);
		\draw [very thick, snake it, color=bleudefrance] (-.5,-.5) -- +(225:1);
		\draw [very thick, snake it, color=bleudefrance] (.5,-.5) -- +(-45:1);
	\end{tikzpicture} =
	\begin{tikzpicture}[vertical align]
		\draw [very thick,snake it, color=bleudefrance] (0,0) -- +(45:1.5);
		\draw [very thick,snake it, color=bleudefrance] (0,0) -- +(-45:1.5);
		\draw [very thick,snake it, color=bleudefrance] (0,0) -- +(135:1.5);
		\draw [very thick,snake it, color=bleudefrance] (0,0) -- +(-135:1.5);
		\draw [very thick,fill=white] (0,0) circle (.5);
		\draw [very thick, snake it] (0,.5) -- (0,-.5);
	\end{tikzpicture}\\
	&\quad =\begin{tikzpicture}[vertical align]
		\draw [very thick, snake it] (-.75,-.75) -- +(60:1.5);
		\draw [very thick, snake it] (.75,-.75) -- +(120:1.5);
		\draw [very thick, snake it] (-.75,-.75) -- +(1.5,-0);
		\draw [very thick, snake it, color=bleudefrance] (-.75,-.75)++(60:1.5) -- +(135:1);
		\draw [very thick, snake it, color=bleudefrance] (-.75,-.75)++(60:1.5) -- +(45:1);
		\draw [very thick, snake it, color=bleudefrance] (-.75,-.75) -- +(225:1);
		\draw [very thick, snake it, color=bleudefrance] (.75,-.75) -- +(-45:1);
		\draw[very thick,fill=white] (-0.75,-0.75) circle (.3);
	\end{tikzpicture}=
	\begin{tikzpicture}[vertical align]
		\draw [very thick] (0,-.5) arc (-90:90:.5);
		\draw [very thick,snake it] (0,.5) arc (90:180:.5);
		\draw [very thick,snake it] (-.5,0) arc (180:270:.5);
		\draw [very thick] (0,.5) -- (0,-.5);
		\draw [very thick,snake it, color=bleudefrance] (-.5,0) -- +(135:1);
		\draw [very thick,snake it, color=bleudefrance] (-.5,0) -- +(225:1);
		\draw [very thick,snake it, color=bleudefrance] (0,0)++(45:.5) -- +(45:1);
		\draw [very thick,snake it, color=bleudefrance] (0,0)++(-45:.5) -- +(-45:1);
	\end{tikzpicture}=
	\begin{tikzpicture}[vertical align]
		\draw [very thick, snake it, color=bleudefrance] (0,0) -- +(135:1);
		\draw [very thick, snake it, color=bleudefrance] (0,0) -- +(225:1);
		\draw [very thick, snake it] (0,0) arc (180:0:.5);
		\draw [very thick, snake it] (0,0) arc (180:360:.5);
		\draw [very thick, snake it, color=bleudefrance] (1.,0) -- +(45:1.4);
		\draw [very thick,fill=white] (1.,0) circle (.4);
		\draw [very thick, snake it, color=white,line width=1.3mm] (1,0)++(-45:1.4) .. controls (1.2,0) .. (.6,0);
		\draw [very thick, snake it, color=bleudefrance] (1,0)++(-45:1.4) .. controls (1.2,0) .. (.6,0);
		\draw [very thick] (1,.4) arc (90:270:.4);
	\end{tikzpicture}  = \mathcal O((gL)^5)\, .
\end{align}

Finally, the four relevant diagrams with two external gluon legs and two external fermion legs are
\begin{align}
	\begin{tikzpicture}[vertical align]
		\draw [very thick, snake it] (-.5,.5) -- (.5,.5);
		\draw [very thick, snake it] (-.5,-.5) -- (-.5,.5);
		\draw [very thick, snake it] (.5,.5) -- (.5,-0.5);
		\draw [very thick] (-.5,-.5) -- +(1,-0);
		\draw [very thick, snake it, color=bleudefrance] (-.5,.5) -- +(135:1);
		\draw [very thick, snake it, color=bleudefrance] (.5,.5) -- +(45:1);
		\draw [very thick, color=bleudefrance] (-.5,-.5) -- +(225:1);
		\draw [very thick, color=bleudefrance] (.5,-.5) -- +(-45:1);
	\end{tikzpicture} &=\frac{2(gL)^3 m}{g}\int d\tau \, D_{abcd}\bar q_c\bar q_d\chi_a^T \gamma^0 \chi_b\,, \\
	\begin{tikzpicture}[vertical align]
		\draw [very thick, snake it] (-.5,-.5) -- +(60:1);
		\draw [very thick, snake it] (.5,-.5) -- +(120:1);
		\draw [very thick] (-.5,-.5) -- +(1,-0);
		\draw [very thick, snake it, color=bleudefrance] (-.5,-.5)++(60:1) -- +(45:1);
		\draw [very thick, snake it, color=bleudefrance] (-.5,-.5)++(60:1) -- +(135:1);
		\draw [very thick, color=bleudefrance] (-.5,-.5) -- +(225:1);
		\draw [very thick, color=bleudefrance] (.5,-.5) -- +(-45:1);
	\end{tikzpicture}&=-\frac{(gL)^3 m}{2g}\int d\tau \, D_{abcd}\bar q_c\bar q_d\chi_a^T \gamma^0 \chi_b\,,\\
	\begin{tikzpicture}[vertical align]
		\draw [very thick, snake it] (-.5,.5) -- (.5,.5);
		\draw [very thick] (-.5,-.5) -- (-.5,.5);
		\draw [very thick, snake it] (.5,.5) -- (.5,-0.5);
		\draw [very thick] (-.5,-.5) -- +(1,-0);
		\draw [very thick, color=bleudefrance] (-.5,.5) -- +(135:1);
		\draw [very thick, snake it, color=bleudefrance] (.5,.5) -- +(45:1);
		\draw [very thick, snake it, color=bleudefrance] (-.5,-.5) -- +(225:1);
		\draw [very thick, color=bleudefrance] (.5,-.5) -- +(-45:1);
	\end{tikzpicture} &= \frac{(gL)^3 m}{g}\int d\tau \, D_{abcd}\bar q_c\bar q_d\chi_a^T \gamma^0 \chi_b\,,\\
	\begin{tikzpicture}[vertical align]
		\draw [very thick] (-.5,.5) -- (.5,.5);
		\draw [very thick] (-.5,-.5) -- (-.5,.5);
		\draw [very thick] (.5,.5) -- (.5,-0.5);
		\draw [very thick, snake it] (-.5,-.5) -- +(1,-0);
		\draw [very thick, snake it, color=bleudefrance] (-.5,.5) -- +(135:1);
		\draw [very thick, snake it, color=bleudefrance] (.5,.5) -- +(45:1);
		\draw [very thick, color=bleudefrance] (-.5,-.5) -- +(225:1);
		\draw [very thick, color=bleudefrance] (.5,-.5) -- +(-45:1);
	\end{tikzpicture} &=\frac{(gL)^3 m}{2g}\int d\tau \, D_{abcd}\bar q_c\bar q_d\chi_a^T \gamma^0 \chi_b\, .
\end{align}

\subsection{Combined effective action}

By adding up all the diagrams from Section~\ref{sec:diagrams}, we find the Euclidean effective action $S_\text{eff} = \int d\tau\,L_\text{eff}$, where 
\begin{equation}\label{SLAgain}
\begin{split}
		L_\text{eff} &=   \frac 12 \dot{q}_a \dot{q}_a   +  \frac{h^\vee g^2}{4\pi} q_a q_a 
	+ \frac 12 \chi_a^T \dot{\chi}_a  + \frac{m}{2}  \chi_a^{T} \gamma^0 \chi_a  \\
	   &{}+ (gL)^{ \frac 32} g^{-\frac 12}\left( \frac{h^\vee g^2}{2 \pi} - \frac{m^2}{2} \right) A_{a} q_a
	   \\
	&{}+ (gL)^2 B_{ab} \left[  \left( \frac{3h^\vee g^2}{4 \pi} - \frac{m^2}{2} \right)  q_a q_b
	+   \frac{m}{2}   \chi_a^{T} \gamma^0 \chi_b \right] \\
	&{}+ (gL)^{\frac 52} g^{ \frac 12 } C_{abc} \left[  \left( \frac{h^\vee g^2}{\pi} - \frac{m^2}{2} \right) q_a q_b q_c
	+   \frac{3m}{2}  q_c  \chi_a^{T} \gamma^0 \chi_b \right] \\ 
	&{}+ (gL)^3 g  D_{abcd} \left[ \left( \frac{5 h^\vee g^2}{4 \pi} - \frac{ m^2}{2} \right)  q_a q_b q_c q_d + 3m  q_c q_d \chi_a^{T} \gamma^0 \chi_b  \right] \\
	&{}+ O((gL)^{7/2}) \,.
\end{split}
\end{equation}
The symmetric tensors appearing here are defined in \eqref{eq:tensors}.

It is straightforward to pass from \eqref{SLAgain} to the effective Hamiltonian:
\begin{equation}\label{Hamiltonian}
\begin{split}
		H_\text{eff} &=   \frac 12 p_a p_a   +  \frac{h^\vee g^2}{4\pi} q_a q_a 
	 + \frac{m}{2}  \chi_a^{T} \gamma^0 \chi_a  \\
	   &{}+ (gL)^{ \frac 32} g^{-\frac 12}\left( \frac{h^\vee g^2}{2 \pi} - \frac{m^2}{2} \right) A_{a} q_a
	   \\
	&{}+ (gL)^2 B_{ab} \left[  \left( \frac{3h^\vee g^2}{4 \pi} - \frac{m^2}{2} \right)  q_a q_b
	+   \frac{m}{2}   \chi_a^{T} \gamma^0 \chi_b \right] \\
	&{}+ (gL)^{\frac 52} g^{ \frac 12 } C_{abc} \left[  \left( \frac{h^\vee g^2}{\pi} - \frac{m^2}{2} \right) q_a q_b q_c
	+   \frac{3m}{2}  q_c  \chi_a^{T} \gamma^0 \chi_b \right] \\ 
	&{}+ (gL)^3 g  D_{abcd} \left[ \left( \frac{5 h^\vee g^2}{4 \pi} - \frac{ m^2}{2} \right)  q_a q_b q_c q_d + 3m  q_c q_d \chi_a^{T} \gamma^0 \chi_b  \right] \\
	&{}+ O((gL)^{7/2}) \,.
\end{split}
\end{equation}
This expression should be viewed as an expansion in the dimensionless parameter $gL$, with the extra powers of $g$ in the various terms appearing for dimensional reasons.  The term proportional to $(gL)^n$ will give a contribution to the energy proportional to $g (gL)^n$. Note that for $\grSU(N)$, the coefficient $A_a$ vanishes due to the center symmetry.

\subsection{Examples}

Let us provide a couple of small $N$ examples where the sums in \eqref{eq:tensors} can be evaluated in closed form. For $G = \grSU(2)$, the effective Hamiltonian can be written in terms of a single bosonic variable $q_1$, its conjugate momentum $p_1$, and a single fermionic variable $\chi_1$:
 \es{SU2Action}{
  H_\text{eff}^{\grSU(2)} &=   \frac 12 p_1^2 
 +  \frac{g^2}{2\pi} q_1^2 
+ \frac{m}{2}  \chi_1^{T} \gamma^0 \chi_1  \\
	&{}+ (gL)^2 \frac{7 \zeta(3)}{4 \pi^3}  \left[  \left( \frac{3 g^2}{2 \pi} - \frac{m^2}{2} \right)  q_1^2 
	+   \frac{m}{2}   \chi_1^{T} \gamma^0 \chi_1 \right] \\
	&{}+ (gL)^3 g  \frac{31 \zeta(5)}{16 \pi^5} \left[ \left( \frac{5 g^2}{2 \pi} - \frac{ m^2}{2} \right)  q_1^4 + 3m  q_1^2 \chi_1^{T} \gamma^0 \chi_1  \right] \\
	&{}+ O((gL)^{7/2}) \,.
 } 

For $G = \grSU(3)$, we have two bosonic and two fermionic variables, and the effective Hamiltonian is
 \es{SU3Action}{
	H_\text{eff}^{\grSU(3)} &=   \frac 12 \left (  p_1^2 + p_2^2 \right )
+  \frac{3 g^2}{4\pi} \left( q_1^2 + q_2^2 \right) 
+ \frac{m}{2}  \left( \chi_1^{T} \gamma^0 \chi_1 + \chi_2^{T} \gamma^0 \chi_2 \right)   \\
	&{}+ (gL)^2 \frac{39 \zeta(3)}{8 \pi^3}  \left[  \left( \frac{9 g^2}{4 \pi} - \frac{m^2}{2} \right)  \left[ q_1^2 + q_2^2 \right] 
	+   \frac{m}{2}   \left[ \chi_1^{T} \gamma^0 \chi_1 + \chi_2^{T} \gamma^0 \chi_2 \right]  \right] \\
	&{}+ (gL)^{\frac 52} g^{ \frac 12 } \frac{3 \left[ \zeta(4, \frac 13) - \zeta(4, \frac 23)  \right]}{64 \pi^4} \biggl[  \left( \frac{3 g^2}{\pi} - \frac{m^2}{2} \right) 
	 \left( - q_1^3+ 3 q_1 q_2^2 \right) \\
	&\qquad \qquad {}+   \frac{3m}{2}  \left( 2 q_2  \chi_1^{T} \gamma^0 \chi_2 + q_1  \chi_2^{T} \gamma^0 \chi_2 - q_1  \chi_1^{T} \gamma^0 \chi_1 \right)  \biggr] \\ 
	&{}+ (gL)^3 g  \frac{1089 \zeta(5)}{128 \pi^5} \biggl[ \left( \frac{15  g^2}{4 \pi} - \frac{ m^2}{2} \right)   \left[q_1^2 + q_2^2 \right]^2  \\
	 &\qquad \qquad {}+ m  
	 \left[ q_1^2 \left( 3 \chi_1^{T} \gamma^0 \chi_1 +  \chi_2^{T} \gamma^0 \chi_2 \right)
	  + 4 q_1 q_2 \chi_1^{T} \gamma^0 \chi_2 + q_2^2 \left( 3 \chi_2^{T} \gamma^0 \chi_2 +  \chi_1^{T} \gamma^0 \chi_1 \right)  \right]  \biggr] \\
	&{}+ O((gL)^{7/2}) \,,
 }
where $\zeta(s, a)$ is the Hurwitz zeta function.

\begin{figure}[t]
	\centering
	\begin{subfigure}[t]{0.444\linewidth}%
		\centering
		\includegraphics[width=\linewidth]{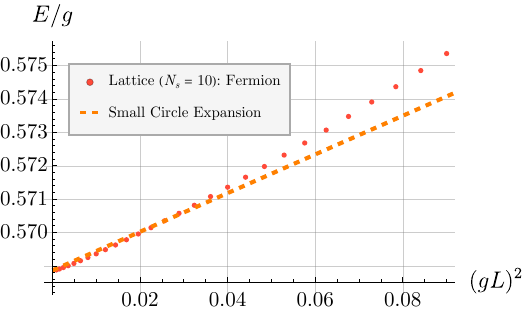}
		\caption{The lowest fermionic excitation for the $\grSU(2)$ lattice theory with $N_s = 10$ sites at $m = 0$, compared with the value derived from \eqref{SU2Action}.}
		\label{fig:lattice_energy_m0}
	\end{subfigure}%
	\hspace{.1\linewidth}%
	\begin{subfigure}[t]{0.44\linewidth}%
		\centering
		\includegraphics[width=\linewidth]{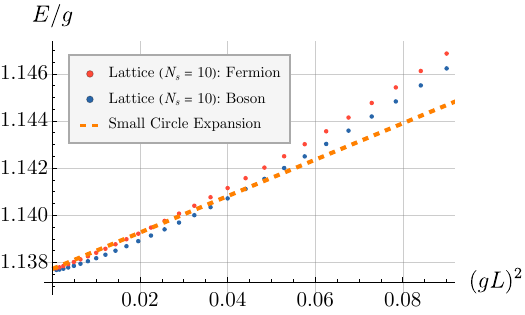}
		\caption{The lowest fermionic and bosonic excitations for the $\grSU(2)$ lattice theory with $N_s = 10$ sites at $m = m_\text{SUSY}$ (with this value slightly adjusted for the finite lattice), compared with the value derived from \eqref{SU2Action}. The slight splitting of the bosonic and fermionic levels is a finite-lattice effect.}
		\label{fig:lattice_energy_msusy}
	\end{subfigure}%
	\caption{}
	\label{fig:lattice}
\end{figure}

We can compare \eqref{SU2Action} with the numerical results obtained using the Hamiltonian lattice formulation of $\grSU(2)$ adjoint QCD$_2$ developed in \cite{Dempsey:2023fvm}. 
From \eqref{SU2Action}, we see that at $m = 0$ we have a harmonic oscillator of frequency
\begin{equation}
	\omega_B^{\grSU(2)}(m = 0) = \sqrt{\frac{g^2}{\pi}\left(1 + \frac{21\zeta(3)}{4\pi^3}(gL)^2\right)}
\end{equation}
plus perturbations of higher order in $gL$.  At least up to order $O(gL)^3$, the fermions completely decouple.  The spectrum is a deformation of that shown in Figure \ref{fig:su2_nonsusy}. 
Thus, in the spectrum of the $p = 0$ universe, the lowest fermionic excitation has energy $E_{F,1} - E_0 = \omega_B^{\grSU(2)}(m = 0)$. In Figure \ref{fig:lattice_energy_m0}, we compare the leading-order correction to this energy difference with lattice data using 10 sites, and find good agreement in the small-circle limit.

We can perform the same check at $m = m_\text{SUSY}$. There we have bosonic and fermionic oscillators of equal frequency
\begin{equation}
	\omega_B^{\grSU(2)}(m = m_\text{SUSY}) = \omega_F^{\grSU(2)}(m = m_\text{SUSY}) = \sqrt{\frac{g^2}{\pi}\left(1 + \frac{7\zeta(3)}{2\pi^3}(gL)^2\right)}\,,
\end{equation}
and the spectrum is a deformation of that shown in Figure \ref{fig:su2_susy}. 
Thus, in the $p = 0$ universe, the lowest bosonic excitation and lowest fermionic excitation each have energy
\begin{equation}
	E_{B,1} - E_0 = E_{F,1} - E_0 = \frac{2g}{\sqrt{\pi}} + \frac{7\zeta(3)}{2\pi^{7/2}}(gL)^2 + \mathcal{O}\left((gL)^3\right)\,.
\end{equation}
Comparing this leading-order correction with $N = 10$ lattice data in Figure \ref{fig:lattice_energy_msusy}, we again find good agreement. The slight splitting of the bosonic and fermionic levels on the lattice is due to the finite lattice size.

An important subtlety in both of these plots is that, on a finite lattice, the leading-order effective potential for the holonomy is not exactly given by \eqref{VeffFundSUN}. In particular, the frequency of the leading-order bosonic oscillator is not $\frac{g}{\sqrt{\pi}}$ but rather $\frac{g}{\sqrt{N_s\sin(\pi/N_s)}}$, where $N_s$ is the number of lattice sites. These of course agree in the large-$N_s$ limit, and for $N_s = 10$ the difference is less than 1\%. Nevertheless, in Figure \ref{fig:lattice} we have made this finite-$N_s$ adjustment so that the constant term in the expansion matches the $gL\to 0$ limit of the lattice data.

\section{Comments on the massless point}
\label{Comments}

As noted in Section~\ref{sec:symmetries} in the $G = \grSU(N)$ case, the adjoint QCD$_2$ theory for arbitrary $m$ is invariant under a $\Z_2$ fermion parity symmetry, a $\Z_2$ charge conjugation symmetry, and a $\Z_N$ one-form center symmetry. Analogous symmetries are also present for an arbitrary gauge group, with the only difference being that the precise center symmetry depends on $G$.  In the effective quantum mechanics, these symmetries of adjoint QCD$_2$ become ordinary symmetries generated by unitary operators ${\cal F}$, ${\cal C}$, and ${\cal U}$, respectively.

The massless point is distinguished in that it also exhibits a $\Z_2$ chiral symmetry, whose analog in the effective quantum mechanics is generated by the unitary operator ${\cal V}$.  However, at least to order $(gL)^3$ in the small circle limit something much stronger is true: as can be seen from \eqref{Hamiltonian}, when $m=0$ the effective Hamiltonian is completely independent of the fermions.  Since there are $\operatorname{rank}(G)$ two-component fermions, and each such fermion corresponds to a two-level system, it follows that every energy level of the effective theory is $2^{\operatorname{rank}(G)}$-degenerate.  One can relate the degenerate states by acting with the fermion zero modes $\chi_a$. 

This picture is reminiscent of the symmetries and vacuum degeneracy of the $G=\grSU(\Nc)$ adjoint QCD$_2$ theory in the infinite-volume limit.   As shown in \cite{Komargodski:2020mxz}, in the infinite-volume limit the adjoint QCD$_2$ theory is invariant under a large set of non-invertible zero-form symmetries, whose consequence is that there are $2^{N - 1}$ vacua related by the action of the corresponding topological defect lines, once the theory is compactified on a large circle.  The algebra of these line defects was determined in \cite{Komargodski:2020mxz} for small values of $N$.  One can take this comparison one step further and show that in the number of vacua in each universe obtained in \cite{Komargodski:2020mxz} (see also \cite{Bergner:2024ttq}) matches the corresponding quantity in our small circle effective theory, and moreover that the number of topological defect lines with a given charge under the center symmetry also matches the analogous count for the operators built out of the fermion zero modes in our effective theory.  

The counting in our effective theory proceeds as follows. Recall from Section~\ref{UFBasis} that instead of working with $\chi_a$, it is more convenient to work with the creation and annihilation operators $c_p^\dagger$ and $c_p$ defined in \eqref{eq:fermi_modes} ($1 \leq p \leq N-1$), which, as per \eqref{UCFVAction}, have charges $p$ (mod $N$) and $N - p$ (mod $N$) respectively, under the generator ${\cal U}$ of the $\Z_N$ center symmetry.  Assuming that the Fock vacuum $\ket{0}$ has vanishing $\Z_N$ charge, we can immediately determine the $\Z_N$ charges of the states obtained by acting with any number of $c^\dagger$'s on $\ket{0}$, and also the charges of the operators built out of the $c$ and the $c^\dagger$ operators.

For $N=2$, we have $2^{N-1} = 2$ vacua, and the only fermionic creation and annihilation operators are $c_1^\dagger$ and $c_1$, both with $\Z_2$ charge $1$.  Thus, the vacua split as $2 = 1 + 1$, with one vacuum ($\ket{0}$) with $\Z_2$ charge $0$, and one vacuum ($c_1^\dagger \ket{0}$) with $\Z_2$ charge $1$.  We have four operators acting on the space of these two vacua, which split as $4 = 2 + 2$, with two operators (${\bf 1}$, $c_1^\dagger c_1$) of $\Z_2$ charge $0$, and two operators ($c_1^\dagger$, $c_1$) of $\Z_2$ charge $1$.

For $N=3$, we have $2^{N-1} = 4$ vacua.  The operators $c_1^\dagger$ and $c_2$ have $\Z_3$ charge $1$, while $c_2^\dagger$ and $c_1$ have $\Z_3$ charge $2$.  The four vacua split as $4 = 2+ 1+ 1$, with two vacua  ($\ket{0}$, $c_1^\dagger c_2^\dagger \ket{0}$) with $\Z_3$ charge $0$, one vacuum ($c_1^\dagger \ket{0}$) with $\Z_3$ charge $1$, and one vacuum ($c_2^\dagger \ket{0}$) with $\Z_3$ charge $2$.  The $16$ operators acting on the space of the $4$ vacua split as $16 = 6 + 5 + 5$ as follows:  $6$ operators (${\bf 1}$, $c_1^\dagger c_1$, $c_2^\dagger c_2$, $c_1^\dagger c_2^\dagger$, $c_1 c_2$, $c_1^\dagger c_2^\dagger c_1 c_2$) with $\Z_3$ charge $0$, $5$ operators  ($c_1^\dagger$, $c_2$, $c_2^\dagger c_1$, $c_1^\dagger c_1 c_2$, $c_1^\dagger c_2^\dagger c_2$) of $\Z_3$ charge $1$, and $5$ operators  ($c_2^\dagger$, $c_1$, $c_1^\dagger c_2$, $c_2^\dagger c_2 c_1$, $c_1^\dagger c_2^\dagger c_1$) of $\Z_3$ charge~$2$.

For $N=4$, we have $2^{N-1} = 8$ vacua and $64$ operators acting on them.  The splitting of the $8$ vacua into the four universes is $8 = 2 + 2 + 2 + 2$, and the splitting of the $64$ operators according to their $\Z_4$ charges is $64 = 16 + 16 + 16 + 16$.

For $N=5$, we have $2^{N-1} = 16$ vacua and $256$ operators acting on them.  The splitting of the $16$ vacua into the five universes is $16 = 4 + 3 + 3 + 3 + 3$, and the splitting of the $256$ operators according to their $\Z_5$ charges is $256 = 52 + 51 + 51 + 51$.

For $N=6$, we have $2^{N-1} = 32$ vacua and $1024$ operators acting on them.  The splitting of the $32$ vacua into the six universes is $32 = 6 + 5 + 5 + 6 + 5 + 5$, and the splitting of the $1024$ operators according to their $\Z_6$ charges is $1024 = 172 + 170 + 170 + 172 + 170 + 170$.  

For $N=7$, we have $2^{N-1} = 64$ vacua and $4096$ operators acting on them.  The splitting of the $64$ vacua into the seven universes is $64 = 10 + 9 + 9 + 9 + 9 + 9 + 9$, and the splitting of the $4096$ operators according to their $\Z_7$ charges is $4096 = 586 + 585 + 585 + 585 + 585 + 585 + 585$.  It is straightforward to generalize this analysis for larger values of $N$.

The counting of the vacua in each universe matches those in Appendix~F of \cite{Komargodski:2020mxz} obtained for $2 \leq N \leq 6$ in the infinite volume limit, suggesting that these degeneracies may persist for any circle length.  Another check of this proposal is the following. Appendix~G of \cite{Komargodski:2020mxz} gives a list of topological lines of the $N=3$ theory.  Using the fact that the lines $\alpha^\pm_{m, n}$ have $\Z_3$ charge $m-n$ (mod $3$), one can deduce that out of the $16$ lines, $6$ have $\Z_3$ charge $0$, $5$ have $\Z_3$ charge $1$, and $5$ have $\Z_3$ charge $2$.  This count matches that of the operators acting on the fermionic Hilbert space given above.

\section{Supersymmetry}
\label{sec:susy}

In Section~\ref{sec:SUNLeading}, we saw explicitly by computing spectra that the leading-order effective theory
\begin{equation}\label{eq:leading_lagrangian}
	L = \frac{1}{2}\dot{q}_a \dot{q}_a + \frac{h^\vee g^2}{4\pi}q_a q_a + \frac{1}{2}\chi_a^T \partial_{\bar\tau} \chi_a + \frac{m}{2}\chi_a^T \gamma^0 \chi_a
\end{equation}
becomes supersymmetric at $m = \pm m_\text{SUSY}$, with $m_\text{SUSY} = g\sqrt{\frac{h^\vee}{2\pi}}$ as in \eqref{mSUSYGen}. We could also see this more abstractly from the supersymmetry transformations
\begin{equation}\label{eq:leading_transformation}
	\delta q_a = \epsilon^T \gamma^0 \chi_a, \qquad \delta \chi_a = \left(\mp \sqrt{\frac{h^\vee}{2\pi}} q_a + (\partial_{\bar\tau} q_a) \gamma^0\right)\epsilon\,.
\end{equation}
One can verify that the transformation of \eqref{eq:leading_lagrangian} vanishes up to a total derivative when $m = \pm m_\text{SUSY}$, and so this transformation leaves the action invariant.

In fact, the effective action in \eqref{SLAgain} is supersymmetric order-by-order, up to the highest order $(gL)^3$ that we have computed. To see this, we can write a more general supersymmetric action in terms of a superpotential $h(q_a)$:
\begin{equation}
	L = \frac{1}{2}\dot{q}_a \dot{q}_a + \left(\frac{\partial h}{\partial q_a}\right)\left(\frac{\partial h}{\partial q_a}\right) + \frac{1}{2}\chi_a^T \dot{\chi}_a - \left(\frac{\partial^2 h}{\partial q_a \partial q_b}\right)\chi^T_a \gamma^0 \chi_b\,.
\end{equation}
This action is invariant under the supersymmetry transformation \cite{Tong}
\begin{equation}
	\delta q_a = \epsilon^T \gamma^0 \chi_a, \qquad \delta \chi_a = \left\lbrack \left(\frac{\partial h}{\partial q_a}\right) + (\partial_{\bar\tau} q_a) \gamma^0\right\rbrack\epsilon\,.
\end{equation}
The action \eqref{eq:leading_lagrangian} and transformations \eqref{eq:leading_transformation} correspond to the choice $h = \mp \frac{m_\text{SUSY}}{g} q_a q_a$. 

We can add terms of higher order in $gL$ to the superpotential in order to reproduce \eqref{SLAgain}. In particular, the choice
\begin{equation}\label{eq:superpotential}
\begin{split}
	h = \mp m_\text{SUSY}\Bigg(&q_a q_a + (gL)^{3/2} g^{-1/2} A_a q_a + \frac{1}{2}(gL)^{2} B_{ab} q_a q_b \\
	&+ \frac{1}{2}(gL)^{5/2} g^{1/2} C_{abc} q_a q_b q_c + \frac{1}{2} (gL)^{3} g D_{abcd} q_a q_b q_c q_d + \mathcal{O}\left((gL)^{7/2}\right)\Bigg)
\end{split}
\end{equation}
in terms of the tensors defined in \eqref{eq:tensors} reproduces \eqref{SLAgain} exactly when $m = \pm m_\text{SUSY}$.

\subsection{Multiple adjoint flavors}\label{sec:flavors}

Arguments from lightcone quantization suggest that if we have 2D QCD with a massive adjoint fermion and massless fermions in arbitrary other irreps, the theory is supersymmetric when the adjoint mass is tuned to a value depending on the total Dynkin index of the (reducible) matter representation \cite{Popov:2022vud}. We can straightforwardly extend our calculation to check this in the case of multiple adjoint flavors, with only one flavor given a nonzero mass.

We start by making the replacement
\begin{align}
	\psi^T (  D_\tau -i  \gamma^5 D_x)  \psi + m \psi^T \gamma^0 \psi \quad\longrightarrow\quad \sum_{f=1}^{N_f}\left[\psi^{fT} (  D_\tau -i  \gamma^5 D_x)  \psi^f + m_f \psi^{fT} \gamma^0 \psi^f\right]
\end{align}
in the action \eqref{eq:SEuc}. When we carry out the perturbative calculation as in Section~\ref{sec:diagrams} but with this modified action, there will be an extra factor of $N_f$ for each diagram with an internal fermion loop. In particular, the first diagram in \eqref{eq:lo_pot_dia} will be multiplied by a factor of $N_f$, which means the leading-order effective potential for the holonomy is similarly rescaled, and thus the location $\vec{a}_*$ of its minimum is unchanged. In addition, for diagrams with external fermion lines, we have to sum over all the fermion flavors.

To order $(gL)^3$, these are the only changes we need to make, and so it is straightforward to modify \eqref{SLAgain} and find the following effective action:
\es{SLNf}{
	L_\text{eff} =   \frac 12 \dot{q}_a \dot{q}_a   +  &{}\frac{h^\vee N_f g^2}{4\pi} q_a q_a 
	+\sum_f\left[ \frac 12 \chi_a^{fT} \dot{\chi}^f_a  + \frac{m_f}{2}  \chi_a^{fT} \gamma^0 \chi^f_a\right]  \\
	+ (gL)^2 &{}\left[  \left( \frac{3h^\vee N_f g^2}{4 \pi} - \frac{\sum_fm_f^2}{2} \right) B_{ab} q_a q_b
	+  \sum_f \frac{m_f}{2} B_{ab}  \chi_a^{fT} \gamma^0 \chi^f_b \right] \\
	+ (gL)^{\frac 52} g^{1/2} &{}\left[  \left( \frac{h^\vee N_f g^2}{\pi} - \frac{\sum_fm_f^2}{2} \right) C_{abc} q_a q_b q_c
	+ \sum_f  \frac{3m_f}{2} C_{abc} q_c  \chi_a^{fT} \gamma^0 \chi^f_b \right] \\ 
	+ (gL)^3 g &{}\left[ \left( \frac{5 h^\vee N_f g^2}{4 \pi} - \frac{ \sum_f m_f^2}{2} \right) D_{abcd} q_a q_b q_c q_d + \sum_f 3m_f D_{abcd} q_c q_d \chi_a^{fT} \gamma^0 \chi_b^f  \right] + O((gL)^{7/2}) \,.
}
If we then set the masses to
\begin{align}
	m_1 = \pm m_\text{SUSY} \equiv \pm g\sqrt{\frac{N_f h^\vee}{2\pi}},\qquad m_{f}=0\qquad \text{for $f=2,\ldots,N_f$}\,,
\end{align}
we see that \eqref{SLNf} is the product of a supersymmetric theory for the holonomy and zero modes of the massive fermion with a free theory for the modes of the other $N_f-1$ fermion flavors.

\section{Discussion}
\label{sec:discussion}

In this work, we studied 2D gauge theory with an arbitrary simply-connected gauge group $G$ coupled to a Majorana fermion in the adjoint representation of $G$, quantized on a spatial circle of length $L$ with periodic boundary conditions. In the limit $gL\ll 1$, most modes of the gauge field and the fermion have energies of order $(gL)^{-1}$. We showed that there are $\text{rank}(G)$ remaining low-energy modes of the gauge field, corresponding to its holonomy around the circle, and also $\text{rank}(G)$ low-energy modes of the fermion. Furthermore, we have found that, when the fermion mass is set to $m = \pm m_\text{SUSY}$, with $m_\text{SUSY} = g\sqrt{\frac{h^\vee}{2\pi}}$, the effective quantum mechanics of these low-energy modes is supersymmetric. This result implies that the supersymmetry of the theory, previously demonstrated for infinite volume using the light-cone techniques, is also present on a circle with periodic boundary conditions. The presence of this extra ``accidental" symmetry of adjoint QCD$_2$ at $m= \pm m_\text{SUSY}$ makes it tempting to ask if the model could be integrable for these values of the mass. The connections between 2D large-$N$ gauge theories and integrability is a fascinating research direction \cite{Fateev:2009jf,Ambrosino:2023dik,Ambrosino:2024prz}, and we hope to return to this question in the future.

It would be interesting to generalize our small-circle perturbation theory to fermions in arbitrary representations of the gauge group. In \cite{Popov:2022vud}, it was found that 2D QCD with a massive adjoint fermion and massless fermions in other irreps has a supersymmetric point where the adjoint mass depends on the Dynkin index of the total (reducible) fermion representation. We expect this finding to also hold in the small-circle limit. More generally, it would be useful to determine the effective action for an arbitrary 2D gauge theory quantized on a small spatial circle.

It is worth contrasting the supersymmetry obtained in the 2D gauge theory with one adjoint Majorana fermion with the manifest $(1, 1)$ supersymmetry present in the Lagrangians of related models.  The first related case is the theory of $\dim G$ massless free fermions obtained by removing the gauge field from the massless adjoint QCD$_2$ theory.  It is known that in this theory one can construct supercurrents that are cubic in the fermions, in both the left-handed and right-handed sectors separately \cite{Antoniadis:1985az}.  Another related theory is the $(1, 1)$-supersymmetric gauge theory, sometimes called SQCD$_2$, that can be obtained from the circle reduction of ${\cal N} = 1$ SQCD in three dimensions \cite{Matsumura:1995kw}.  The matter content of the latter theory consists of a 3D gauge field and a Majorana gluino.  Upon reduction to two dimensions, the 3D gauge field gives rise to a 2D gauge field and a real adjoint scalar, while the 3D adjoint Majorana fermion gives rise to the 2D adjoint Majorana. The Lagrangian of SQCD$_2$ also includes a Yukawa coupling between the fermion and the scalar with a coefficient related to $g$. In order to preserve $(1, 1)$ supersymmetry, the adjoint fermion and scalar must be massless in this case. This theory was studied using light-cone quantization in \cite{Matsumura:1995kw,Antonuccio:1998jg} (see \cite{Lunin:1999ib} for a review) and, in contrast to the adjoint QCD$_2$, its spectrum appears to be gapless.  Interestingly, the light-cone supersymmetry of adjoint QCD$_2$ is a mix of the light-cone supersymmetries of the two other theories we just mentioned:  the supercharge $Q^+$ has the same form as in the free fermion theory \cite{Kutasov:1993gq}, while $Q^-$ has the same form as in SQCD$_2$ after setting the scalar field to zero \cite{Lunin:1999ib}. Studying SQCD$_2$ on a small circle could provide further insight into this model.

A remarkable feature of our quantum mechanical effective action (\ref{SLAgain}) is that, for $m=0$, the fermions do not appear in the Hamiltonian at all. Therefore, they encode the topological sector of the model. For $G=\grSU(\Nc)$, this gives rise to $2^{\Nc-1}$ degenerate vacua distributed among the $\Nc$ universes. In our small-circle expansion, these vacua appear directly as the Hilbert space generated by the $\Nc - 1$ zero modes of the fermion.
It would be interesting to study, following our Section~\ref{Comments}, the precise connection of these results with the non-invertible symmetries of adjoint QCD$_2$ uncovered in \cite{Komargodski:2020mxz}. Combining these perspectives on the symmetries of this model may lead to further insights, and we hope to develop this connection further in the future.

\section*{Acknowledgments}

We are grateful to Ofer Aharony for useful discussions.  This work was supported in part by the Simons Foundation Grant No.~917464 (Simons Collaboration on Confinement and QCD Strings), and by the US National Science Foundation under Grants No.~PHY-2111977 and PHY-2209997. RD was also supported in part by an NSF Graduate Research Fellowship and a Princeton University Charlotte Elizabeth Procter Fellowship. We thank the SwissMAP Research Station in Les Diablerets, where some of this work was carried out, for the warm hospitality. 

\appendix

\section{One-loop effective potential for holonomy}\label{app:1-loop_pot}
To evaluate the effective potential in \eqref{VeffSUN}, we need to integrate over the quantum fluctuations for the heavy fields while keeping the light holonomy mode fixed. At leading order, we can restrict the gauge field to have a constant value corresponding to a given holonomy $\Omega$, and then compute the partition function over a Euclidean time interval $\Delta \tau$ and extract the potential via
\begin{equation}\label{eq:ZtoV}
	Z[\Omega]=e^{-\Delta \tau\,  V_\text{eff}(\Omega)} \,.
\end{equation}
The constant gauge field $A^\Omega_\mu$ that corresponds to a non-trivial holonomy $\Omega$ is
 \es{BackgroundGauge}{
	A^\Omega_\mu = \frac{X}{L}\delta_{\mu x}\,, \qquad \text{with}\qquad P e^{i\int_0^L dx\, A^\Omega_x}=e^{iX}=\Omega\,.
 }
We decompose the gauge field as $A_\mu = A^\Omega_\mu +\delta A_\mu$, where $\delta A_\mu$ are the fluctuations to be integrated out.  For the fermion field $\psi$, we simply integrate over all heavy modes.  Furthermore, we can take the heavy fermion modes to be massless, since corrections proportional to the fermion mass are suppressed by $(gL)^2$.

To evaluate the partition function, we compute the path integral of the thoery \eqref{SLor} with a gauge-fixing prescription. For our purposes it will be convenient to work in the Feynman gauge with respect to the background field gauge, which amounts to adding a gauge fixing term and Faddeev-Popov ghosts to the Euclidean action \eqref{eq:SEuc}
 \es{eq:gaugeAction}{
	S = \frac{1}{2h^\vee}\int d\tau\, dx \, \tr_\text{adj} \biggl(&\frac{1}{2 g^2} F_{\mu\nu} F^{\mu\nu}- \frac{1}{g^2} D^\Omega_\mu \delta A_\nu D^{\Omega\mu} \delta A^\nu + \bar \psi \gamma_E^\mu D_\mu \psi \\
	 &{}+\bar c D^\Omega_\mu D^{\Omega \mu} c-i\bar cD^\Omega_\mu[\delta A^\mu,c]\biggr)  \,,
 }
with $D^\Omega_\mu \delta A^\mu= \partial_\mu \delta A^\mu -i [A^\Omega_\mu, \delta A^\mu]=0$ being the background covariant derivative, $c,\bar c$  the anti-commuting scalar ghosts transforming in the adjoint representation, and $\gamma^\mu_E$ the Euclidean gamma matrices.\footnote{We take $\mu = 1, 2$, and $\gamma^0 = \gamma^2_E$, $\gamma^1 = i \gamma^1_E$, such that $\{\gamma^\mu_E, \gamma^\nu_E\} = 2 \delta^{\mu\nu}$.  In \eqref{eq:gaugeAction}, $\bar{\psi} = \psi^T \gamma^2_E$. }

We then expand the Euclidean action \eqref{eq:gaugeAction} to second order in $\delta A_\mu$, $\psi$, and the ghost fields. The background gauge field \eqref{BackgroundGauge} is flat, $F^\Omega_{\mu\nu}=0$, so the full field strength is $F_{\mu\nu} = D^\Omega_\mu \delta A_\nu-D^\Omega _\nu \delta A_\mu-i[\delta A_\mu,\delta A_\nu]$, which implies that the only quadratic terms in the expanded action are
\begin{align}
	S = \frac{1}{2h^\vee}\int d\tau\, dx \, \tr_\text{adj} \left( \frac{1}{g^2} \delta A^\nu D^\Omega_\mu D^{\Omega \mu}\delta A_\nu + \bar \psi \gamma_E^\mu D^\Omega_\mu \psi +\bar c D^\Omega_\mu D^{\Omega \mu} c\right) +\text{higher order} \,,
\end{align}
In general, in $d$-dimensional QCD with one direction compactified into a circle of length $L$ with a holonomy $\Omega$ and a fermion in a real representation $R$, the path integral evaluates to  \cite{Gross:1980br}
\es{logZ}{
	-\log Z[\Omega] = \frac {d-2}2 \tr' \log \left(-D_\text{adj}^2 \right)
	- \frac 12 \tr' \log  \left( \slashed{D}_\text{R} \right)\,,
}
where $D_\text{adj}^2 = (\partial_\mu - i A_\mu^\Omega)^2$ is the background gauge-covariant Laplacian acting on a scalar field in the adjoint representation, $\slashed{D}_\text{R} = \slashed{\partial} - i \slashed{A}^\Omega$ is the background gauge-covariant Dirac operator acting on a spinor field in representation $R$, and $\tr'$ is the trace over only the heavy modes. In \eqref{logZ}, the first term comes from the gauge field and the ghosts, and the second term comes from the fermions.  When $d=2$ the first term vanishes, and because the background gauge field $A^\Omega_\mu$ has vanishing field strength, we have $\slashed{D}_R^2={\bf 1} D^2_R$, where $D_R^2$ is now the background gauge covariant Laplacian acting on a {\em scalar} field transforming in the representation $R$ and ${\bf 1}$ is the $2 \times 2$ identity matrix.  The logarithm of the partition function can then be simplified to
\es{logZAgain}{
	-\log Z[\Omega] = -\frac14 \log (-\slashed{D}_R^2)=- \frac 12  \tr' \log  \left( -D_R^2 \right) \,.
}
From \eqref{eq:ZtoV} we then find
\es{logZAgain2}{
	V_\text{eff}(\Omega) &= -\frac 12  \frac{1}{\Delta \tau} \tr' \log \left(- (\partial_\mu - i L^{-1}X \delta_{\mu x})^2 \right)\\
	&= -\frac 12  \frac{1}{\Delta \tau} \tr' \log [-\partial^2_\tau-(\partial_x-iL^{-1}X)^2]\,. 
}
The eigenvalues corresponding to the zero-modes of $D_x=\partial_x-iL^{-1}X$ are independent of $X$, and so we can replace $\tr'$ with $\tr$ by adding a constant shift to the potential. We then find
\es{trLog}{
	V_\text{eff}(\Omega) &= \frac{1}{\Delta \tau} \tr \log \left(- (\partial_\mu - i L^{-1} X \delta_{\mu x})^2 \right) \\
	&= \int \frac{d\omega}{2 \pi} \sum_{n \in \Z} \tr_R\log \left(\omega^2 + L^{-2}(2 \pi n - X)^2 \right) \\
	&=-\frac{2}{\pi L}\Re\sum_{n=1}^\infty \frac{1}{n^2}\tr_R(e^{inX})\,,
}
which coincides with \eqref{VeffSUN} when choosing $R$ to be the adjoint representation. 

\section{Feynman diagrams}\label{app:feynman}
The computation of the effective action for the zero modes is mostly a straightforward application of the Feynman rules described in Section~\ref{sec:feyn_rules} to evaluate the diagrams in Table~\ref{tab:diagrams}. However, a few diagrams requires extra care. In this appendix, we give more detailed evaluations of some representative diagrams.

\subsection{Tadpole Diagram}\label{sec:tadpole_diagram}

Many of the loop integrals appearing in the diagrams in Table~\ref{tab:diagrams} are ill-defined until we use the point-splitting procedure \eqref{eq:point_splitting}. One example appears in
\begin{align}\label{eq:LO_linear}
	\begin{tikzpicture}[vertical align]
		\draw [very thick, snake it, color=bleudefrance] (-1,0) -- (0,0);
		\draw[very thick] (1,0) arc (0:360:0.5);
	\end{tikzpicture}=\frac{1}{\sqrt{gL}} \sum_{\vec\alpha>0}(\bar{q}_a(\omega=0)\alpha_a)\sum_n\int\frac{d\omega'}{2\pi} e^{i\epsilon\sgn(\vec\alpha)\omega'}\tr [\gamma^5 G_{\vec{a}_* \cdot \vec \alpha}(\omega',n)]\,,
\end{align}
where the factor of $e^{i\epsilon \sgn(\vec{\alpha})\omega'}$ follows from \eqref{eq:point_splitting}.
We have
\begin{equation}
	\tr\left\lbrack \gamma^5 G_{\vec{a}_*\cdot\vec{\alpha}}(\omega', n)\right\rbrack = \frac{2gL(2\pi n - \vec{a}_*\cdot\vec{\alpha})}{(gL \omega')^2 + (2\pi n-\vec{a}_*\cdot\vec{\alpha})^2 + (mL)^2}\,,
\end{equation}
and thanks to the point-splitting the integral over $\omega'$ converges and is given by
\begin{equation}
	\int \frac{d\omega'}{2\pi} e^{i\epsilon\sgn(\vec\alpha)\omega'}\tr [\gamma^5 G_{\vec{a}_* \cdot \vec \alpha}(\omega',n)] = (2\pi n - \vec{a}_* \cdot \vec \alpha)\frac{e^{-\epsilon E_n/gL}}{E_n}\,,
\end{equation}
where $E_n\equiv \sqrt{(2\pi n- \vec{a}_*\cdot\vec{\alpha})^2+(mL)^2}$. After performing the sum over $n$, we can take the regulator $\epsilon$ to 0 and find
\begin{equation}
	\lim_{\epsilon\to 0} \sum_{n=-\infty}^{\infty} (2\pi n - \vec{a}_* \cdot \vec \alpha)\frac{e^{-\epsilon E_n/gL}}{E_n} = -1 + 2\left\lbrace \frac{\vec{a}_* \cdot\vec{\alpha}}{2\pi}\right\rbrace + \mathcal{O}\left((gL)^2\right)\,.
\end{equation}
Thus,
\begin{equation}
	\begin{tikzpicture}[vertical align]
		\draw [very thick, snake it, color=bleudefrance] (-1,0) -- (0,0);
		\draw[very thick] (1,0) arc (0:360:0.5);
	\end{tikzpicture}=\frac{1}{\sqrt{gL}} \bar{q}_a(\omega = 0) \sum_{\vec{\alpha}>0} \left\lbrack\alpha_a \left(2\left\lbrace \frac{\vec{a}_* \cdot\vec{\alpha}}{2\pi}\right\rbrace - 1\right)\right\rbrack + \mathcal{O}\left((gL)^{3/2}\right)\,.
\end{equation}
Finally, using \eqref{VMin}, we find that this diagram vanishes to leading order.

\subsection{Leading $q^2$ diagram}\label{sec:propagator_correction_diagram}

Another diagram whose evaluation is somewhat subtle is the leading quadratic term
\begin{align}\label{eq:Sqq}
	\begin{tikzpicture}[vertical align]
		\draw [very thick, snake it, color=bleudefrance] (-1,0) -- (0,0);
		\draw [very thick, snake it, color=bleudefrance] (1,0) -- (2,0);
		\draw[very thick] (1,0) arc (0:360:0.5);
	\end{tikzpicture}=\frac{1}{2gL}&\int \frac{d\omega}{2\pi}\sum_{\vec\alpha>0}(\alpha_a \bar{q}_a(\omega))(\alpha_b \bar{q}_b(-\omega))\\
	&\quad \times\sum_n \int \frac{d\omega'}{2\pi}e^{i\epsilon\sgn(\vec\alpha)(\omega+2\omega')}\tr[\gamma^5 G_{\vec a \cdot \vec \alpha}(\omega+\omega',n)\gamma^5 G_{\vec a \cdot \vec \alpha}(\omega',n)]\nonumber
\end{align}
To evaluate it, we first perform the integral over $\omega'$ and expand to leading order in $gL$:
\begin{align}
	\int \frac{d\omega'}{2\pi}e^{i\epsilon\sgn(\vec\alpha)(\omega+2\omega')}\tr[\gamma^5 G_{u}(\omega+\omega',n)\gamma^5 G_{u}(\omega',n)]&= \frac{2\sin \epsilon\omega}{\omega} e^{-\frac{2\epsilon|2\pi n - u|}{gL}}+\mathcal{O}((gL)^3)\,.
\end{align}
After summing over $n$, we can take the regulator $\epsilon$ to 0 and obtain a finite limit:
\begin{align}
	\lim_{\epsilon\to 0}\frac{2\sin \epsilon\omega}{\omega}\sum_n e^{-\frac{2\epsilon|2\pi n - u|}{gL}} = \frac{gL}{\pi}\,.
\end{align}
Using this result, and evaluating the integral over $\omega$ by returning to position space, we find a contribution
\begin{align}
	&\begin{tikzpicture}[vertical align]
		\draw [very thick, snake it, color=bleudefrance] (-1,0) -- (0,0);
		\draw [very thick, snake it, color=bleudefrance] (1,0) -- (2,0);
		\draw[very thick] (1,0) arc (0:360:0.5);
	\end{tikzpicture}= \frac{1}{2\pi}\int d\tau\sum_{\vec\alpha>0}(\alpha_a \bar{q}_a(\tau))^2 +\mathcal{O}((gL)^2)
\end{align}
to the effective action.

\subsection{Two-loop Diagram}\label{sec:two_loop_diagram}
To illustrate the calculation of a two-loop diagram, we take a representative example:
\begin{align}
	&\begin{tikzpicture}[vertical align]
		\draw [very thick, snake it, color=bleudefrance] (-1,0) -- (0,0);
		\draw[very thick,snake it] (.5,0) circle (0.5);
		\filldraw[very thick,fill=white,draw=black] (1,0) circle (0.25);
	\end{tikzpicture} = \frac{1}{\sqrt{gL}}\bar{q}_a(\omega=0)\cdot \sum_{\vec \alpha>0}\alpha_a\int \frac{d\omega}{2\pi}\frac{d\omega'}{2\pi}e^{i\epsilon \sgn(\vec \alpha)\omega}\sum_{n}(2\pi n-\vec a\cdot \vec \alpha)D_{\vec a\cdot \alpha}(\omega,n)^2\\
	&\times\tr\Bigg[2\vec\alpha^2\sum_{n'\neq 0}G_0(\omega',n')G_{\vec a\cdot \vec \alpha}(\omega+\omega',n+n')\nonumber\\
	&\qquad\qquad+\sum_{n'}\sum_{\vec \alpha'\neq-\vec \alpha}N_{\vec \alpha,\vec \alpha'}^{\vec \alpha+\vec \alpha'} N_{-\vec \rt, \vec \rt + \vec \rt'}^{\vec \rt'}G_{\vec a \cdot \vec \alpha'}(\omega',n')G_{\vec a \cdot \vec \alpha'}(\omega+\omega',n+n')\Bigg]\, .\nonumber
\end{align}
In the first term on the second line, one of the fermion lines carries a mode in a Cartan direction, while in the second term on the second line both fermion lines carry modes in root directions. The point-splitting factor $e^{i\epsilon \sgn(\vec \alpha)\omega}$ is important when carrying out the frequency integrals using residues. After performing the integrals, the diagram takes the form
\begin{align}
	\begin{tikzpicture}[vertical align]
		\draw [very thick, snake it, color=bleudefrance] (-1,0) -- (0,0);
		\draw[very thick,snake it] (.5,0) circle (0.5);
		\filldraw[very thick,fill=white,draw=black] (1,0) circle (0.25);
	\end{tikzpicture} &= - \frac{(gL)^{\frac 32}}{2}\int d\tau\, \sum_{\vec \alpha >0} (\bar{q}_a(\tau)\alpha_a) \tilde C_{\vec{a}_*\cdot \vec \alpha} + \mathcal O((gL)^{\frac 52})\, ,
\end{align}
where we have defined
\begin{align}\label{eq:sum}
	\tilde C_{\vec{\hol} \cdot \vec{\rt}} = 2\vec \alpha^2 C_{\vec{\hol} \cdot \vec{\rt}, 0} 
	+   \sum_{\vec \rt' \neq - \vec{\rt}} N_{\vec \rt, \vec \rt'}^{\vec \rt + \vec \rt'} N_{-\vec \rt, \vec \rt + \vec \rt'}^{\vec \rt'} C_{\vec{\hol} \cdot \vec{\rt}, \vec{\hol} \cdot \vec{\rt}'} 
\end{align}
and
\begin{align}
	C_{Q,Q'} =  \sum_{n, n'} \frac{ \sgn (2 \pi n' - Q') \sgn (2 \pi (n + n') - (Q + Q'))-1}{ (2 \pi n - Q)^3} \,.
\end{align}
The sign functions appearing here arise from the $\epsilon\to 0$ limit in the form $\lim_{\epsilon\to 0}\frac{x}{\sqrt{x^2+\epsilon^2}} = \sgn x$, so we should take $\sgn 0 = 0$. In Appendix \ref{app:sum} we show that
\begin{align}\label{eq:sum_res}
	\tilde C_{\vec{\hol} \cdot \vec{\rt}}  = -\frac{h^\vee}{\pi}\sum_n\frac{\sgn(2 \pi n - \vec{\hol} \cdot \vec{\rt})}{\abs{2 \pi n - \vec{\hol} \cdot \vec{\rt}}^2}.
\end{align}
Thus, putting everything together, we find
\begin{align}\label{eq:2loopRef}
	\begin{tikzpicture}[vertical align]
		\draw [very thick, snake it, color=bleudefrance] (-1,0) -- (0,0);
		\draw[very thick,snake it] (.5,0) circle (0.5);
		\filldraw[very thick,fill=white,draw=black] (1,0) circle (0.25);
	\end{tikzpicture} &= \frac{(gL)^{\frac 32}h^\vee}{2 \pi}\int d\tau\, \bar{q}_a(\tau)\cdot\sum_{\vec \alpha >0} \alpha_a \sum_n \frac{\sgn(2 \pi n - \vec{\hol} \cdot \vec{\rt})}{\abs{2 \pi n - \vec{\hol} \cdot \vec{\rt}}^2} +\mathcal O((gL)^{7/2})\, .
\end{align}

\section{A group theory sum}\label{app:sum}
In this appendix, we will carry out the sum needed in \eqref{eq:sum}. We start by recalling a few facts about Lie algebra representations.

In the adjoint representation, we choose the normalization
\begin{align}
	\braket{H^a|H^b}=\delta^{ab},\qquad \braket{E^{\vec\alpha}|E^{-\vec\alpha'}}=\delta^{\vec\alpha\vec\alpha'}\,,
\end{align}
with $\braket{X|Y}= \frac{1}{h^\vee}\tr_\text{adj}(XY)$ denoting the Killing form. As pointed out in Footnote~\ref{FDFootnote}, the inner product between the leading order potential minimum $\vec a_*$ and any root $\vec \alpha$ is bounded by
\begin{align}\label{eq:dot_bounds}
	-2\pi < \vec a\cdot \vec\alpha <2\pi.
\end{align}
Lastly, the quadratic Casimir in adjoint representation is
\begin{align}
	C_2=\sum_a H^a H^a + \sum_{\vec\alpha>0}(E^{\vec\alpha} E^{-\vec\alpha}+E^{-\vec\alpha}E^{\vec\alpha}) = h^\vee \mathds{1}\, .
\end{align}

\subsection{Doing the sums}
To evaluate \eqref{eq:sum}, we need to do the sums
\begin{align}
	S_1 &= 2\vec\alpha^2\sum_{n'}\left[\sgn(2\pi n')\sgn(2\pi(n+n')-\vec{a}_*\cdot\vec\alpha)-1\right],\\
	S_2 &= \sum_{\vec\alpha'\neq -\vec\alpha}\sum_{n'} N_{\vec\alpha,\vec\alpha'}^{\vec\alpha+\vec\alpha'}N^{\vec\alpha'}_{-\vec\alpha,\vec\alpha+\vec\alpha'}\left[\sgn(2\pi n'-\vec{a}_*\cdot \vec\alpha')\sgn(2\pi (n+n')-\vec{a}_*\cdot (\vec\alpha+\vec\alpha'))-1\right]\,,\nonumber
\end{align}
where $\vec \alpha$ is held constant.
The first sum $S_1$ is can done using
\begin{align}
	\sum_n[1-\sgn(n)\sgn(n-x)] = (2\lfloor x\rfloor +1)\sgn x \,,
\end{align}
with $\sgn 0=0$ and $x\notin \mathbb Z$. This leads to
\begin{align}
	S_1 &= 2\vec\alpha^2\left(2\left\lfloor\frac{\vec{a}_*\cdot\vec\alpha}{2\pi}\right\rfloor-2n+1\right)\sgn(2\pi n-\vec{a}_*\cdot \vec\alpha)\nonumber\\
	&=2\vec\alpha^2(2\theta(\vec\alpha>0)-2n-1)\sgn(2\pi n-\vec{a}_*\cdot \vec\alpha) \,,
\end{align}
where used the bounds \eqref{eq:dot_bounds} on $\vec{a}_*\cdot \vec\alpha$ to swap the floor function for an indicator function on positive roots.

For the second sum $S_2$, we can also immediately do the $n'$ sum using
\begin{align}
	\sum_n [1-\sgn (n-x)\sgn (n-y)]=2(\lfloor x \rfloor -\lfloor y\rfloor)\sgn (x-y) \,,
\end{align}
where $x,y\notin \mathbb Z$, which yields
\begin{align}
	S_2 &=2\sum_{\vec\alpha'\neq -\vec\alpha}N_{\vec\alpha,\vec\alpha'}^{\vec\alpha+\vec\alpha'}N^{\vec\alpha'}_{-\vec\alpha,\vec\alpha+\vec\alpha'}(\theta(\vec\alpha'>0) -\theta(\vec\alpha+\vec\alpha'>0)+n)\sgn(\vec{a}_*\cdot \vec\alpha-2\pi n) \,.
\end{align}
To proceed, we need to unpack the product of structure constants 
\begin{align}
	N_{\vec\alpha,\vec\alpha'}^{\vec\alpha+\vec\alpha'}N^{\vec\alpha'}_{-\vec\alpha,\vec\alpha+\vec\alpha'}
	&=\braket{E^{-\vec\alpha}|E^{-\vec\alpha'}E^{\vec\alpha'}|E^{\vec\alpha}}\,,
\end{align}
where $E^\alpha$ is understood to act in the adjoint. This resembles the Casimir and can indeed be related to it, but takes a bit of manipulations of each of the three terms in $S_2$.

The last term in $S_2$ only depends on $\vec{\alpha}'$ through the structure constants, thus we need to solve
\begin{align}
	\sum_{\vec\alpha'\neq -\vec\alpha}N_{\vec\alpha,\vec\alpha'}^{\vec\alpha+\vec\alpha'}N^{\vec\alpha'}_{-\vec\alpha,\vec\alpha+\vec\alpha'} &= \sum_{\vec\alpha'\neq -\vec\alpha}\braket{E^{-\vec\alpha}|E^{-\vec\alpha'}E^{\vec\alpha'}|E^{\vec\alpha}}=\sum_{\vec\alpha'}\braket{E^{-\vec\alpha}|E^{-\vec\alpha'}E^{\vec\alpha'}|E^{\vec\alpha}}-\vec\alpha^2\nonumber\\
	&= \braket{E^{-\vec\alpha}|\sum_iH^iH^i+\sum_{\vec\alpha'>0}[E^{-\vec\alpha'}E^{\vec\alpha'}+E^{\vec\alpha'}E^{-\vec\alpha'}]| E^{\vec\alpha}}-{2}\vec\alpha^2\\
	&=\braket{E^{-\vec\alpha}|C_2| E^{\vec\alpha}}-2\vec{\alpha}^2=h^\vee-2\vec\alpha^2\,.
\end{align}
The first two terms of $S_2$ can be put into the form of a commutator. To do this we first need to rewrite the two terms as follows
\begin{align}
	\sum_{\vec\alpha'\neq -\vec\alpha}N_{\vec\alpha,\vec\alpha'}^{\vec\alpha+\vec\alpha'}N^{\vec\alpha'}_{-\vec\alpha,\vec\alpha+\vec\alpha'}\theta(\vec\alpha'>0) &=\sum_{0<\vec\alpha'\neq-\vec\alpha} \braket{E^{-\vec\alpha'}|E^{-\vec\alpha}E^{\vec\alpha}|E^{\vec\alpha'}}\\
	&=\sum_{\vec\alpha'>0} \braket{E^{-\vec\alpha'}|E^{-\vec\alpha}E^{\vec\alpha}|E^{\vec\alpha'}}-\vec\alpha^2\theta(\vec\alpha<0)\,,\nonumber
\end{align}
and
\begin{align}
	\sum_{\vec\alpha'\neq -\vec\alpha}N_{\vec\alpha,\vec\alpha'}^{\vec\alpha+\vec\alpha'}N^{\vec\alpha'}_{-\vec\alpha,\vec\alpha+\vec\alpha'}\theta(\vec\alpha+\vec\alpha'>0)&=\sum_{\vec \alpha',\vec\alpha''>0} \braket{E^{-\vec\alpha}|E^{-\vec\alpha'}|E^{\vec\alpha''}}\braket{E^{-\vec\alpha''}|E^{\vec\alpha'}|E^{\vec\alpha}}\\
	&=\sum_{\vec \alpha'\vec\alpha''>0} \braket{E^{-\vec\alpha''}|E^{\vec\alpha}|E^{\vec\alpha'}}\braket{E^{-\vec\alpha'}|E^{-\vec\alpha}|E^{\vec\alpha''}}\nonumber\\
	&=\sum_{\vec\alpha''>0} \left(\braket{E^{-\vec\alpha''}|E^{\vec\alpha}E^{-\vec\alpha}|E^{\vec\alpha''}}-\sum_i \braket{E^{-\vec\alpha''}|E^{\vec\alpha}|H^{i}}\braket{H^{i}|E^{-\vec\alpha}|E^{\vec\alpha''}}\right)\nonumber\\
	&=\sum_{\vec\alpha''>0} \braket{E^{-\vec\alpha''}|E^{\vec\alpha}E^{-\vec\alpha}|E^{\vec\alpha''}}-\theta(\vec\alpha>0)\vec \alpha^2\,.\nonumber
\end{align}
Their difference can now be evaluated as a commutator
\begin{align}
	\sum_{\vec\alpha'\neq -\vec\alpha}N_{\vec\alpha,\vec\alpha'}^{\vec\alpha+\vec\alpha'}N^{\vec\alpha'}_{-\vec\alpha,\vec\alpha+\vec\alpha'}[\theta(\vec\alpha'>0)-\theta(\vec\alpha+\vec\alpha'>0)]&=\sum_{\vec\alpha'>0} \braket{E^{-\vec\alpha'}|[E^{-\vec\alpha},E^{\vec\alpha}]|E^{\vec\alpha'}}-\vec\alpha^2+2\theta(\vec\alpha>0)\vec \alpha^2\nonumber\\
	&=-2\vec\rho \cdot \vec\alpha-\vec\alpha^2+2\theta(\vec\alpha>0)\vec \alpha^2\,.
\end{align}
Finally, we can add up $S_1$ and $S_2$ and arrive at the final result
\begin{align}
	S_1+S_2&=2\left[\vec\alpha^2(2\theta(\vec\alpha>0)-2n-1)-(-2\vec\rho \cdot \vec\alpha-\vec\alpha^2+2\theta(\vec\alpha>0)\vec \alpha^2)-n(h^\vee-2\vec\alpha^2)\right]\sgn(2\pi n-\vec{a}_*\cdot \vec\alpha)\nonumber\\
	&=-\frac{h^\vee}{\pi}\left[2\pi n -\frac{4\pi \vec\rho\cdot \vec \alpha}{h^\vee}\right]\sgn(2\pi n-\vec{a}_*\cdot \vec\alpha) = -\frac{h^\vee}{\pi}|2\pi n-\vec{a}_*\cdot \vec\alpha|\, ,
\end{align}
which leads to the result quoted in \eqref{eq:sum_res}.

\bibliographystyle{ssg}
\bibliography{SmallCircleDraft}

\end{document}